\documentclass[final,3p,times,twoside]{elsarticle}
\usepackage{graphicx}
\usepackage{subcaption}
\usepackage{geometry}
\usepackage{dsfont}
\usepackage{amsmath}
\usepackage{cancel}	
\usepackage{amssymb}
\usepackage{amsbsy}
\usepackage{amsthm}
\usepackage{amsfonts}
\usepackage{latexsym}
\usepackage{bm}
\usepackage{verbatim}
\usepackage{mathrsfs}
\usepackage{graphicx}
\usepackage{color}
\usepackage{amsmath} 
\usepackage{wrapfig}
\usepackage{fmtcount}
\usepackage{color}
\usepackage{booktabs}
\usepackage{lscape}
\usepackage{mathrsfs}
\usepackage[ruled,vlined]{algorithm2e}
\usepackage{amsmath}
\usepackage{float} 
\usepackage{lipsum}
\usepackage[normalem]{ulem}
\usepackage{bm}
\usepackage{hyperref} 
\journal{Computer Methods in Applied Mechanics and Engineering}

\definecolor{red}{rgb}{1,0,0}								
\definecolor{blue}{rgb}{0,0,1}							

\begin{document}

\begin{frontmatter}

\title{CM-GAI: Continuum Mechanistic Generative Artificial Intelligence Theory for Data Dynamics}
\author[MECH,COMP]{Shan Tang\corref{cor}}
\author[MECH]{Ziwei Cao}
\author[MECH]{Zhenling Yang}
\author[NORTH]{Jiachen Guo}
\author[MECH]{Yicheng Lu}
\author[DALIAN,NORTH]{Wing Kam Liu\corref{cor}}
\author[MECH,COMP]{Xu Guo\corref{cor}}

\cortext[cor]{Corresponding authors.
\\Email addresses: shantang@dlut.edu.cn;w-liu@northwestern.edu; guoxu@dlut.edu.cn}

\address[MECH]{State Key Laboratory of Structure Optimization and CAE Software, Department of Engineering Mechanics, Dalian University of Technology, Dalian, 116023, PR China}
\address[COMP]{International Research Center for Computational Mechanics, Dalian University of Technology, PR China}
\address[DALIAN]{Co-founder of HIDENN-AI, LLC}
\address[NORTH]{Department of Mechanical Engineering, Northwestern University, Evanston, IL, USA}
\begin{abstract}
Generative artificial intelligence (GAI) plays a fundamental role in high-impact AI-based systems such as SORA and AlphaFold. Currently, GAI shows limited capability in the specialized domains due to data scarcity. In this paper, we develop a continuum mechanics-based theoretical framework to generalize the optimal transport theory from pure mathematics, which can be used to describe the dynamics of data, realizing the generative tasks with a small amount of data. The developed theory is used to solve three typical problem involved in many mechanical designs and engineering applications: at material level, how to generate the stress-strain response outside the range of experimental conditions based on experimentally measured stress-strain data; at structure level, how to generate the temperature-dependent stress fields under the thermal loading; at system level, how to generate the plastic strain fields under transient dynamic loading. Our results show the proposed theory can complete the generation successfully, showing its potential to solve many difficult problems involved in engineering applications, not limited to mechanics problems, such as image generation. The present work shows that mechanics can provide new tools for computer science. The limitation of the proposed theory is also discussed.
\end{abstract}
\begin{keyword}
Data-driven; Continuum mechanics; Probability space; Generative artificial intelligence; Optimal transport
\end{keyword}
\end{frontmatter}

\section{Introduction}
Generative artificial intelligence (GAI) is profoundly transforming numerous domains of human society, a development driven by the synergistic evolution of several core generative models. These models exhibit significant differences in theoretical foundations, architectural characteristics, and application scenarios, collectively advancing the field of AI-generated content (AIGC).

Auto-regressive (AR) models, represented by Generative Pre-trained Transformers (GPT) and various other large language models (LLMs), are fundamentally based on the concept of treating data as sequences and predicting subsequent elements sequentially based on the existing context \citep{vaswani2017attention,brown2020language}. These models demonstrate exceptional performance in natural language processing (NLP) tasks, which is capable of generating fluent text sequences. However, their sequential generation mechanism leads to lower computational efficiency and presents limitations in modeling long-range dependencies \citep{vaswani2017attention}. In terms of data requirements, AR models rely on large-scale, high-quality text corpora for pre-training \citep{brown2020language}. Furthermore, as they generate content entirely based on statistical patterns learned from training data, they may produce outputs containing factual inaccuracies or physically implausible content \citep{vaswani2017attention}. Generative Adversarial Networks (GANs) learn to generate data through an adversarial game between a generator and a discriminator. The generator strives to produce realistic samples capable of deceiving the discriminator, while the discriminator aims to accurately distinguish between real and generated samples \citep{goodfellow2014generative}. GANs have shown remarkable success in image generation, producing samples with high visual fidelity \citep{kynkaanniemi2019improved}. Nevertheless, their training process is often unstable and susceptible to mode collapse, and they typically require substantial amounts of training data \citep{goodfellow2014generative,kynkaanniemi2019improved}. Additionally, GANs generally exhibit weak physical interpretability, as their generated results often lack explicit physical constraints \citep{yang2020physics}.
Diffusion models, inspired by non-equilibrium thermodynamics, learn the data distribution through a process of progressively adding and removing noise \citep{ho2020denoising}. These methods have rapidly emerged as a leading paradigm in image generation, praised for their high sample quality, diversity, and training stability, therefore has been extended to other fields \citep{shi2025diffusion,xiao2025geometric}. A primary limitation is their slow inference speed due to the iterative denoising process, and they similarly depend on large-scale, high-quality training datasets \citep{ho2020denoising,dhariwal2021diffusion}.
Variational Autoencoders (VAEs) learn a low-dimensional latent representation of data based on an encoder-decoder architecture. Their advantages include stable training, an interpretable latent space that supports smooth interpolation, and facilitated controllable generation \citep{kingma2013auto}. A main limitation is that the quality of generated samples is often inferior to that of GANs and diffusion models. VAEs require less amount of training data compared to other models \citep{kingma2019introduction}. In terms of physical interpretability, their structured latent space can be advantageous for embedding and interpreting physical laws. Flow-based models utilize invertible transformations to enable exact likelihood computation, possessing theoretical advantages for probability density estimation. Although they impose strict constraints on network architecture design, they can provide accurate probability density estimates \citep{dinh2016density}. Their data requirements are similar to those of VAEs, making them suitable for medium-scale datasets \citep{kingma2013auto,dinh2016density}. Regarding physical plausibility, their capacity for exact likelihood computation holds significant value for physical modeling and uncertainty quantification.

With generative models as the core architecture and integrating physical laws, data-driven approaches, and domain knowledge, researchers are gradually reshaping the research paradigm of computational mechanics. These methods not only inherit the powerful data representation and distribution mapping capabilities of generative models, but also, through ingenious physics-informed guidance and structural design, apply them to a series of challenging problems in mechanics—ranging from efficient forward dynamic simulations, end-to-end prediction from micro-structures to macroscopic responses, multi-fidelity simulation fusion, reduced-order modeling of nonlinear dynamical systems, to solving uncertain inverse problems. Below, we will illustrate through several representative works how these five types of generative models are implemented in specific applications within the field of mechanics.

Liao et al. \citep{2025Mechanics} proposed a Mechanics-Informed Transformer‑GCN (MI‑TGCN), which embeds physical mechanisms into the deep learning architecture: they incorporated the principle of modal superposition into the multi-head attention mechanism of the Transformer, and replaced the adjacency matrix in the graph convolutional network with the stiffness matrix, enabling the model to naturally satisfy deformation compatibility conditions during training. This method ensures accuracy while improving computational efficiency by an order of magnitude compared to traditional finite element methods, demonstrating the advantages of physics-informed generative architectures in forward dynamic simulations. Yang et al. \citep{2021End} utilized conditional generative adversarial networks to achieve an end-to-end mapping from composite micro-structures to full-field strain and stress tensors. Their approach not only delivered high prediction accuracy, but also validated its physical consistency through superposition principles and mixture rules, highlighting the strong representational power of generative models in predicting mechanical responses of complex structures. Shi et al. \citep{shi2025diffusion} introduced a diffusion-based surrogate model and proposed an innovative multi-fidelity simulation calibration framework. This framework uses low-cost simulations as conditional inputs to the generative process, while transforming high-cost simulations into physics-informed signals during the sampling stage. As a result, it effectively integrates deterministic physical laws with statistical patterns in the data, supported by only a small amount of high-fidelity data. This method significantly enhanced the spatial detail authenticity and temporal evolution coherence of predictions in fluid buoyancy field reconstruction and metal additive manufacturing thermal process prediction, while effectively correcting systematic biases in traditional simulation models. Lopez and Atzberger \citep{2020Variational} proposed a variational autoencoder (VAE)-based dynamical learning framework for reduced-order modeling of nonlinear physical systems. By encoding the system states into a latent manifold space with specific geometries (such as a torus), this method explicitly embeds physical constraints and topological priors, effectively improving the stability and generalization capability of the model in long-term predictions. Experiments on the Burgers equation and constrained mechanical systems showed that this approach can more accurately capture nonlinear evolution features compared to traditional linear reduction methods and enhance model interpretability. Invertible neural networks, as a special type of flow-based model, provide a complete Bayesian posterior estimation tool for inverse problems—such as inferring system parameters or loads from observed responses—through bidirectional reversible transformations and the introduction of latent variables. The work of Ardizzone et al. \citep{2018Analyzing} demonstrated that INNs can implicitly learn inverse mappings via forward training from parameters to observations, while using latent variables to retain information lost during the forward process. This enables the generation of multimodal, highly correlated posterior distributions of parameters during inverse inference.

In general, the training of these generative models is typically complicated, involving complex neural network architectures with billions of parameters. This is inherently challenging due to the curse of dimensionality \citep{bengio2003} and results in extremely high computational costs for online prediction. On the other hand, although they require massive data, it should be noted that for many standard problems—such as text and image generation tasks, where training data can be readily obtained from the internet, books, and news—or for certain standard scenarios in scientific computing where data may come from public simulation databases, benchmark experiments, or well-documented material property libraries, data acquisition is not a major bottleneck. However, in real-world engineering applications, this is often not the case, especially under novel or extreme conditions—such as extremely high or low temperatures, high pressure, or high loading rates. Under such circumstances, not only is experimental testing difficult, costly, or even infeasible to conduct, but finite element simulations also frequently face significant challenges, including long computation times, difficulties in achieving convergence, or an inability to accurately model the physical behavior. Moreover, even when experiments or simulations are possible, the obtained data may still be incomplete. Experimental data can suffer from gaps due to limitations in measurement equipment, while simulation results are often constrained by computational costs and model simplifications, making it difficult to provide complete, high-fidelity datasets. Together, these factors lead to a severe shortage of data available for training models. Moreover, the proposed generative models do not consider the underlying physics of the engineering application as mentioned above. For example, most available generative models can potentially generate low-quality images/videos that disobey the physical laws. Consequently, researchers in the field of computer science recently proposed the concept of a world model \citep{ma2025building}, trying to incorporate the underlying physics into the generative task.

\begin{figure}[htbp]
	\centering
       \includegraphics[width=0.6\textwidth]{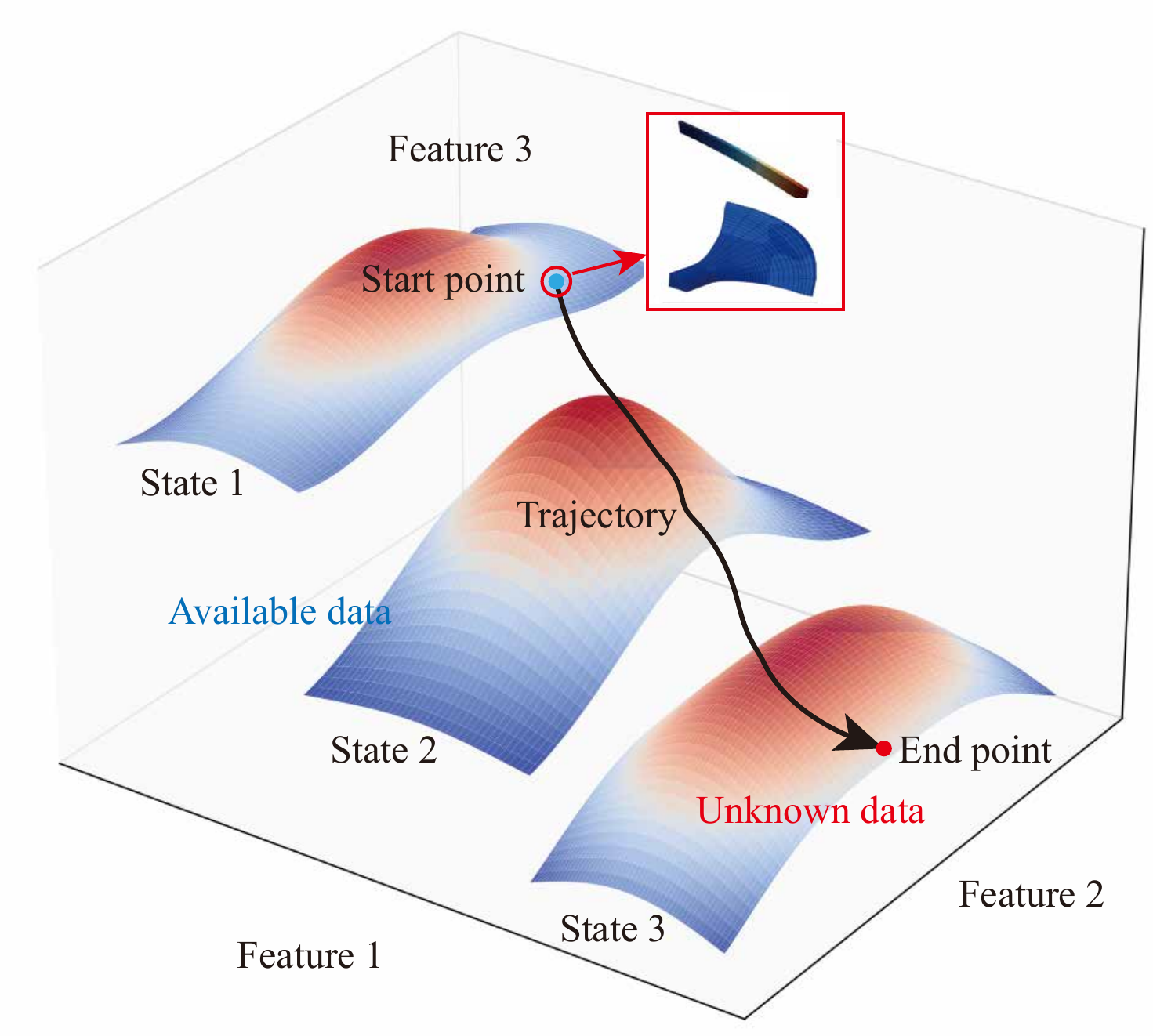}
	\setlength{\abovecaptionskip}{1mm}
	\setlength{\belowcaptionskip}{-2mm}
	\caption{Conceptual framework of CM-GAI in unknown data generation. The general problem for generating the relationship between features under the specified state when the data under the other states are known.}
	\label{cmgai_diagram}
\end{figure}

GAI can be used to generate mechanics-related physical fields, which need to be predicted under the special condition based on the easily obtained data under real service conditions in the physical feature space (Figure \ref{cmgai_diagram}). Evidently, the easily available data can serve as the foundation for simulation modeling and design specifications. However, under the special conditions with the extreme environments, complex loads, and dynamic loading, the traditional physics-based experimental and numerical methods often face bottlenecks in efficiency, cost, or feasibility. As discussed before, directly applying generative methods such as LLMs and diffusion to solve the mechanics-related physical fields generation problem is challenging due to data scarcity. Nevertheless, the core idea of GPT — learning the joint probability distributions of word sequences — can still be leveraged. If measurements or calculations at the working conditions can be treated as a stochastic process, the observed relevant mechanics-related physical data may follow a probability distribution. By learning this distribution from limited but trustworthy data, we could potentially generate relevant mechanics-related physical fields at target conditions. Here, time-dependent optimal transport theory \citep{Villani2008, gu2021optimal}, as a way to describe transforming a probability distribution with minimal effort and finally being able to generate probability distributions of data, may serve as a powerful tool for predicting such distributions, especially in high-dimensional feature space of the data.

On the other hand, continuum mechanics, including mechanics of materials \citep{Timoshenko1and2}, fluid mechanics \citep{fluidLandau}, etc., is closely related to the physics of the deformation and kinematics of solids and fluids. Since kinematics in continuum mechanics is closely related to optimal transport theory, the natural question is whether the continuum mechanics theory can be used to generalize optimal transport.

In this paper, we propose a continuum mechanics–based generative method that learns unknown probability distributions by establishing mappings between conditional data distributions via optimal transport, thereby generating data at the target condition. The objective of the proposed generative model is consistent with denoising probabilistic diffusion models \citep{ddpm} and flow matching models \citep{flowmatching}, that is, to approximate the target distribution. Unlike previous methods, the proposed approach integrates continuum mechanics with optimal transport constraints in the flow of probability distribution. This integration enhances the physical consistency, particularly under conditions of data scarcity. It should be emphasized that our interest is in merging the merits of both optimal transport and continuum mechanics. As a result, mathematically grounded optimal transport can be generalized on the basis of physical understanding, making the axiomatic-abstract theory easier to understand and apply. Based on our development, the mathematical optimal transport theory can be generalized from different perspectives in mechanics and be more powerful. With the proposed physics-informed theory, the training for generative tasks only needs a small amount of data, and offer a potential solution for the world model of a generative task.
\section{Theory}
In this section, we will first review the theory of optimal transport. Then, we will develop the continuum mechanics perspective on optimal transport. Generally, the principle of optimal transport states that, for transporting one probability distribution to another, the solution minimizes the total transportation cost \citep{Villani2003}. This concept is analogous to the principle of least action in continuum mechanics \citep{berdichevsky2009principle}. 
\subsection{Optimal transport}
Let us first review the theory of optimal transport. Optimal transport was first formulated by the French mathematician Monge at the end of the 18th century \citep{monge1781}. Monge's problem was rediscovered by the Russian mathematician, Kantorovich (laureate of Nobel Memorial Prize in Economic Sciences) \citep{kantorovich}. By the end of 1980s, optimal transport theory has been extensively studied and applied to solve many problems. For further details on the theoretical foundations and numerical implementation methods, readers may refer to the books from the viewpoints of pure mathematics \citep{Villani2003, Villani2008}, applied mathematics \citep{santambrogio2015} or numerical computation \citep{peyre2020}.
\begin{figure}[htbp]
	\centering
	\begin{minipage}[t]{0.45\textwidth}
		\centering
		\includegraphics[width=\textwidth]{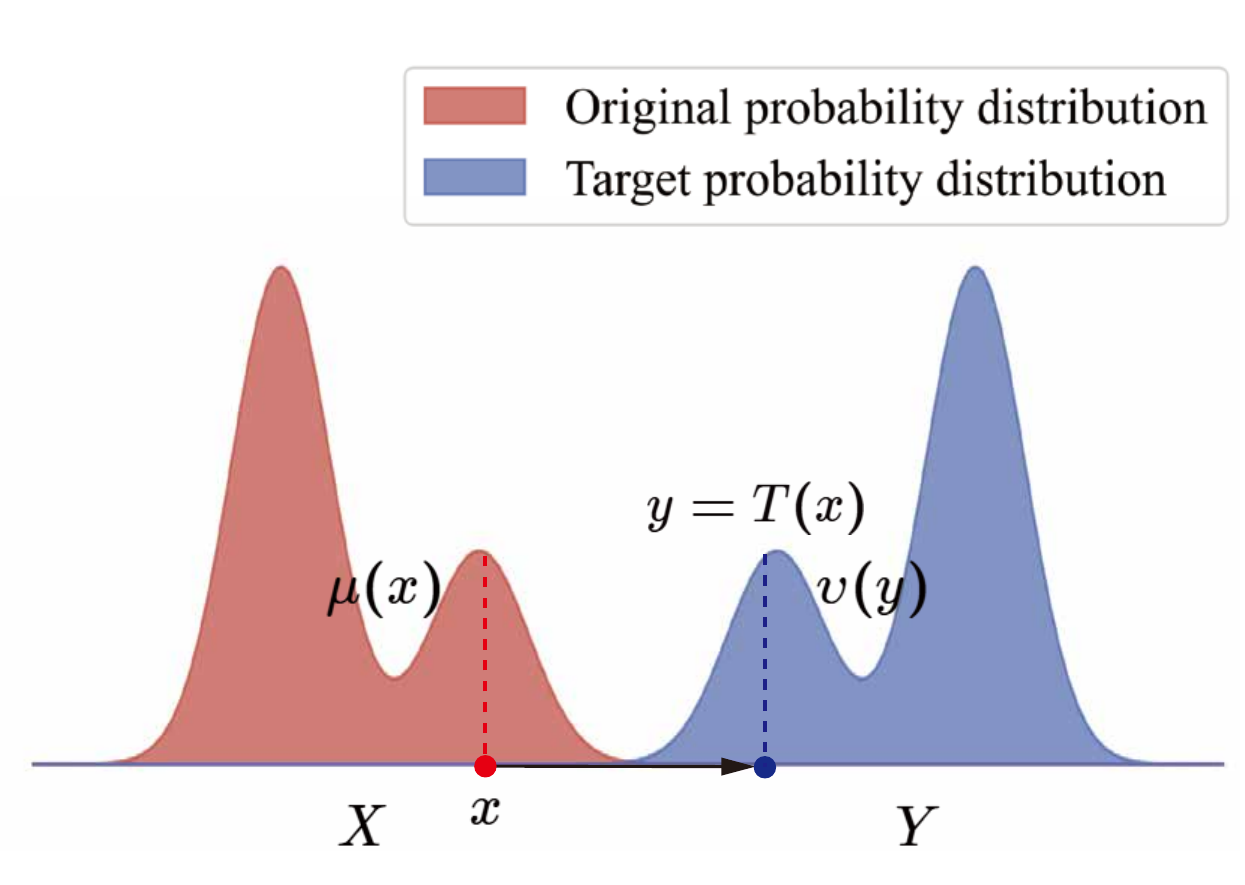}
		\setlength{\abovecaptionskip}{1mm}
		\setlength{\belowcaptionskip}{-2mm}
		\caption{Time-independent optimal transport. The original probability distribution in the space $X$ and the target probability distribution in the space $Y$ are colored by red and blue, respectively.}
		\label{fig:Fig 2}
	\end{minipage}
	\hfill 
	\begin{minipage}[t]{0.45\textwidth}
		\centering
		\includegraphics[width=\textwidth]{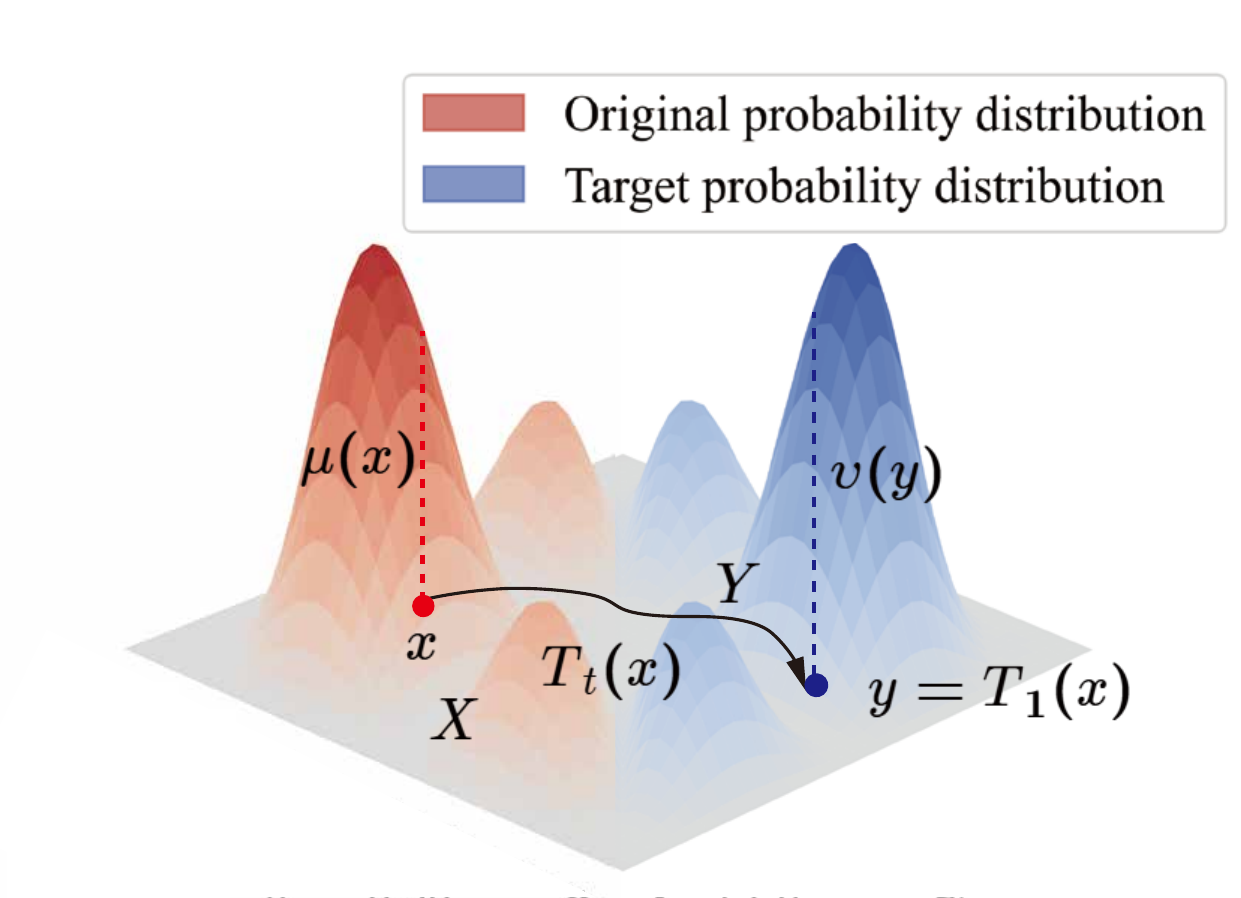}
		\setlength{\abovecaptionskip}{1mm}
		\setlength{\belowcaptionskip}{-2mm}
		\caption{Time-dependent optimal transport. The mapping $T$ is a function of time $t$, which can move the mass from $t=0$ to $t=1$ along the trajectory to the target.}
		\label{fig:Fig 3}
	\end{minipage}
\end{figure}

We adopt the terminology and mathematical notation of Villani \citep{Villani2003}. The transportation problem can be visualized as moving a pile of sand (with a given distribution) to fill a hole (with a different target distribution). As illustrated in Figure \ref{fig:Fig 2}, the original and target distributions are represented by red and blue coloring, respectively. Let us define two probability spaces $X$ for the sand pile and $Y$ for the hole with the probability measures $\mu$ and $\nu$, respectively. In the following sections, we assume these probability spaces are continuous and differentiable, consistent with the framework of continuum mechanics theory. However, the mathematical theory of optimal transport is more general and mathematically rigorous. The transportation process is characterized by a transport map $y=T\left(x\right)$, which means a unit mass at location $x$ is relocated to position $y$. This mapping $T$ effectively performs a change from measure $\mu$ to $\nu$. The movement of a unit mass from $x$ to $y$ incurs a transportation cost, quantified by a cost function $c\left(x, y\right)$. The cost function typically takes the following form:
\begin{equation}
	c\left(x,y\right) =\left\vert x-y\right\vert^{2}
	\label{costfunction}
\end{equation}%
The power index $2$ often exhibits a good mathematical property. The physical meaning is also clear, representing the squared distance between $x$ and $y$. The optimal transport problem can then be formulated as Monge's optimal transport problem:
\begin{equation}
\inf_{T} \int_{X}c\left( x,T\left(x\right)\right) d\mu\left(x\right) 
\label{Monge}
\end{equation}%
That is, minimizing transportation cost on all sets of measurable maps $T$. The mapping $T$ must satisfy $T\# \mu = \nu$. Here, $\#$ represents the push-forward operator. This measure mapping ($T\#$) can be understood as follows. For the continuous probability distribution, 
\begin{equation}
d\mu=f\left(x\right)dx \qquad d\nu=g\left(y\right)dy
\label{continous distribution}
\end{equation}%
Under the transport map,
\begin{equation}
	dy=\det \left( \frac{dy}{dx}\right) dx
	\label{Jacob transformation}
\end{equation}%
Substituting into Eq. \ref{continous distribution} and equating the measures, the $\mu$-almost surely
\begin{equation}
	f\left(x\right)=g(T\left(x\right))\det \left( \frac{dy}{dx}\right)
	\label{continous distribution f and g}
\end{equation}%

\begin{figure}[H]
	\centering
       \includegraphics[width=0.6\textwidth]{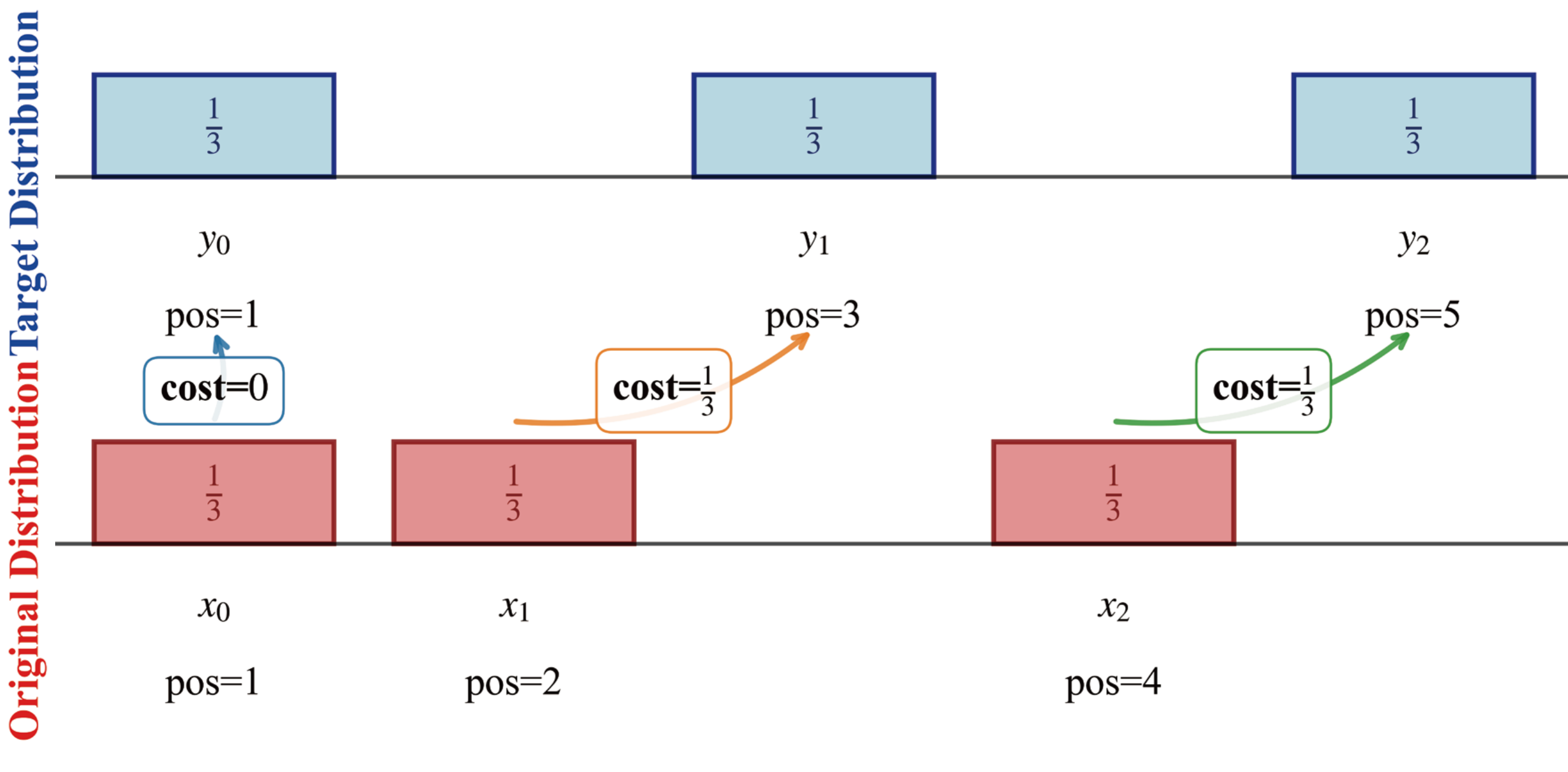}
	\setlength{\abovecaptionskip}{1mm}
	\setlength{\belowcaptionskip}{-2mm}
	\caption{A simple example for time-independent optimal transport.}
	\label{simple example for time-independent}
\end{figure}

To help readers understand the concept of time-independent optimal transport, we then present a simple discrete example. For pedagogical clarity, we consider two discrete uniform probability distributions, denoted as \(\mu\) (original) and \(\nu\) (target), noting that the theory mentioned above is continuous one. The original distribution \(\mu\) is defined over positions \(\{1, 2, 4\}\), and the target distribution \(\nu\) over positions \(\{1, 3, 5\}\), with each point carrying a probability mass of \(1/3\), as shown in Figure \ref{simple example for time-independent}. The objective is to find a transport map \(T\) that transforms \(\mu\) into \(\nu\) while minimizing the total transportation cost, where the unit cost is defined as the squared Euclidean distance as mentioned above. The total cost for a candidate map \(T\) is given by:

\begin{equation}
\sum_{i} c(x_i, T(x_i)) \cdot \mu(x_i) = \sum_{i} |x_i - T(x_i)|^2 \cdot \mu(x_i).
\end{equation}

The map \(T\) must satisfy the push-forward condition \(T_\#\mu = \nu\), which ensures that the probability mass of \(\mu\) is fully redistributed to form \(\nu\) under \(T\). By enumerating all six possible bijective mappings satisfying \(T_\#\mu = \nu\) and evaluating their respective costs, we determine that the optimal transport map is defined by \(T(x_0) = y_0\), \(T(x_1) = y_1\), and \(T(x_2) = y_2\). Specifically, this maps the original point at position 1 to the target at 1, the point at 2 to 3, and the point at 4 to 5, yielding a total cost of \(2/3\), which is the minimum among all permutations. In summary, this optimal transport map achieves the transformation from the original to the target distribution with minimal cost, illustrating the core principles of time-independent optimal transport.

In fact, Monge's minimization problem can be relaxed to the Kantorovich optimal transport problem, which always has a solution, whereas Monge's problem may not. Specifically, the Monge problem requires moving "mass" like moving piles of sand, piece by piece, seeking an optimal one-to-one map for which a perfect solution might not even exist. The Kantorovich problem, however, allows the mass to be "split" and distributed "probabilistically" to different destinations, seeking an optimal transport plan; this more flexible approach always guarantees a solution. The interested reader can refer to Villani's classical book for more details \citep{Villani2003}. We retain Monge's formulation because, as will be seen shortly, it connects directly with continuum mechanics.

The above discussion concerns the time-independent optimal transport problem. That is, the cost function for transporting a unit mass from $x$ to $y$ depends solely on the initial and final positions ($x$ and $y$), without considering the transportation path history. In fact, for generative artificial intelligence applications, the time-dependent transportation formulation is more appropriate. This will become clear in the subsequent numerical examples.

For time-dependent transportation, the optimal transport problem must be formulated in terms of trajectories. Consider a trajectory illustrated in Figure \ref{fig:Fig 3}, beginning at time $t=0$ and ending at $t=1$. The time-dependent transport map ($T_{t}$) can be expressed as $z\left(t\right)=T_{t}\left(x\right)$ where $z\left(0\right)=T_{0}(x)=\text{Id}(x)=x$ and $z\left(1\right)=T_{1}\left(x\right)=y$. Id is the identity tensor, which maps itself to itself and therefore can be used for the initial boundary condition.

The total transportation cost should be accumulated along the trajectory. We first compute the incremental cost from time $t$ to $t+\Delta t$, where the transportation occurs between positions $z\left(t\right)$ and $z\left(t+\Delta t\right)$. This cost is given by:
\begin{equation}
c\left( z\left( t\right) ,z\left( t+\Delta t\right) \right) \approx \frac{%
	\left\vert z\left( t\right) -z\left( t+\Delta t\right) \right\vert ^{2}}{%
	\Delta t}=\left\vert \frac{dz}{dt}\right\vert ^{2}\Delta t
\end{equation}%
The total transportation cost along the trajectory can be expressed as:
\begin{equation}
C(T_t(x))=\sum_{trajectory}\left\vert \frac{dz}{dt}\right\vert ^{2}\Delta
t=\int_{0}^{1}\left\vert \frac{dz}{dt}\right\vert ^{2}dt
\end{equation}%
The time-dependent minimization problem can be formulated as:
\begin{equation}
\inf_{T_{t}}\int_{X}C(T_t(x))d\mu\left(x\right)\qquad T_{0}=\text{Id\qquad }T_{1}\#\mu =\nu 
\label{time-dependent transportation}
\end{equation}%
The optimal transport problem requires minimizing this integral over all trajectories generated by the time-dependent transport map $T_{t}$. In this framework, the quantity ${dz}/{dt}$ is referred to as the differential cost in optimal transport theory.

\begin{figure}[H]
	\centering
       \includegraphics[width=0.95\textwidth]{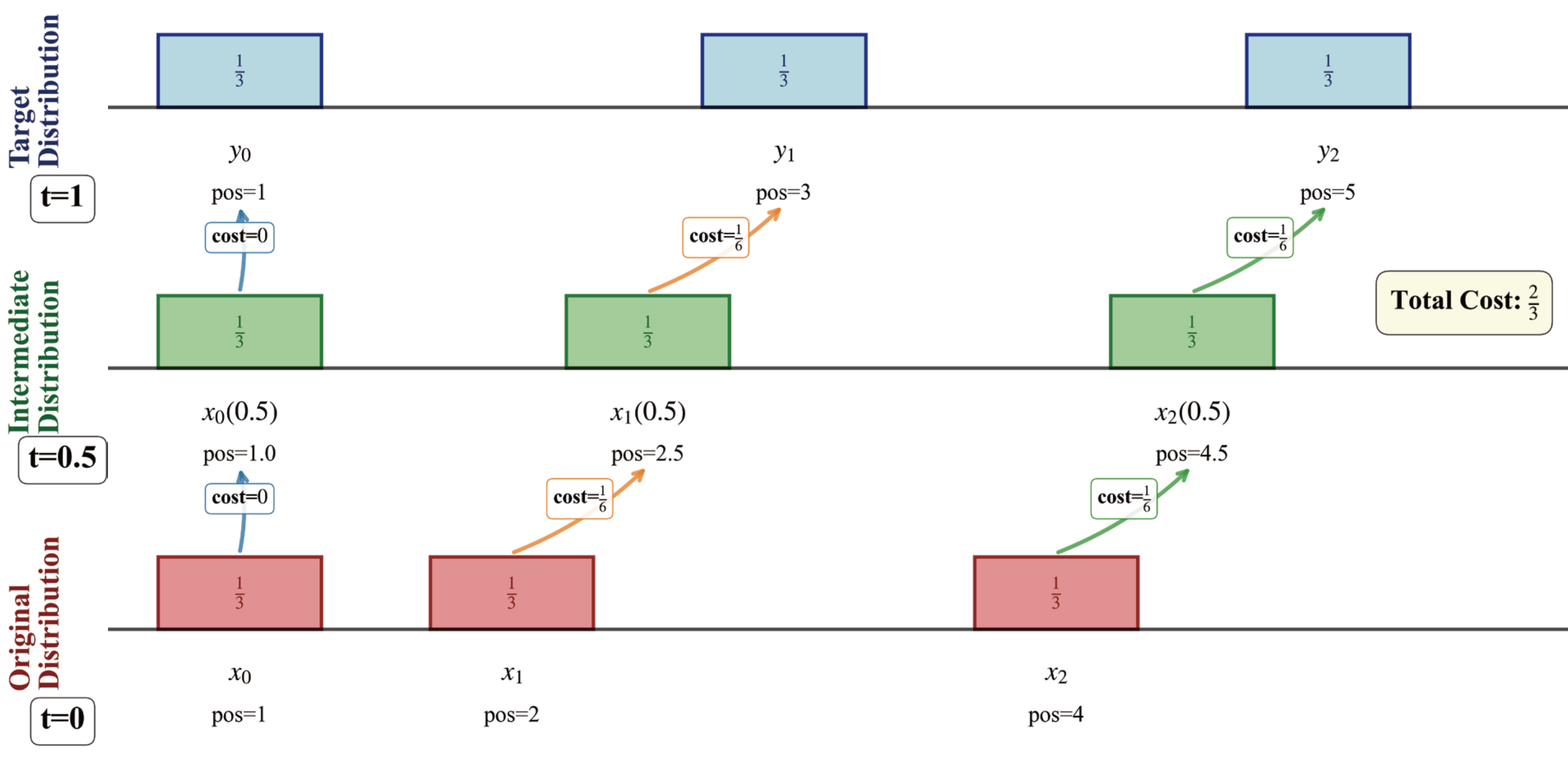}
	\setlength{\abovecaptionskip}{1mm}
	\setlength{\belowcaptionskip}{-2mm}
	\caption{A simple example for time-dependent optimal transport.}
	\label{simple example for time-dependent}
\end{figure}

To help readers understand the concept of time-dependent optimal transport, we present a simple discrete example that illustrates the dynamic evolution of probability distributions. We consider an original distribution $\mu$ at time t=0 defined over positions {1, 2, 4} and a target distribution $\nu$ at time t=1 over positions {1, 3, 5}, with each point carrying a probability mass of 1/3, as shown in Figure \ref{simple example for time-dependent}. The objective is to find a time-dependent transport mapping that transforms $\mu$ into $\nu$ while minimizing the total transportation cost:

\begin{equation}
\sum_i \int_0^1 \left|\frac{dz_i(t)}{dt}\right|^2 dt \cdot \mu(x_i).
\end{equation}

The mapping must satisfy the push-forward condition $T_0 = \text{Id}$ and $T_1{\#}\mu = \nu$ at the initial and final time moment. Through optimization of the continuous paths, we obtain the optimal transport: point \(x_0\) remains stationary at position 1, point \(x_1\) moves linearly from 2 to 3, and point \(x_2\) moves linearly from 4 to 5. This yields a total cost of \(2/3\), demonstrating that the optimal time-dependent transport achieves the distribution transformation while minimizing the total cost, illustrating the fundamental principles of time-dependent optimal transport.

\subsection{High-dimension continuum mechanics}
In optimal transport theory, the process involves either mass transfer from $x$ to $y$ or flow along trajectories in probability spaces. This bears a strong resemblance to the deformation of physical bodies (solids or fluids) as described by continuum mechanics. In this section, we employ continuum mechanics theory \citep{marsden1994mathematical} to characterize the transportation process. \textit{It is crucial to emphasize that transportation typically involves large deformations, necessitating formulation within the finite deformation framework of continuum mechanics.} (Finite deformation problems can be solved using the Arbitrary Lagrangian-Eulerian method.\citep{belytschko2014nonlinear}) We first present a terminology correspondence table between optimal transport theory and continuum mechanics theory. Readers may consult this table throughout to understand the conceptual relationships between these theories better.

\begin{figure}[htbp]
	\centering
	\begin{minipage}[t]{0.35\textwidth}
		\centering
		\includegraphics[width=\textwidth]{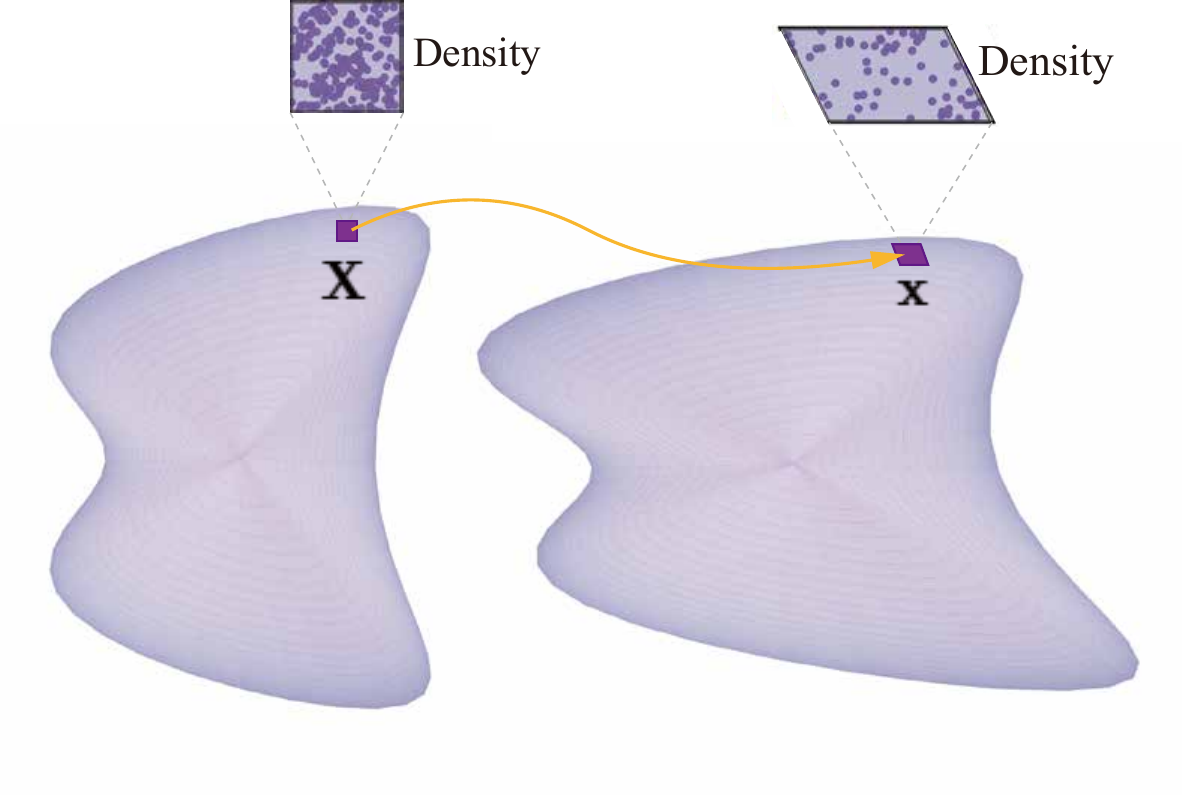}
		\caption{When deformation occurs, a material point initially located at $\mathbf{X}$ in the reference configuration moves to a new position $\mathbf{x}$ in the current configuration. Consequently, the density of the material point at $\mathbf{X}$ in the reference configuration differs from its density at $\mathbf{x}$ in the deformed configuration.}
		\label{fig:Fig 4}
	\end{minipage}
	\hfill 
	\begin{minipage}[t]{0.58\textwidth}
		\centering
		\includegraphics[width=\textwidth]{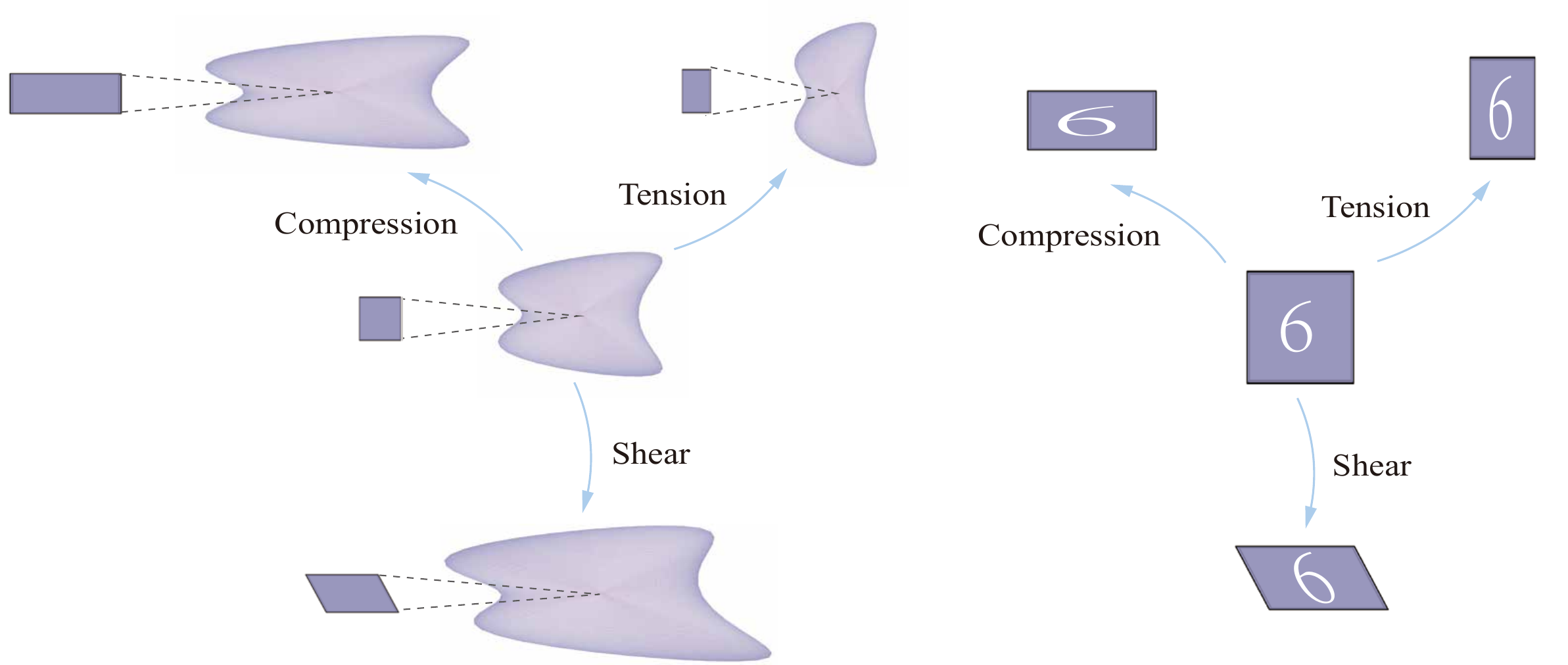}
		\caption{By analogy with continuum mechanics, mass transport may induce deformations in the probability measure. Fundamental deformation modes — such as tension, compression, and shear — can analogously manifest in this framework. When generating handwritten digit images using a generative model, the standard handwritten digit images are processed through deformations to obtain different handwritten digit images.}
		\label{fig:Fig 5}
	\end{minipage}
\end{figure}

\begin{table}[H]
	\centering
	\footnotesize	
	\caption{Translation of terminology}
	\begin{tabular}{p{0.4\textwidth}p{0.5\textwidth}} 
		\toprule
		\textbf{Optimal transport terminology} & \textbf{Continuum
			mechanics terminology} \\
		\midrule
$X$ probability space & $\Omega_{0} $ continuum body at time $t=0$ \\ 
                      & $\Omega_{t} $ continuum body at time $t$ \\ 
$Y$ probability space & $\Omega_{1} $ continuum body at time $t=1$ \\ 
$x$ (random variable in $X$ space) & $\mathbf{X}$ the coordinate at the reference configuration ($t=0$) \\ 
 & $\mathbf{x}(t)$  the coordinate at the current configuration \\ 
$y$ (random variable in $Y$ space) & $\mathbf{x}$ the coordinate at the current configuration ($t=1$)  \\ 
- & $\mathbf{u}$ displacement $\left( \mathbf{x}(t)-\mathbf{X}\right) $ \\ 
${dz}/{dt}$ (differential cost) & $\mathbf{\dot{x}}$ velocity of the
particle \\ 
- & $\mathbf{F}_{b}$ external force \\ 
$f\left( x\right)$ the probability distribution in the $X$ space & $\rho \left( \mathbf{X}\right) $ the material density
at the reference configuration at time $t=0$ \\ 
& $\rho \left( \mathbf{x,}t\right) $ the material density at the current
configuration at time $t$ \\ 
$g(y)$ the probability distribution in the $Y$ space & $\rho \left( \mathbf{x,}1\right) $ the material density at the
current configuration at time $t=1$ \\ 
- & $W$ the strain energy of unit mass \\ 
$\frac{1}{2}\left({dz}/{dt}\right) ^{2}$ transportation cost & $K_{0}=%
\frac{1}{2}\mathbf{\dot{x}}\mathbf{\dot{x}}$ the kinematic energy of unit mass\\
		\bottomrule
	\end{tabular}
\end{table}
Because the probability space can have dimension $N>3$, let us consider an $N$-dimensional physical body $\Omega_{0}$ at time $t=0$ and formulate it within the framework of high-dimensional continuum mechanics theory. The visualization of a conventional three-dimensional physical body from traditional continuum mechanics theory can provide an intuitive understanding. When transportation occurs, the point  $\mathbf{X}$ in the configuration at $t=0$ moves to point $\mathbf{x}$ in the configuration at time $t$ (see Figure \ref{fig:Fig 4}). This continuum body at time $t$ is denoted as $\Omega_{t}$. In optimal transport theory, there is no need to distinguish between reference and current configurations (typically, the configuration at $t=0$ is chosen as the reference configuration and that at $t$ as the current configuration). However, we find that defining the original (reference) configuration following continuum mechanics theory makes it easier to understand optimal transport theory.

At material point $\mathbf{X}$ in the reference configuration of the continuum, there exists a material density $\rho\left(\mathbf{X}\right)$, which depends solely on the original position $\mathbf{X}$ and not on time. When transportation occurs, the material density in the current configuration becomes $\rho\left(\mathbf{x},t\right)$, now a function of both the current position $\mathbf{x}$ and $t$. In fact, the key concept to understand the optimal transport theory is in analogy the density function of the material in the continuum mechanics theory with the probability distribution function in the probability space (denoted by $f\left(x\right)$ and $g\left(y\right)$ in the previous section).

The movement of mass points involves deformation (excluding rigid displacement). Such material deformation is primarily analyzed in continuum mechanics theory, including tension, compression, and shear (see Figure \ref{fig:Fig 5}). While the strain energy changes induced by this deformation are not considered as costs in optimal transport theory, incorporating the concept of probability space deformation can enhance the theory's capabilities for generative tasks. To describe deformation in probability space, analogous to continuum mechanics theory, we may introduce the displacement of the $N$-dimensional physical body at point $\mathbf{X}$.

\begin{equation}
	u_i=x_i-X_i
	\label{displacement}
\end{equation}%
where $i$ represents the $i$-th component, ranging from $1$ to $N$. Then the deformation gradient can be defined:
\begin{equation}
	F_{ij}=\frac{\partial x_{i}}{\partial X_{j}}=\delta_{ij}+\frac{\partial u_{i}}{\partial X_{j}}
	\label{deformation gradient}
\end{equation}%
where $\delta_{ij}$ is the Kronecker delta. Note that this high-dimensional deformation gradient is not confined to three-dimensional space but exists in $N$-dimensional space. The symmetric left Cauchy-Green tensor can then be introduced:
\begin{equation}
	C_{ij}=F_{ik}F_{jk}
	\label{Cauchy-Green}
\end{equation}%
Even in high-dimensional space, the deformation energy of a unit mass material can still be defined through constitutive modeling. The constitutive models for different kinds of material are systematically studied by the mechanician, such as Truesdell, and Erigen etc., triggering an important research direction in the past half century. Recently, data-driven approaches for building constitutive models that incorporate machine learning techniques have also developed rapidly \citep{map123, map123ep, deeplearningConstitutive}. For illustration purposes, we assume the physical body (or probability space) consists of a hyper-elastic material. Analogous to the widely used neo-Hookean model \citep{abaqus}, the deformation energy density can be similarly defined:
\begin{equation}
	W=\frac{1}{2}G\left[\text{Tr}\left[C_{ij}\right]-N\right] 
	\label{W-neoHooke}
\end{equation}%
where $G$ is the shear modulus and $\text{Tr}$ represents the trace operator of the matrix. 

In optimal transport theory, the differential cost is introduced. Direct comparison shows that $1/2$ times the squared differential cost corresponds exactly to the kinetic energy of a unit mass in continuum mechanics theory:
\begin{equation}
	K_{0}=\frac{1}{2}{\frac{\partial x_i}{\partial t}\frac{\partial x_i}{\partial t}}=\frac{1}{2} \frac{\partial u_{i}\left(%
		\mathbf{X},t\right) }{\partial t}\frac{\partial u_{i}\left(%
		\mathbf{X},t\right) }{\partial t}
	\label{Kinematic-energy}
\end{equation}%

The work done by external forces must also be considered. Consequently, we introduce the body force $F_{B}^{i}$, although this term is not typically included in optimal transport theory. The justification for incorporating external forces will be provided in the following section, as it is closely related to the geometry of the data space. 

Now, we can define the total kinematic energy $K$, the total strain energy $U$, and the potential energy of the body force $A$ as:
\begin{equation}
K=\int_{\Omega _{0}}K_{0}\rho _{0}\left( \mathbf{X}%
\right) dV
\end{equation}%
\begin{equation}
U=\int_{\Omega _{0}}W\rho _{0}\left( \mathbf{X}\right) dV	
\end{equation}%
\begin{equation}
\delta A=-\int_{\Omega _{0}}F_{b}^{i}\delta u^{i}\rho _{0}\left( \mathbf{X}%
\right) dV
\end{equation}%
respectively. $\delta$ is the variational symbol. Note that all the volume integration is carried out in the reference configuration. This can bring great convenience for later derivation. The Hamilton principle can be applied to give the solution for the optimal transport by defining the Lagrangian $L=U-K+A$:
\begin{equation}
\delta \int_{0}^{1}Ldt=\delta \int_{0}^{1}(U-K+A)dt=0
\label{Hamilton}
\end{equation}
where $\delta$ represents the variation. In fact, the time-dependent optimal transport problem (Eq. \ref{time-dependent transportation}) can be solved by the same variational approach \citep{fengYC}. Derived in the Appendix A, the variations lead to:
\begin{equation}
	\frac{\partial ^{2}u_{k}\left( \mathbf{X},t\right) }{\partial t^{2}}-G\frac{%
		\partial ^{2}u_{k}\left( \mathbf{X},t\right)}{\partial X_{i}^{2}}-F_{b}^{k}=0
\label{equilirium}
\end{equation}
In addition to the variation equations (Eq. \ref{Hamilton}), a constraint must be imposed for the mass conservation:
\begin{equation}
\rho_0\left(\mathbf{X}\right)dV=\rho\left(\mathbf{x},t\right)dv	
\label{mass-conservation-solid}
\end{equation}
where $dV$ and $dv$ represent the differential volumes in the reference and current configurations, respectively. This equation is mathematically equivalent to the probability conservation law given by Eq. (\ref{continous distribution f and g}). In fact, it is also called the mass conservation in the optimal transport theory. It is well known in continuum mechanics,
\begin{equation}
	JdV=dv\qquad J=\text{det}\left({F_{ij}}\right)
	\label{volume relation}
\end{equation}
where $\text{det}$ denotes the determinant of the tensor. This formulation can be generalized to high-dimensional spaces \citep{moderngeometryII}. By substituting Eq. \ref{volume relation} into Eq. \ref{mass-conservation-solid}, the constraint equation simplifies to:
\begin{equation}
\rho_0\left(\mathbf{X}\right)=\rho\left(\mathbf{x},t\right)\text{det}\left({F_{ij}}\right)
\label{Monge-Ampere Equation-Equi}
\end{equation}
In optimal transport theory, mass conservation is typically expressed as:
\begin{equation}
\bigtriangledown \cdot \left( \rho\left(\mathbf{x},t\right) \mathbf{v}\right) \mathbf{+}\frac{%
	\partial\rho\left(\mathbf{x},t\right) }{\partial t}=0
\label{mass-conservation-fluid}
\end{equation}
This is a well-known conservation of mass equation in continuum mechanics. The equation is expressed in the current configuration using $\bigtriangledown$, which denotes the gradient operator with respect to the current coordinates. In fact, this equation is equivalent to Eq. \ref{mass-conservation-solid}. However, it can be seen that Eq. \ref{mass-conservation-solid} is simpler than Eq. \ref{mass-conservation-fluid}. In terms of the numerical implementation, Eq. \ref{mass-conservation-fluid} is much more complicated due to the convective term involved, see \citep{NFECS}.
Substituting Eq. \ref{deformation gradient} into Eq. \ref{Monge-Ampere Equation-Equi}, it leads to: 
\begin{equation}
	\rho_0\left(\mathbf{X}\right)=\rho\left(\mathbf{x},t\right)\text{det}\left(\frac{\partial x_{i}}{\partial X_{j}}\right)
	\label{Monge-Ampere Equation-Equi-new}
\end{equation}
If the deformation is curl-free, then the displacement field can be expressed as the gradient of a potential function:
\begin{equation}
	x_i=\frac{\partial \psi\left(\mathbf{X},t\right)}{\partial X_{i}}
	\label{potential function}
\end{equation}
Eq. \ref{Monge-Ampere Equation-Equi-new} becomes:
\begin{equation}
   \rho_0\left(\mathbf{X}\right)=\rho\left(\mathbf{x},t\right)\text{det}\left(\frac{\partial^2 \psi\left(\mathbf{X},t\right)}{\partial X_{i} \partial X_{j}}\right)
\end{equation}
This leads to the famous Monge-Ampère equations, which have been extensively studied in mathematics. However, in our formulation, the displacement field $\bm{u}$ serves as the primary unknown variable, and we do not assume the existence of a potential function for the displacement.
Brenier's polar factorization theorem \citep{brenier1991polar} establishes that any nondegenerate vector-valued mapping can be expressed as the gradient of a convex potential function. This is a fundamental mathematical result applicable to high-dimensional spaces. For three-dimensional problems—such as void growth in plastic solids—velocity potential functions are commonly employed to determine deformation fields \citep{rice1969, gurson1977}.
\subsection{Non-Euclidean space}

\begin{figure}[htbp]
	\centering
	\begin{minipage}[t]{0.45\textwidth}
		\centering
		\includegraphics[width=\textwidth]{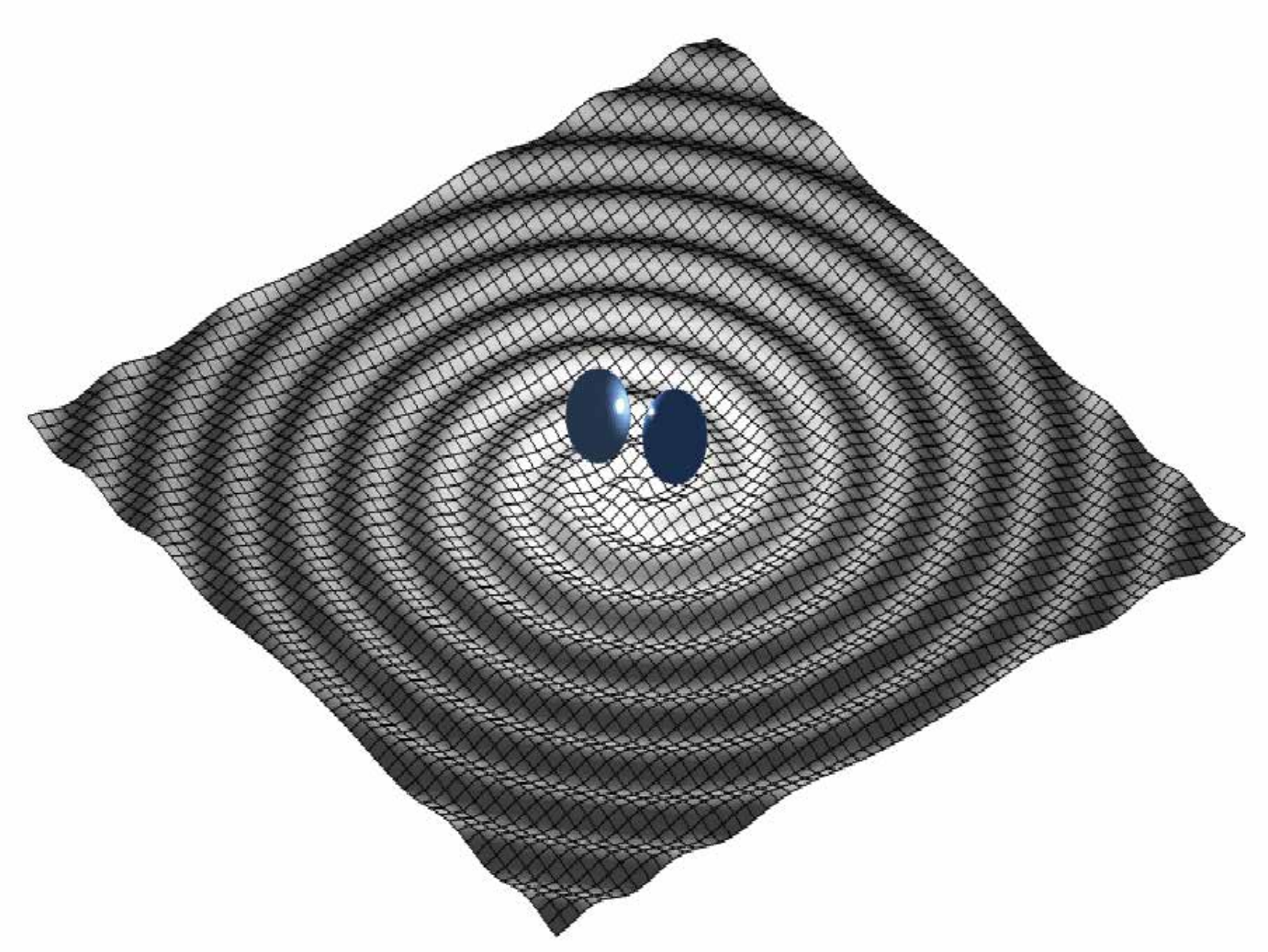}
		\caption{The intrinsic non-flat geometry of the space with the data living on can cause the external force, which is discussed in Einstein's general relativity theory.}
		\label{fig:Fig 6}
	\end{minipage}
	\hfill 
	\begin{minipage}[t]{0.45\textwidth}
		\centering
		\includegraphics[width=\textwidth]{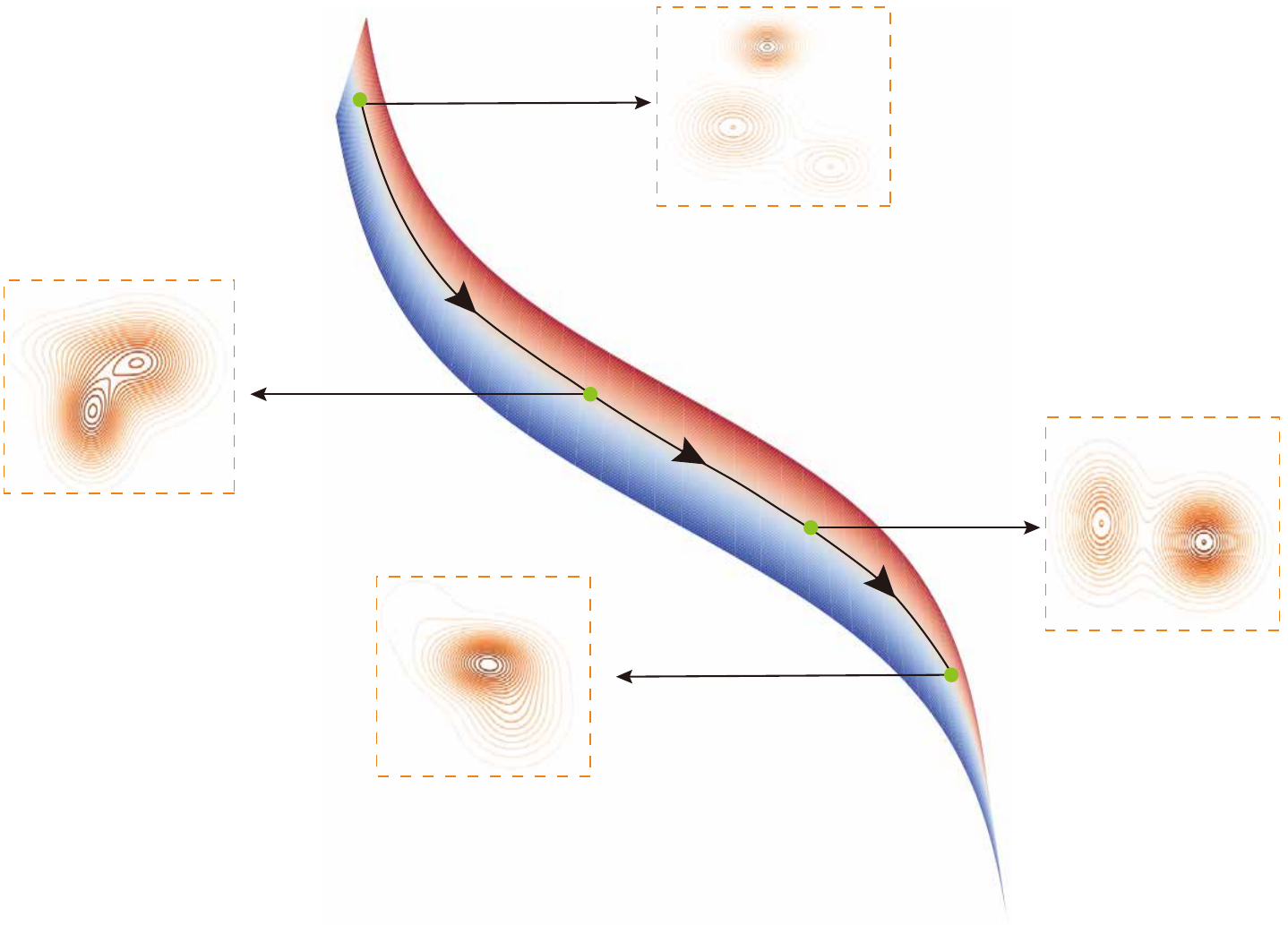}
		\caption{When the data is in the non-Euclidean space, the metric tensor of the space can be learned by solving the inverse problem of data dynamics.}
		\label{flow concept}
	\end{minipage}
\end{figure}

In general, data resides in a non-Euclidean space. The time-dependent minimization problem (Eq. \ref{time-dependent transportation}) in such spaces is equivalent to finding the transportation trajectory with minimal kinetic energy. Notably, the current classic formulation does not account for deformation energy induced by transportation. If the deformation energy in the probability space is considered, a similar NeoHookean hyper-elastic material in the non-Euclidean space can also be introduced. Due to the non-flat geometry, the deformation energy density per unit mass must be reformulated as:%
\begin{equation}
W=\frac{1}{2}G\left( g_{0}^{ij}g_{ij}-N\right) 
\end{equation}
where $g_{0}^{ij}$ denotes the covariant metric tensor at the initial point $\mathbf{X}$, and $g_{ij}$ represents the contravariant metric tensor at points $\mathbf{x}$ along the trajectory. The general solution for the transportation trajectory can be obtained by minimizing the integral of Lagrangian \citep{1984Modern,green1992theoretical}:
\begin{equation}
\frac{d^{2}x^{j}}{dt^{2}}-G\left(\Gamma
_{ik}^{i}g_{0}^{kj}+\Gamma _{ik}^{j}g_{0}^{ik}\right) +\Gamma _{ki}^{j}\frac{%
	dx^{k}}{dt}\frac{dx^{i}}{dt}=0
\label{motion-non-Elucidean}
\end{equation}
where $\Gamma_{ki}^{j}$ denotes the connection on the Riemann surface, defined through the metric tensor $g_{ij}$ (contravariant form) and $g^{ij}$ (covariant form) at point $\mathbf{x}:$
\begin{equation}
\Gamma _{ij}^{k}=\frac{1}{2}g^{kl}\left( \frac{\partial g_{lj}}{\partial
	x^{i}}+\frac{\partial g_{il}}{\partial x^{j}}-\frac{\partial g_{ij}}{%
	\partial x^{l}}\right)
\label{riemannn connection equation}
\end{equation}
We omit the detailed derivation here to focus on the Euclidean space formulation, which is more accessible for engineering applications. This formulation is also what we adopt for numerical implementation.

Compared with Eq. \ref{equilirium}, it can be seen that the non-flat geometry with the nonzero connection can cause the term $\Gamma _{ki}^{j}\frac{dx^{k}}{dt}\frac{dx^{i}}{dt}$ not the external force $F_{B}^{i}$. This means that in the non-Euclidean space, the force can be induced by the curvature. It is similar to Einstein's general relativity theory in which the curved time-space can cause gravity (Figure \ref{fig:Fig 6}). Consequently, we introduce external forces in Euclidean space to compensate for curvature effects originating from the non-flat data space.

In addition to minimizing the Lagrangian, mass conservation must also be enforced:
\begin{equation}
	\rho\left(\mathbf{X}\right)dV=\rho\left(\mathbf{x}\right)dv
\end{equation}
For the Riemann manifold,
\begin{equation}
dV=\sqrt{g_0\left( \mathbf{X}\right)}dx_1...dx_N \qquad dv=\sqrt{g\left( \mathbf{x}\right)}dx_1...dx_N
\end{equation}
where $x_1\cdot\cdot\cdot x_N$ are local coordinates.
\begin{eqnarray}
	g_{0}\left( \mathbf{X}\right)  &=&\det \left(g_{ij}\left( \mathbf{X}\right)
	\right)  \\
	g\left( \mathbf{x}\right)  &=&\det \left(g_{ij}\left( \mathbf{x}\right)
	\right) 
\end{eqnarray}%
Finally, it obtains:
\begin{equation}
\rho _{0}\left( \mathbf{X}\right) \sqrt{g_{0}\left( \mathbf{X}\right) }=\rho
\left( \mathbf{x},t\right) \sqrt{g\left( \mathbf{x}\right)}
\label{nonE-mass conservation}
\end{equation}

Manifold learning, as an important machine learning technique, is closely related to data in non-Euclidean space \citep{improvedflowmatching, tsne}. The proposed continuum mechanics framework enables direct learning of the non-Euclidean manifold's metric tensor (Figure \ref{flow concept}). This allows for the determination of optimal transport trajectories through the solution of Eq. \ref{motion-non-Elucidean} while strictly satisfying the mass conservation constraints (Eq. \ref{nonE-mass conservation}). Because the data are observed at some time points, it is assumed that the optimal transport trajectory is observed at those time points. The metric tensor can then be learned through inverse solving. The dynamics of data on the Lobachevsky plane are discussed in Appendix B, which provides insight into how to learn and generate in non-Euclidean space.

\section{Numerical implementation}
In this section, we propose numerical methods for time-dependent optimal transport based on continuum mechanics theory, which can be applied to generative tasks. These methods are designed for easy adaptation to various applications. The generative task then becomes: inferring the probability distribution of the data under some challenging conditions using results from reliable experiments or numerical experiments that are feasible to conduct. To accomplish the generative task, two key aspects should be clarified further.

First, the variable $t$ in the transportation process may not represent real time but rather a pseudo time. It can correspond to any varying parameter. In our example in result and discussion, the temperature, loading rate and the impact velocity can serve as a "time" in the transportation process. In this generative task, the pseudo time plays a role analogous to time step in the other generative models, such as those of diffusion model. The pseudo time can guide the generative process. Under this framework, we assume that the probability distribution of physical field data evolves physically with this pseudo time (i.e., temperature or loading rate). Additionally, the time variable in the theoretical formulation spans from $0$ to $1$. This implies that, for numerical implementation, the pseudo time must be normalized to the interval $\left(0\cdot\cdot1\right)$.

Second, as shown in Eq. \ref{Monge-Ampere Equation-Equi-new}, the probability distribution at any time $\rho{\left(\mathbf{x},t\right)}$ can be computed if the displacement field $u_k$ and the probability distribution at the reference configuration $\rho{\left(\mathbf{X}\right)}$ are known. To determine the displacement field $u_k$, we must first solve Eq. \ref{equilirium}. This requires assuming a specific form for the body force. Based on the derivation in Section 2.3, the body force can be treated as a function of the current coordinate $\mathbf{x}$ and time $t$. With this assumed form, simultaneously solving Eq. \ref{equilirium} and Eq. \ref{Monge-Ampere Equation-Equi-new} can yield the solution for $F_b$ and $u_k$ from available data. Once the displacement field is obtained, the probability density at an unknown time can be computed using Eq. \ref{Monge-Ampere Equation-Equi-new}.

\begin{figure}[H]
	\centering
       \includegraphics[width=1\textwidth]{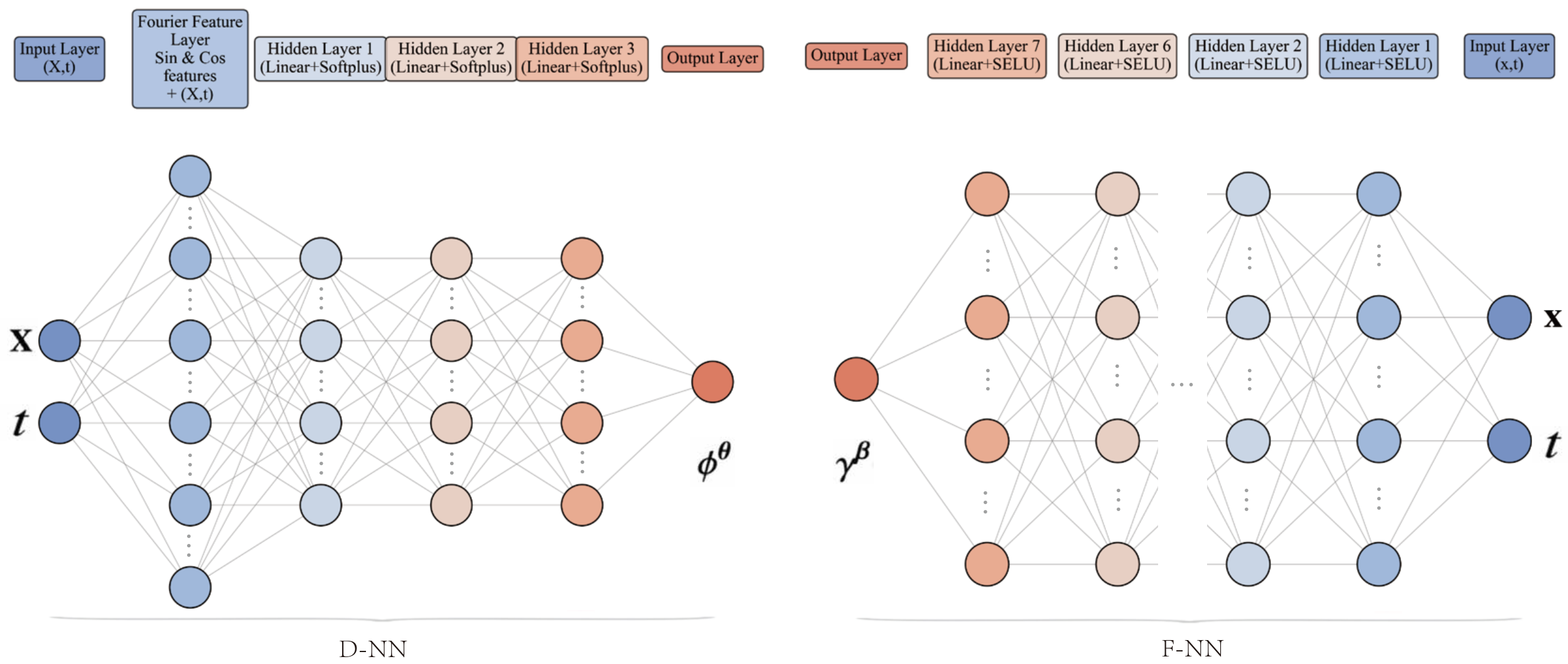}
	\setlength{\abovecaptionskip}{1mm}
	\setlength{\belowcaptionskip}{-2mm}
	\caption{Two neural networks are used to represent the displacement and the body force field, respectively.}
	\label{fig:Fig 8}
\end{figure}

Based on the above analysis, the generative task in this study can be formulated as identifying the evolution of both body force and displacement with respect to pseudo time using available experimental or numerical experimental data. In essence, the above problem constitutes a typical pattern recognition task, with the additional requirement that the entire process must satisfy the physical constraint given by Eq. \ref{equilirium}. The widespread adoption of physics-informed neural networks (PINNs) provides an effective approach to address this challenge \citep{jcpPinn}, motivating our proposed one neural network architecture to capture the time-dependent characteristics of the displacement field and another neural network to parameterize the body force, as illustrated in Figure \ref{fig:Fig 8}. In this architecture, both the spatial component of and the temporal component of $u_k$ are parameterized by an artificial neural network (D-NN) $\phi^{\bm{\theta}}$. Besides, body force $F_b$ is parameterized by an artificial neural network (F-NN) $\gamma^{\bm{\beta}}$.
\begin{equation}
    u_{\bm{\theta}}(\mathbf{X},t)=
    \phi^{\bm{\theta}}(\mathbf{X},t)
    \label{ann function}
\end{equation}
\begin{equation}
    F_{b,\bm{\beta}}(\mathbf{x},t)=\gamma^{\bm{\beta}}(\mathbf{x},t)
    \label{bodyforcefunction}
\end{equation}

The loss function design constitutes a core component of this study and directly determines model performance. As shown in Eq. \ref{eq:loss_function}, the loss function comprises three key terms: (1) The probability distribution loss $L_{1}$ ensures the transformed data $\mathbf{X}+u_{\bm{\theta}}(\mathbf{X}, t_{i})$ follow a probability distribution that closely approximates the target distribution after the optimal transport from the initial data, which means the motion of data satisfies Eq. \ref{Monge-Ampere Equation-Equi-new}; (2) The boundary matching loss $L_{2}$ enforces consistency between the transformed boundary $\partial X+u_{\bm{\theta}}(\partial X, t_{i})$ and the target boundary $\partial \mathbf{x}_{i}$, which is the boundary condition of Eq. \ref{Monge-Ampere Equation-Equi-new}; (3) The equation of motion loss $L_{3}$ enforces the motion of data satisfying Eq. \ref{equilirium}. The three components of the overall loss function can be adjusted via weighting parameters according to the typical scale of each component, which facilitates smoother convergence during training.
\begin{equation}
\begin{aligned}
L(\bm{\theta},\bm{\beta}) = \underbrace{ E_{\mathbf{X}}\sum_{i=1}^{n}\frac{1}{n}\left\| \frac{\rho(\mathbf{X})}{J_{i}} - \rho_i\left(\mathbf{X} + u_{\bm{\theta}}(\mathbf{X}, t_{i})\right) \right\|^2 }_{L_{1}}
+ \underbrace{ E_{\partial \mathbf{X}} \sum_{i=1}^{n}\frac{1}{n}\left\| \partial \mathbf{X} + u_{\bm{\theta}}(\partial \mathbf{X}, t_{i}) - \partial \mathbf{x}_i \right\|^2 }_{L_2}\\+\underbrace{E_{\mathbf{X}}\sum_{j=1}^{n'}\frac{1}{n'}\left\|\frac{\partial ^{2}{u_{\bm{\theta}}}\left( \mathbf{X},t’_j\right)}{\partial t^{2}}-F_{b,\bm{\beta}}\left( \mathbf{X}+u_{\bm{\theta}}(\mathbf{X}, t'_{j}),t’_j\right)\right\|^2}_{L_{3}}
\label{eq:loss_function}
\end{aligned}
\end{equation}
where $E$ represents the expectation and $\left\|*\right\|^2$ represents the squared distance, which can be adapted according to the specific problem. $n$ is the number of "time" moments of the available data and $n'$ is the number of "time" moments uniformly sampled over the time interval (0, 1). $t_i$ is the pseudo time snapshots of the available data and $t'_j$ is the pseudo time snapshots uniformly sampled over the time interval (0, 1). $\rho$ is the initial probability distribution, and $\rho_i$ is the probability distribution at target time point $t_i$. $\partial \mathbf{X}$ is the boundary of the initial domain, and $\partial \mathbf{x}_i$ is the boundary of the domain at target time point $t_i$. In the example in our result and discussion section, the probability density is assumed to be Gaussian and can be computed by adding Gaussian noise perturbations to the mean values obtained from experimental or numerical experiments. However, it does not mean our approach is limited to handling Gaussian distributions.

The Jacobian matrix $J_{i}$ is computed as: 
\begin{equation}
J_{i} =\text{det}\left[\frac{\partial (\mathbf{X}+u_{\bm{\theta}}(\mathbf{X},t_{i})} {\partial \mathbf{X}}\right]
\label{Jacobi}
\end{equation}

The optimal model parameters $\bm{\theta}^{*}$ and $\bm{\beta}^{*}$ are obtained by minimizing the loss function $L(\bm{\theta},\bm{\beta})$ using the backpropagation method \citep{barckforward},
\begin{equation}
\bm{\theta}^*,\bm{\beta}^* = \arg\min_{\bm{\theta},\bm{\beta}} L(\bm{\theta},\bm{\beta})
\label{minimum-dynamics}
\end{equation}

By minimizing the loss function $L(\bm{\theta},\bm{\beta})$, the neural network has learned the evolution of both body force and displacement with respect to pseudo time from available data. Substituting the displacement field into Eq. \ref{Monge-Ampere Equation-Equi-new} can complete the generation task of the probability distribution at other pseudo time. By taking the mean of this distribution, we obtain the final generated physical quantities. In summary, once the training of the neural networks is completed, we can input probability distribution at the initial pseudo time step, and obtain the physical quantities we need at the target pseudo time step through calculations. The pseudo-code of CM-GAI is shown in Algorithm 1.

\begin{algorithm}[htbp]
\caption{CM-GAI's training and generating process}
\KwIn{Probability distributions of the features in available states, the corresponding state parameters, and the state parameter to be generated}
\KwOut{Features and their probability distributions in the state to be generated}
Normalize all the state parameters (i.e., temperature or loading rate) to obtain "time" moment $t$

\While{D-NN and F-NN model have not converged}{
    Calculate the loss function $L(\bm{\theta},\bm{\beta})$ 
    
    Calculate the gradient based on this loss function
    
    Update the parameters $\bm{\theta}$ and $\bm{\beta}$ of D-NN, and F-NN model using the gradient descent method
}

Feed the final moment (normalized to 1) into D-NN to predict the displacement at "time" 1

Obtain the probability density distribution of the features generated at "time" 1 from the mass conservation equation (Eq. \ref{Monge-Ampere Equation-Equi-new})

Taking the mean of this probability density distribution yields the features in target state 

    \Return The generated probability density distribution of the features in target state and the features in target state
\end{algorithm}

Due to the stochastic nature of neural network training, not every training session yields satisfactory results. We selected neural networks that performed well on the training set to generate the target probability distribution. When sufficient experimental data is available, readers may reserve a portion as a validation set for selecting trained models. The detailed neural network model architectures and training hyperparameters are summarized in Appendix C.

\section{Results and discussion}
\begin{figure}[htbp]
	\centering
	\includegraphics[width=1\textwidth]{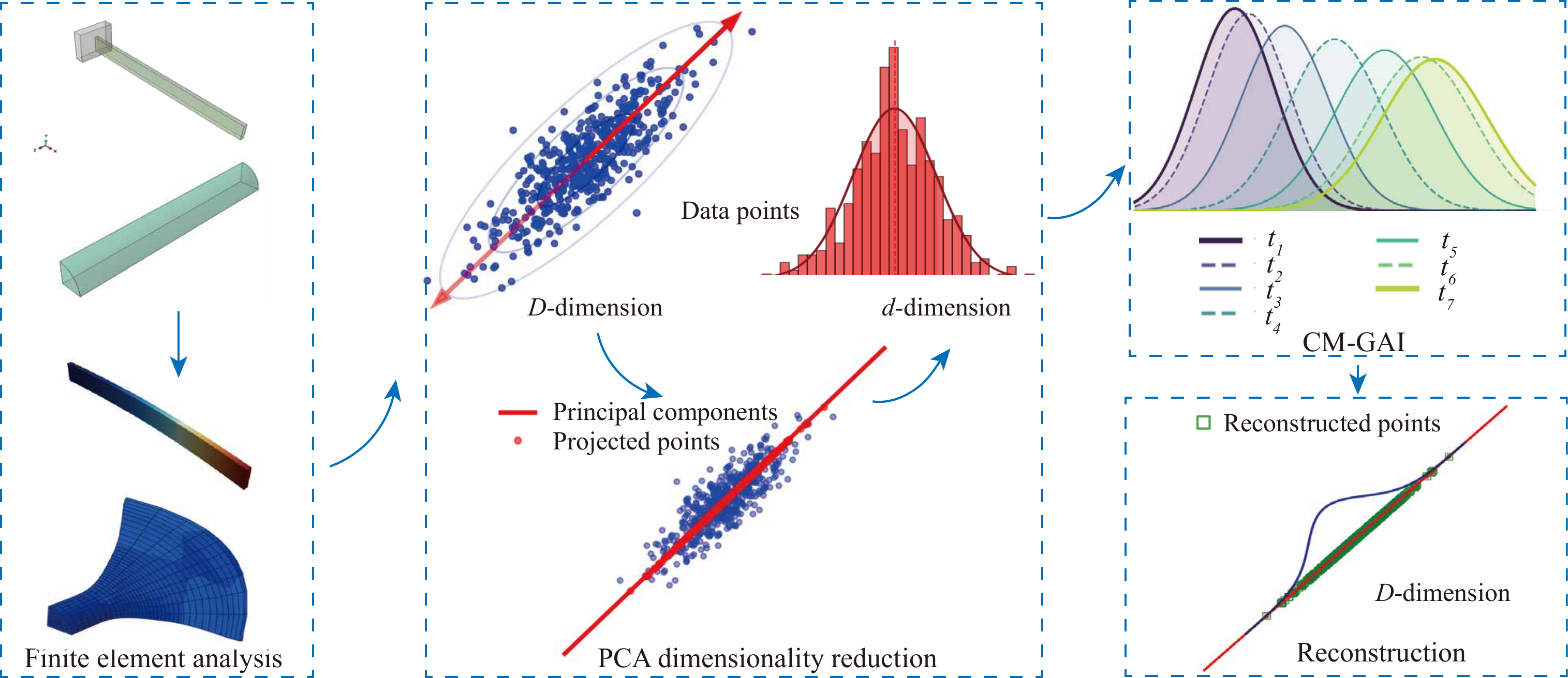}
	\setlength{\abovecaptionskip}{1mm}
	\setlength{\belowcaptionskip}{-2mm}
	\caption{Generation of stress field based on CM-GAI.}
	\label{fig:method}
\end{figure}

This section demonstrates the capability of CM-GAI to generate key material and structural responses. Specifically, it is used to generate stress-strain responses under varying temperatures and strain rates (constitutive relationships), determine temperature-induced stress fields, and analyze transient strain fields under dynamic loading. The presentation of results in this subsection follows a progression from simple to complex. First, to demonstrate the effectiveness of the proposed method for generation of two-dimensional probability density distribution problems and to provide readers with a straightforward case for better understanding, we apply the CM-GAI to generate temperature or strain rate-dependent material stress-strain responses. This example offers a novel approach for obtaining constitutive data beyond experimentally measurable sets. Then, to illustrate the capability of our method in handling complex high-dimensional probability density distribution generation problems, we employ CM-GAI to generate the temperature-dependent stress field response of a cantilever beam and the transient dynamic response of a copper Taylor rod subjected to high-velocity axial impact against a rigid wall. (Figure \ref{fig:method}). The implementation of this example bypasses full-process simulations of thermo-mechanical coupling and transient dynamics to some extent, providing a novel pathway for acquiring relevant physical field variables under varying loading conditions.

\subsection{Generation of Temperature- and Strain-Rate-Dependent Stress-Strain Response of Materials}
In this section, we examine the effectiveness of CM-GAI in generating the stress-strain response of materials under different temperatures or strain rates across nine distinct types of materials (results for four materials are presented in this section, while the remaining results are provided in Appendix D). We adopt two-dimensional stress–strain curve generation as our experimental task for two reasons. 

On the one hand, the problem is ubiquitous in engineering practice, spanning materials characterization and design assessment, with clear practical relevance. Obtaining data under extreme temperatures or strain rates poses a significant challenge usually. Taking hot melt adhesives (e.g., Fuller EH9821B) as an example, their stress-strain response near the operating temperature (approximately 40°C) is critical for assessing the reliability of bonded joints. However, within the temperature range of 20°C to 50°C, the material undergoes a drastic transition from a purely solid state to a purely fluid state. Within the transition interval (30°C to 50°C), the material exists in a complex solid-liquid mixed state, making effective measurement difficult using conventional rheometer or solid testing machines. Although customized non-standard specimens can be explored for testing, the resulting data—particularly the response at the critical operating temperature of 40°C—are of questionable reliability due to the non-standard measurement approach. Consequently, a core question arises: is it possible to use generative models, combined with readily available conventional temperature data, to reliably generate the complete stress-strain response of materials at extreme or difficult-to-measure temperatures?
 
On the other hand, given the theoretical abstraction of our framework, this representative test case isolates the core mechanism without extraneous complexity and provides a reproducible basis for subsequent extensions to higher-dimensional settings and more complex operating conditions. It should be noted that, due to manufacturing variability and measurement noise, the stress–strain relation at a given temperature or strain rate is better regarded as a probability distribution rather than a deterministic function.
 
\begin{figure}[htbp]
	\centering
	\includegraphics[width=0.4\textwidth]{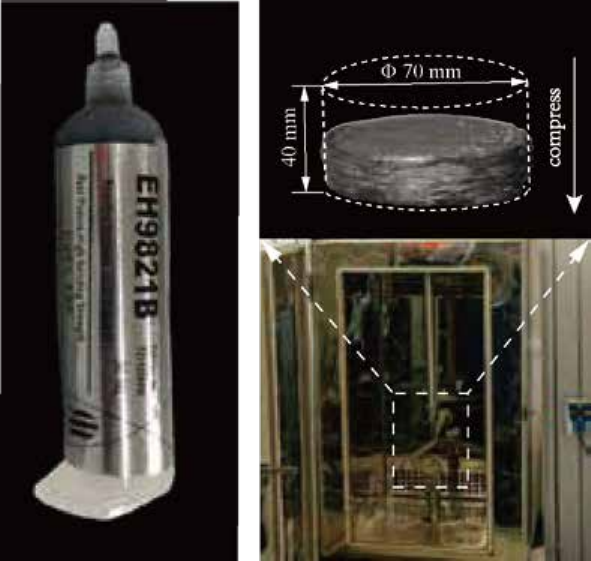}
	\setlength{\abovecaptionskip}{1mm}
	\setlength{\belowcaptionskip}{-2mm}
	\caption{Fuller EH9821B adhesive. Schematic diagram of the specimen compression experiment under constant temperature conditions.}
	\label{fig:fullerexperiment}
\end{figure}

First, we apply the CM-GAI approach to our experimental data for the adhesives introduced first. Recent years have witnessed significant advancements in industrial adhesive development. Through careful formulation adjustments—including the incorporation of fillers, tackifiers, stabilizers, and antioxidants—researchers have achieved enhanced melting temperatures, improved adhesive strength, and increased flexibility, along with superior heat resistance and flame retardancy. For our experiments, we used Fuller EH9821B adhesive (Figure \ref{fig:fullerexperiment}), a commercially available product widely used in industry. This adhesive remains solid at room temperature.

To use the adhesive for bonding components, it must first be heated to $200\,^\circ\mathrm{C}$, at which point it becomes fluid. After heating, the adhesive needs to cool to the working temperature of approximately $40\,^\circ\mathrm{C}$. At this temperature, obtaining accurate stress-strain responses under uniaxial compression is crucial for controlling the adhesive's bonding morphology and strength. However, the adhesive at $40\,^\circ\mathrm{C}$ exhibits properties intermediate between fluids and solids, making it extremely difficult to prepare standard cylindrical specimens as with solid materials. Any attempt to fabricate standard specimens results in collapse due to the adhesive's self-weight under these conditions. As an alternative, we prepared shorter nonstandard specimens through trial and error, allowing for approximate measurement of the stress-strain response. In contrast, cylindrical specimens could be fabricated at $29\,^\circ\mathrm{C}$, $31\,^\circ\mathrm{C}$, $32\,^\circ\mathrm{C}$, $35\,^\circ\mathrm{C}$ (shown in Figure \ref{fig:fullerexperiment}) with greater reliability. Using a temperature-controlled incubator, we measured the uniaxial compression stress-strain responses at these temperatures.

\begin{figure}[htbp]
    \centering
    \begin{subfigure}[t]{0.45\textwidth}
        \centering
        \includegraphics[width=\textwidth]{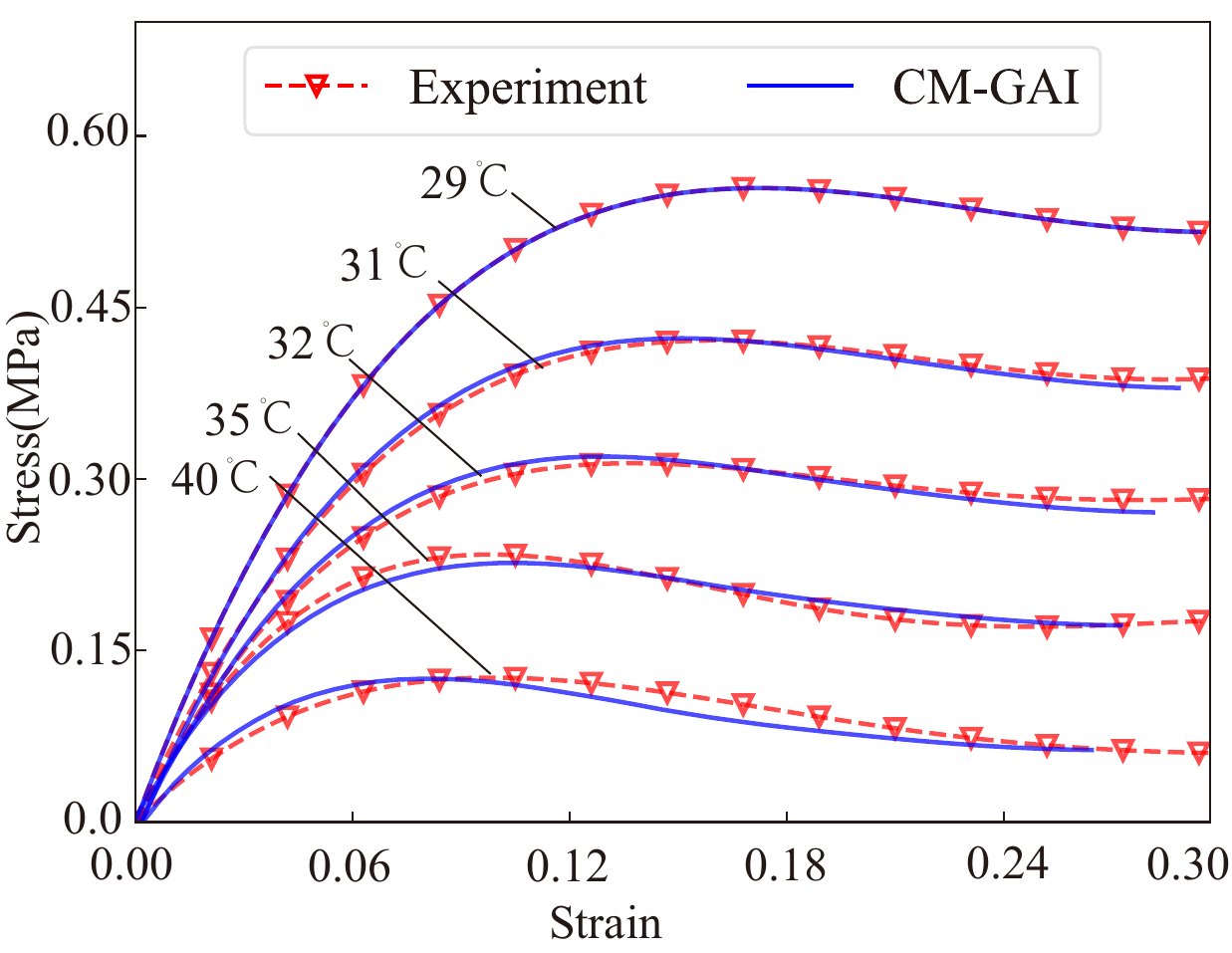}
        \caption{The stress-strain curves at the different temperatures.}
        \label{fig:adhesivea}
    \end{subfigure}
    \begin{subfigure}[t]{0.45\textwidth}
        \centering
        \includegraphics[width=\textwidth]{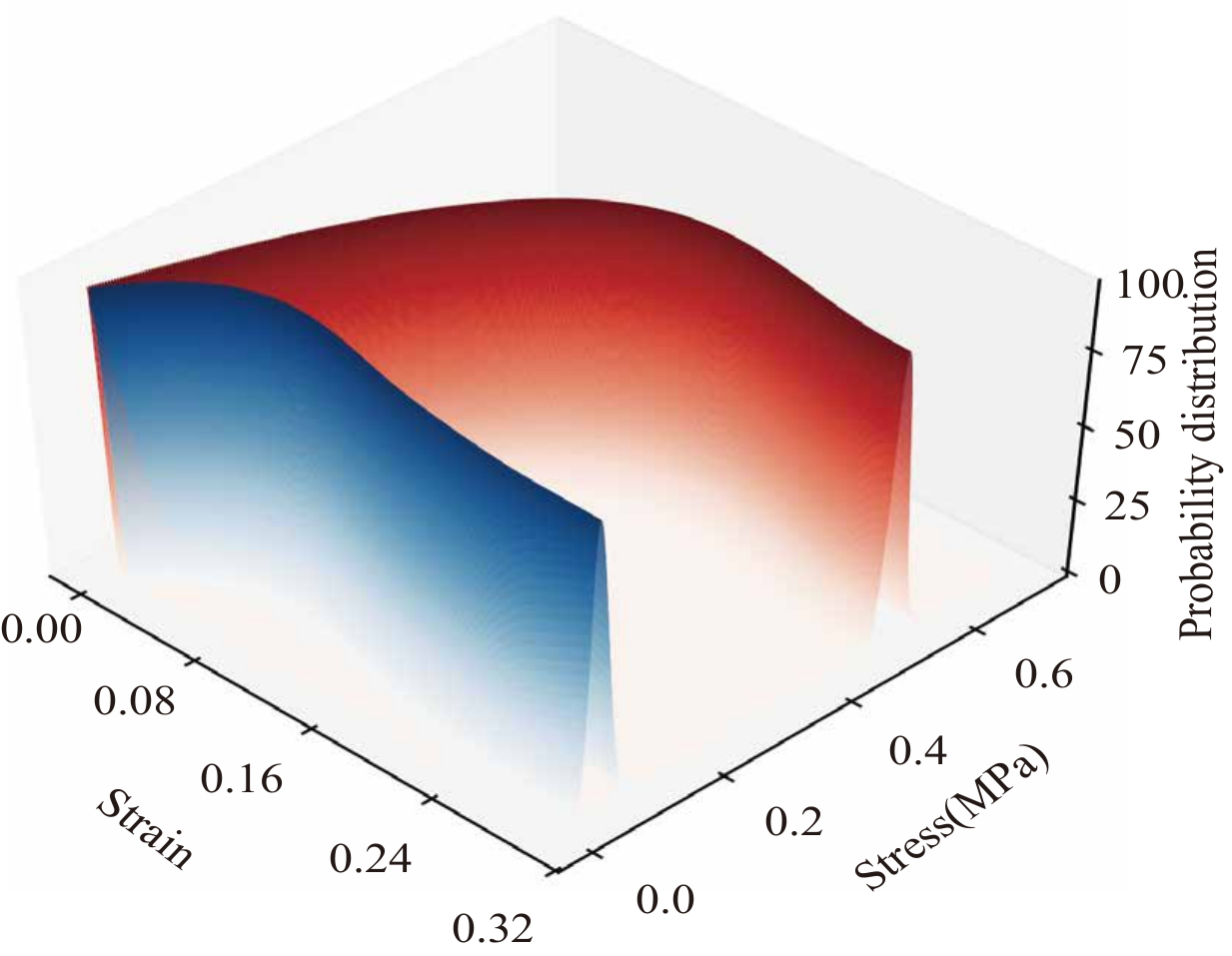}
        \caption{Original probability distribution and the final probability distribution after transportation calculated by CM-GAI.}
        \label{fig:adhesiveb}
    \end{subfigure}
    \caption{The comparison between the experiment data of hot melt adhesive and CM-GAI predictions.}
    \label{fig:adhesive}
\end{figure}

In this problem, the stress-strain data varies with temperature. We therefore treat temperature as the time variable for optimal transport. The temperature range is normalized to $t=0$ (corresponding to $29^{\circ}$) to $t=1$ (corresponding to $40\,^\circ\mathrm{C}$). Using CM-GAI, we then predict the stress-strain response at $t=1$ ($40\,^\circ\mathrm{C}$). This prediction assumes known stress-strain data at other temperatures ($29\,^\circ\mathrm{C}$, $31\,^\circ\mathrm{C}$, $32\,^\circ\mathrm{C}$, $35\,^\circ\mathrm{C}$). The governing equations are solved through neural network training. The predicted results are presented in Figure \ref{fig:adhesive}. As shown in Figure \ref{fig:adhesivea}, CM-GAI achieves reasonably accurate predictions of the stress-strain response.

\begin{figure}[htbp]
    \centering
    \begin{subfigure}[t]{0.45\textwidth}
        \centering
        \includegraphics[width=\textwidth]{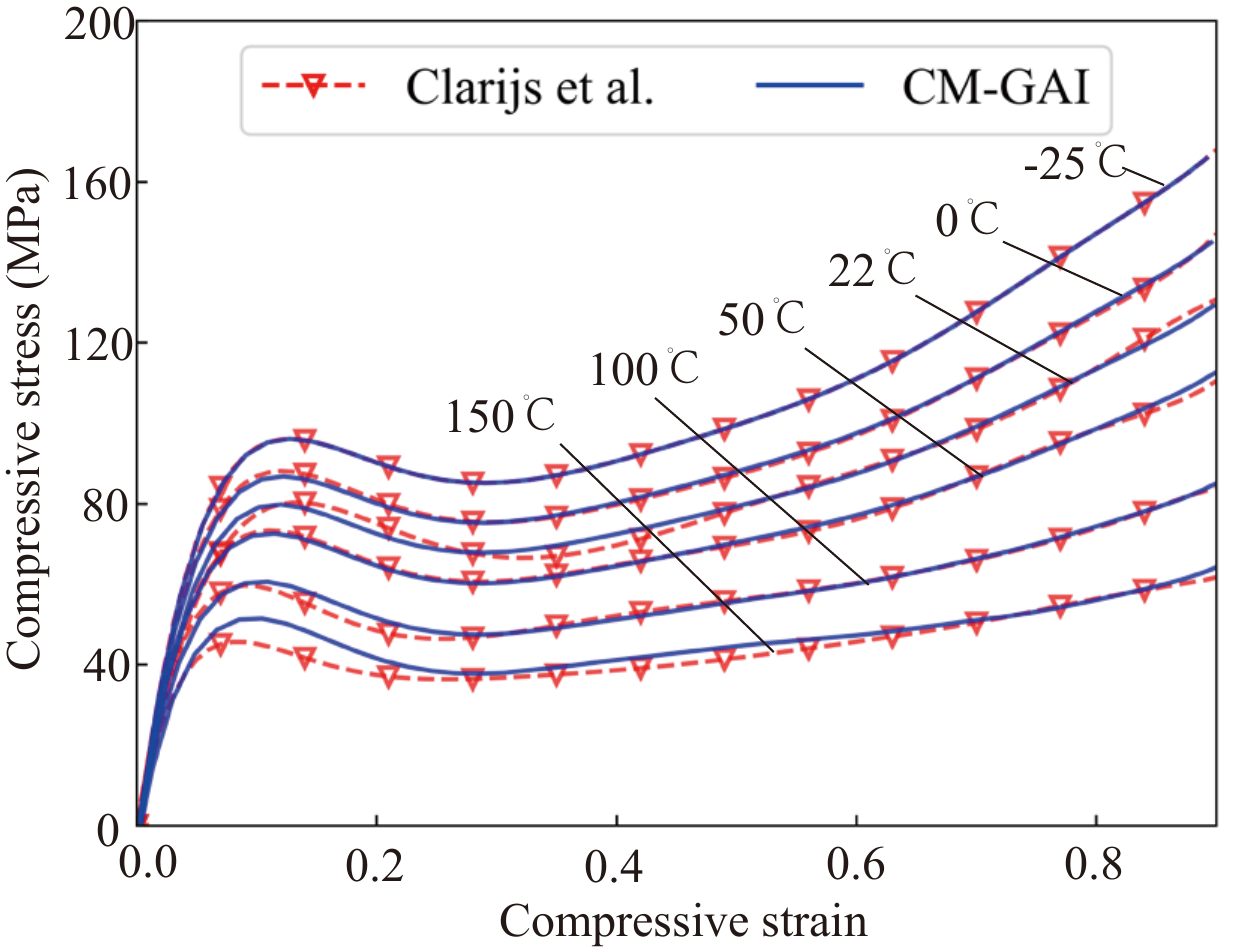}
        \caption{The stress-strain curves at the different temperatures.}
        \label{fig:clarijsa}
    \end{subfigure}
    \begin{subfigure}[t]{0.45\textwidth}
        \centering
        \includegraphics[width=\textwidth]{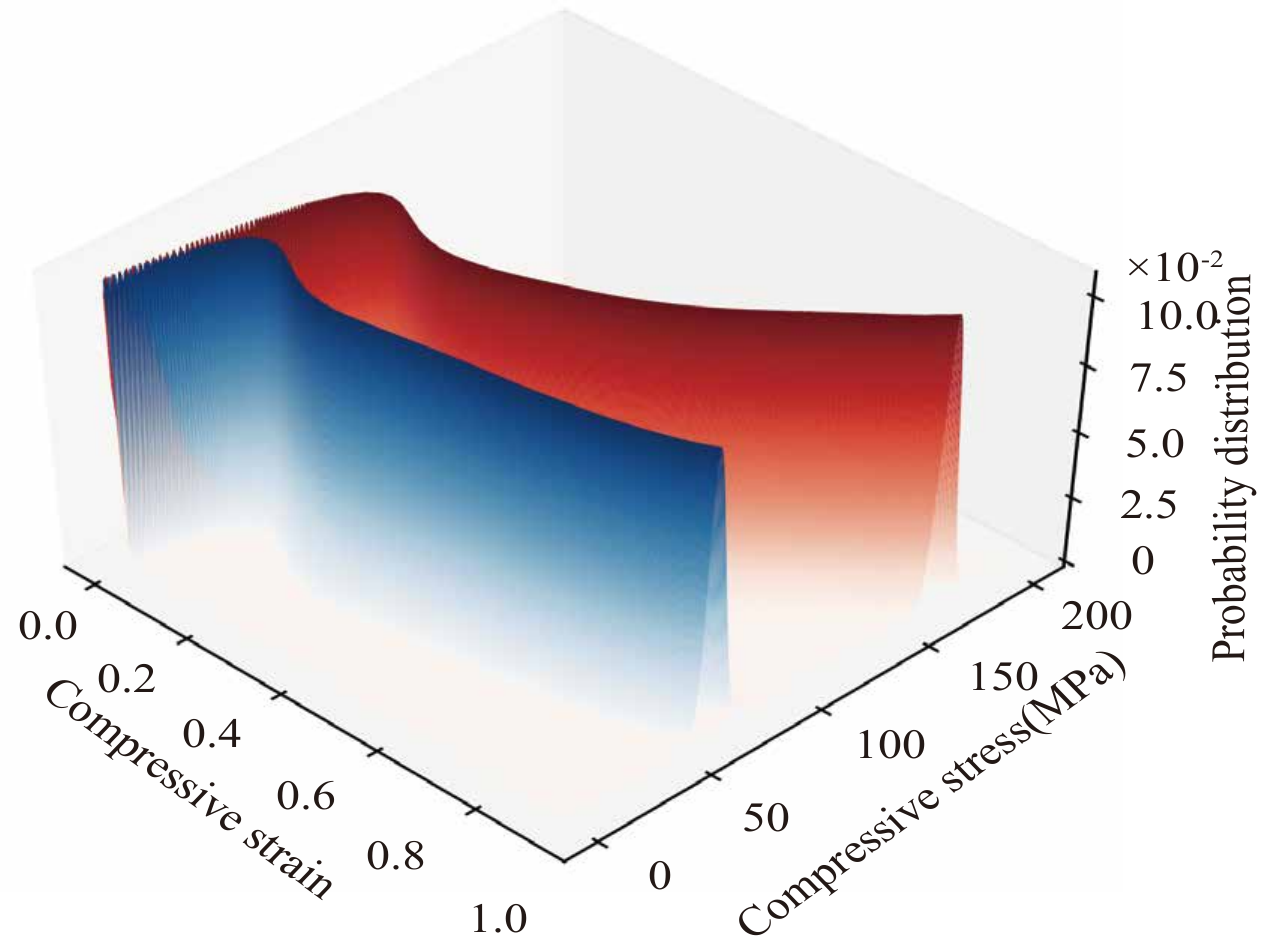}
        \caption{Original probability distribution and the final probability distribution after transportation calculated by CM-GAI.}
        \label{fig:clarijsb}
    \end{subfigure}
    \caption{The comparison between Clarijs et al.'s experiment \citep{clarijs2019} and CM-GAI predictions.}
    \label{fig:clarijs}
\end{figure}

\textbf{Glassy polymers} are widely used in many engineering applications and often exhibit complex deformation behavior. A thorough understanding of the stress-strain response over a wide temperature range under uniaxial loading is the first step for constitutive modeling, which can be used to design engineering structures composed of glassy polymers. The temperature-dependent deformation behavior of a glassy polymer (PPSU, polyphenylsulfone) was measured by Clarijs and Govaert \citep{clarijs2019}. Quasi-static uniaxial compression tests were performed on cylindrical specimens ($\Phi$6 mm $\times$ 4 mm) using a Zwick tester equipped with a temperature chamber. In addition to ambient temperature ($22\,^\circ\mathrm{C}$), tests were conducted at $-25\,^\circ\mathrm{C}$, $0\,^\circ\mathrm{C}$, $50\,^\circ\mathrm{C}$, $100\,^\circ\mathrm{C}$, and $150\,^\circ\mathrm{C}$. The stress-strain data at these temperatures (see Figure 6 in the reference) is reproduced in Figure \ref{fig:clarijsa}, plotted with red dashed lines and inverted triangle symbols.

As in the first example, temperature is treated as the time variable for optimal transport analysis. The temperatures are normalized to $t=0$ (corresponding to $-25\,^\circ\mathrm{C}$) to $t=1$ (corresponding to 150°C). CM-GAI is then employed to predict the stress-strain response at $t=1$ ($150\,^\circ\mathrm{C}$) using the known stress-strain data at other temperatures ($-25\,^\circ\mathrm{C}$, $0\,^\circ\mathrm{C}$, $22\,^\circ\mathrm{C}$, $50\,^\circ\mathrm{C}$, $100\,^\circ\mathrm{C}$). The governing equations in CM-GAI are solved through neural network training.

Figure \ref{fig:clarijsa} shows the predicted stress-strain response at $150\,^\circ\mathrm{C}$ using CM-GAI, along with predictions for the training data ($-25\,^\circ\mathrm{C}$, $0\,^\circ\mathrm{C}$, $22\,^\circ\mathrm{C}$, $50\,^\circ\mathrm{C}$, $100\,^\circ\mathrm{C}$). The trained neural networks also predict the probability distribution at $150\,^\circ\mathrm{C}$, shown in Figure \ref{fig:clarijsb}. Red represents the original probability density distribution, while blue represents the final probability distribution
after transportation calculated by CM-GAI. Subsequent probability distribution figures follow the same color convention as defined in this figure. The stress-strain response exhibits strong nonlinearity across temperatures - stress typically increases initially with strain, then decreases before increasing again. Despite this complexity, CM-GAI provides good agreement with the test data.

\begin{figure}[htbp]
    \centering
    \begin{subfigure}[t]{0.45\textwidth}
        \centering
        \includegraphics[width=\textwidth]{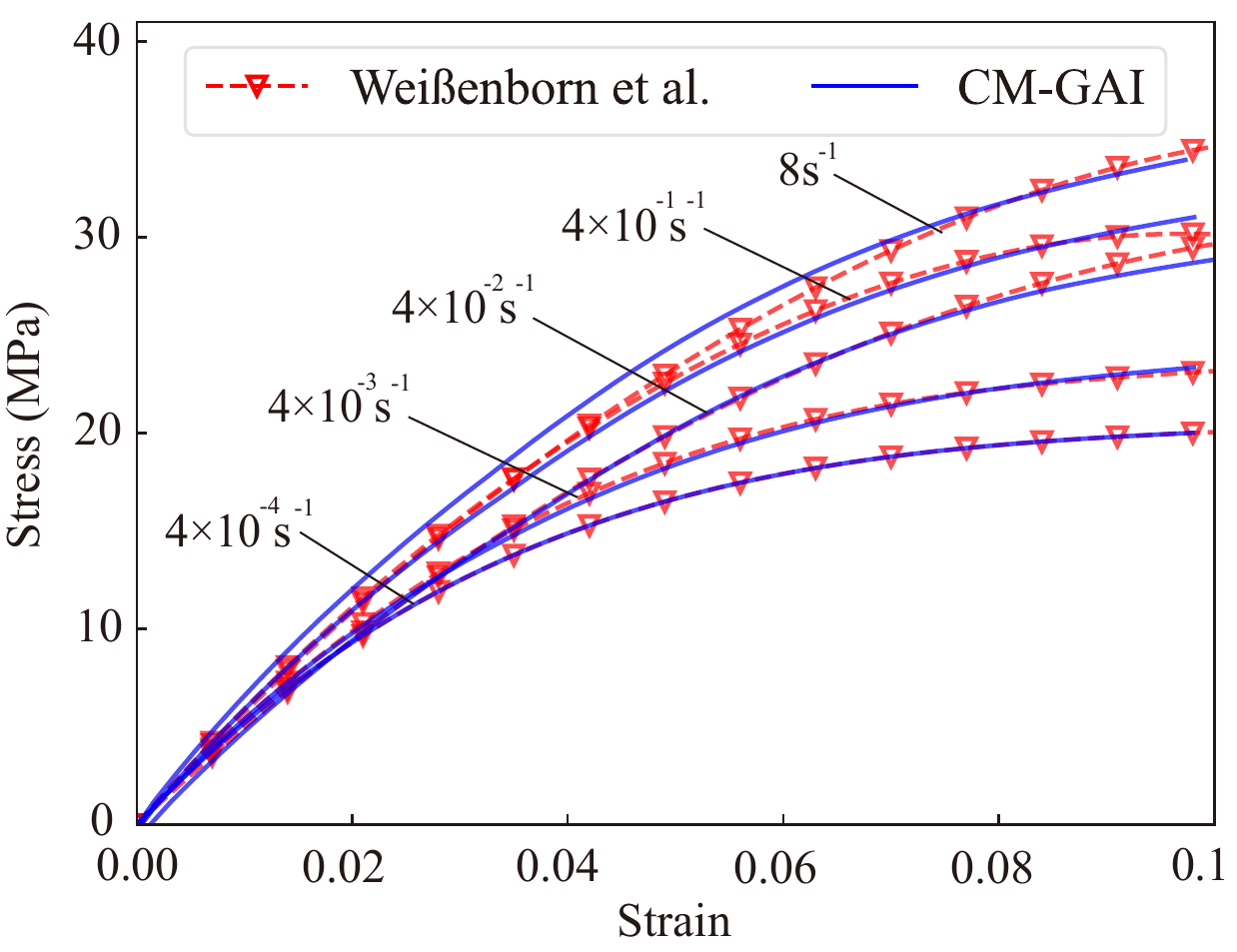}
        \caption{The stress-strain curves at the different strain rates.}
        \label{fig:weissenborna}
    \end{subfigure}
    \begin{subfigure}[t]{0.45\textwidth}
        \centering
        \includegraphics[width=\textwidth]{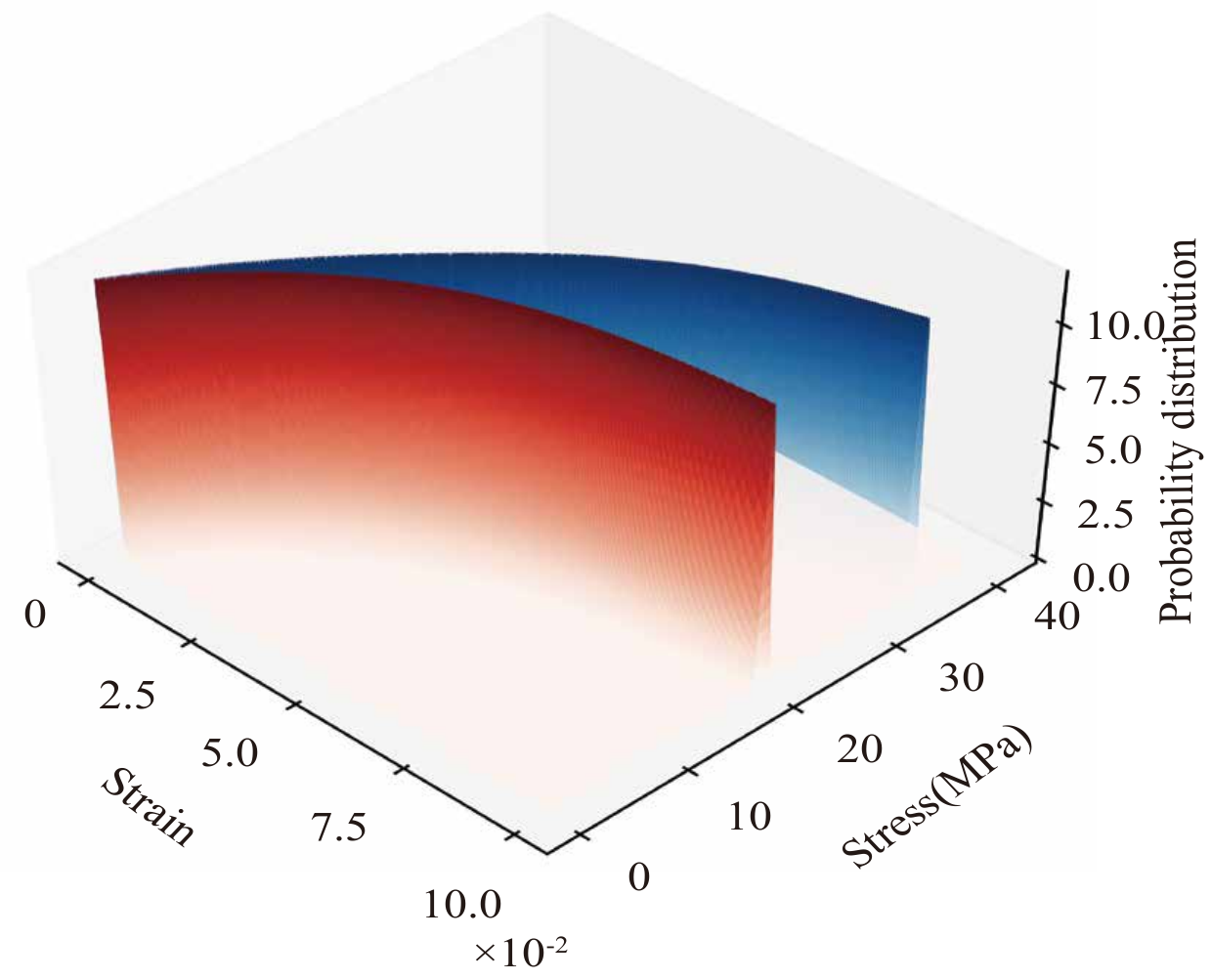}
        \caption{Original probability distribution and the final probability distribution after transportation calculated by CM-GAI.}
        \label{fig:weissenbornb}
    \end{subfigure}
    \caption{The comparison between Weißenborn et al.'s experiment \citep{weissenborn2016} and CM-GAI predictions.}
    \label{fig:weissenborn}
\end{figure}

\textbf{Polymer foams} are widely used in impact and crash components of lightweight structures. Understanding their stress-strain response at various strain rates is essential for designing these structures. Weißenborn et al. \citep{weissenborn2016} investigated the strain-rate-dependent deformation behavior of polyurethane foams. They first fabricated polymeric foams with densities of ($230$ kg/m$^3$, 420 kg/m$^3$ and 610 kg/m$^3$). Specimens measuring $25$$\times$$25$$\times$$25$ mm were prepared, and compressive experiments were conducted using a high-speed testing machine (Instron VHS 160/20, Instron GmbH, Darmstadt, Germany). The testing machine allowed velocities ranging from $0.0001$ to $20$ m/s. Different strain rates (the average one) can be computed by taking into account the specimen size and the braking distance of the upper clamp into account. Due to complex deformation phenomena during compression, obtaining accurate true stress measurements was challenging. Therefore, engineering stress and strain were used instead. The stress-strain data of the polymer foam with the density of 610 kg/m$^3$ for five strain rates, ranging from 0.0004 s$^{-1}$ and 8 s$^{-1}$ (see Figure 4 in the original study), is reproduced in Figure \ref{fig:weissenborna}, represented by red dash lines with inverted triangle symbols.

Due to the limited specimen size and micro-structural defects, the stress-strain data exhibit a certain degree of scatter that depends on the foam density. These scattering characteristics are shown in Figures 5-8 of the reference. We therefore assume the stress-strain data follows a Gaussian distribution, consistent with their approach. The distribution function at the strain rate of $0.0004$ s$^{-1}$ is shown in Figure \ref{fig:weissenbornb}.

In this problem, the stress-strain data varies with strain rate. We therefore use the strain rate as the time variable for optimal transport analysis. Since the strain rate spans a wide range, we first take the logarithm of the strain rate. The strain rates are then normalized to $t=0$ (corresponding to $0.0004$ s$^{-1}$) to $t=1$ (corresponding to $8$ s$^{-1}$). Using CM-GAI, we predict the stress-strain response at $t=1$ ($8$ s$^{-1}$) given the stress-strain data at other strain rates ($0.0004$ s$^{-1}$, $0.004$ s$^{-1}$, $0.04$ s$^{-1}$, $0.4$ s$^{-1}$). The governing equations are solved through neural network training.

Figure \ref{fig:weissenborna} shows the stress-strain data predicted by CM-GAI. The blue lines represent predictions for the training data (strain rates of $0.0004$ s$^{-1}$, $0.004$ s$^{-1}$, $0.04$ s$^{-1}$, $0.4$ s$^{-1}$), and the testing data (strain rates of $8$ s$^{-1}$), demonstrating good agreement. The trained neural networks also predict the probability distribution for $8$ s$^{-1}$, shown in Figure \ref{fig:weissenbornb}. The CM-GAI prediction provides a reasonable approximation compared to the measured stress-strain data at $8$ s$^{-1}$.

\begin{figure}[H]
    \centering
    \begin{subfigure}[t]{0.47\textwidth}
        \centering
        \includegraphics[width=\textwidth]{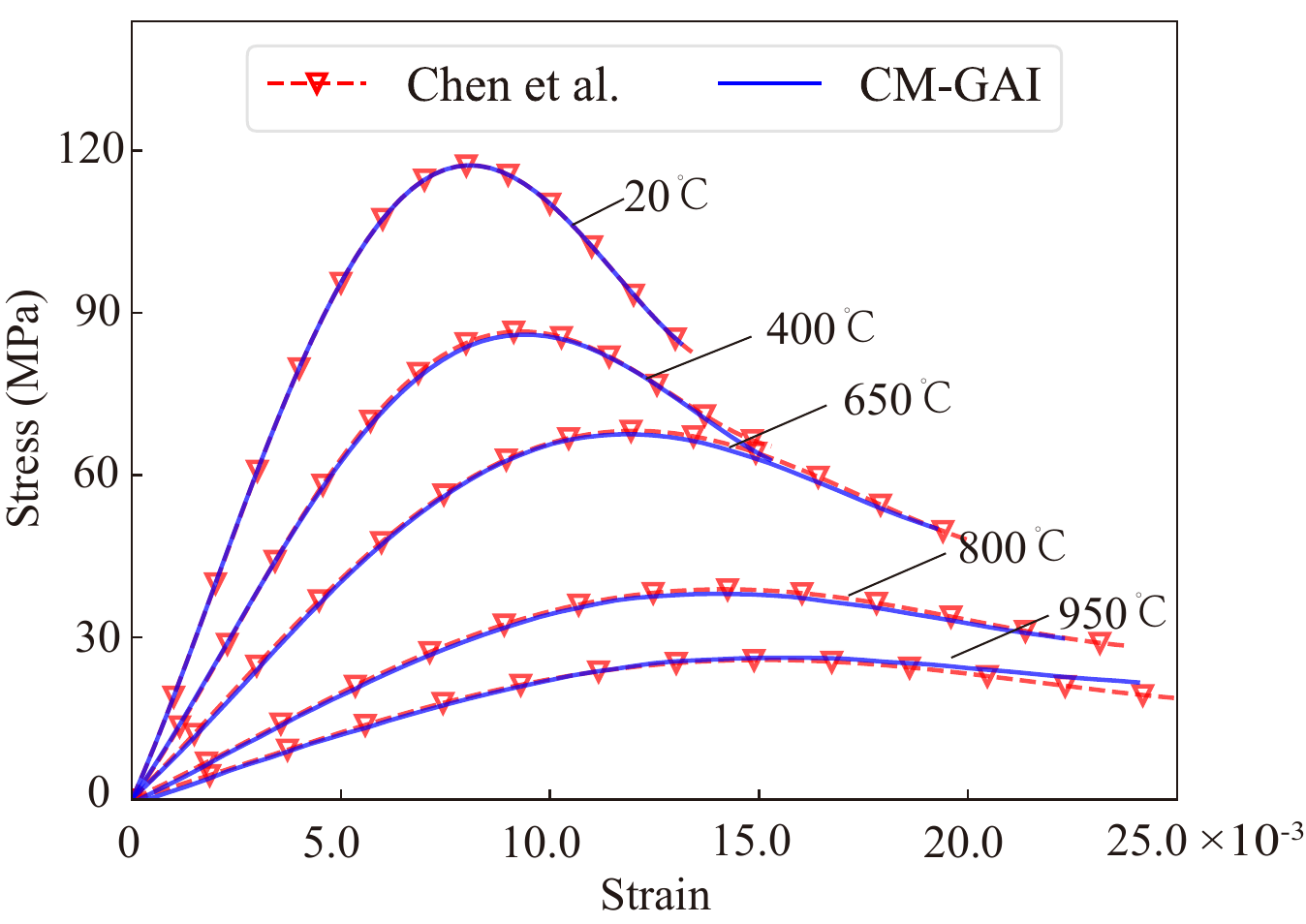}
        \caption{The stress-strain curves at different temperatures.}
        \label{fig:chena}
    \end{subfigure}
    \begin{subfigure}[t]{0.43\textwidth}
        \centering
        \includegraphics[width=\textwidth]{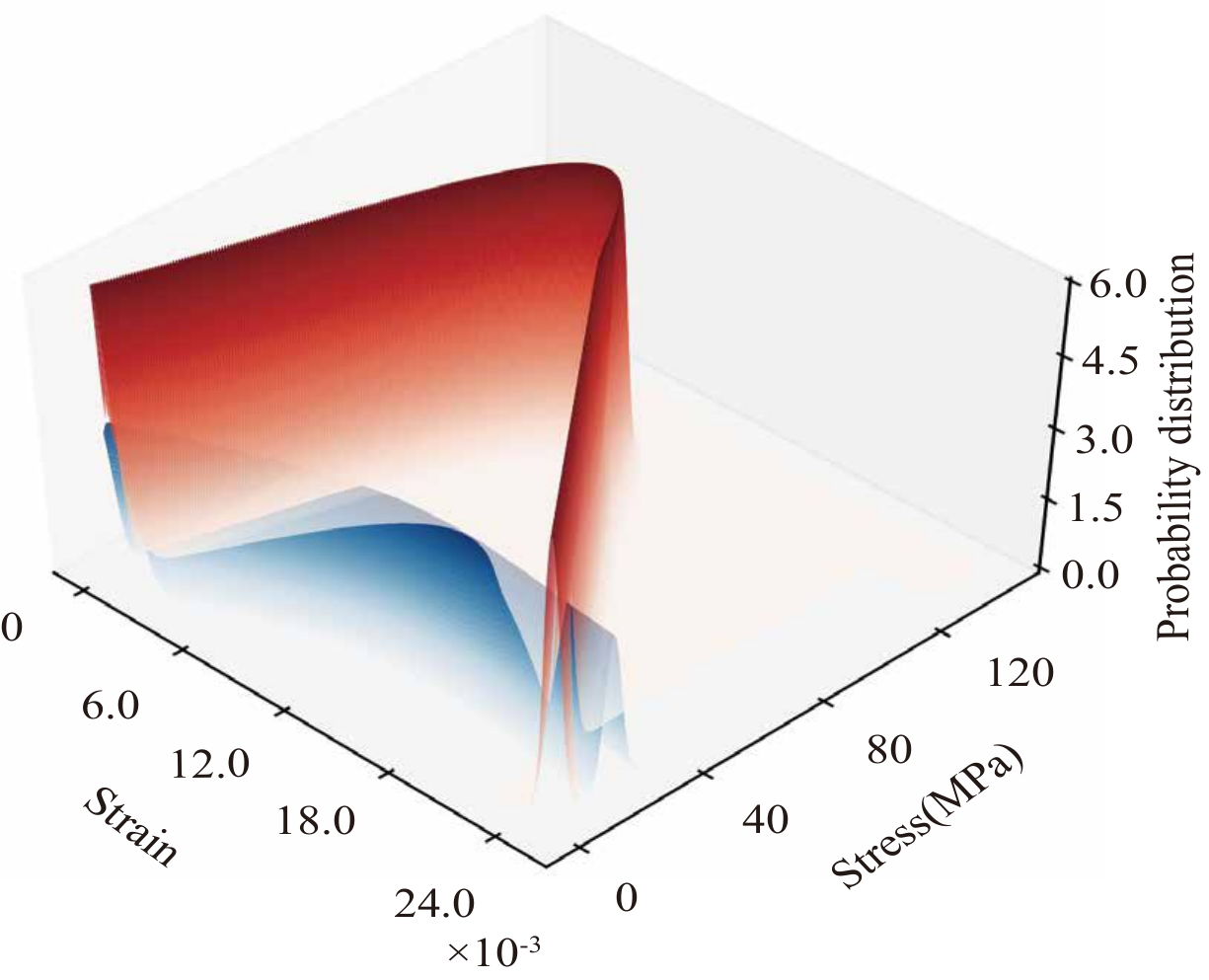}
        \caption{Original probability distribution and the final probability distribution after transportation calculated by CM-GAI.}
        \label{fig:chenb}
    \end{subfigure}
    \caption{The comparison between Chen et al.'s experiment \citep{chen2015} and CM-GAI predictions.}
    \label{fig:chen}
\end{figure}

\textbf{Concrete material} is widely used in civil engineering applications such as high-rise buildings and bridges. A thorough understanding of the stress-strain response over a wide temperature range is required to properly design structures subjected to blast and fire conditions. The temperature-dependent deformation behavior of concrete was measured by Chen et al. \citep{chen2015}. Quasi-static uniaxial compression tests were performed on cylindrical specimens ($\Phi$70 mm $\times$ 40 mm) using a 200-ton MTS hydraulic testing machine. The specimens were first heated to predetermined temperatures in an industrial microwave oven and then wrapped with insulating material during compression testing. In addition to ambient temperature ($20\,^\circ\mathrm{C}$), elevated temperatures of $400\,^\circ\mathrm{C}$, $650\,^\circ\mathrm{C}$, $800\,^\circ\mathrm{C}$, and $950\,^\circ\mathrm{C}$ were investigated. The stress-strain data at these temperatures (see Figure 17 in the reference) is reproduced in Figure \ref{fig:chena}, plotted with red dash lines and inverted triangle symbols.

In this problem, the stress-strain data varies with temperature. Therefore, temperature is treated as the time variable for optimal transport analysis. The temperatures are normalized to $t=0$ (corresponding to the temperature $20^{\circ}$) to $t=1$ (corresponding to the temperature $950\,^\circ\mathrm{C}$). CM-GAI is then employed to predict the stress-strain response at $t=1$ ($950\,^\circ\mathrm{C}$) using the known stress-strain data at other temperatures ($20\,^\circ\mathrm{C}$, $400\,^\circ\mathrm{C}$, $650\,^\circ\mathrm{C}$, $800\,^\circ\mathrm{C}$). The governing equations are solved through neural network training.

Figure \ref{fig:chena} shows the predicted stress-strain response at $950\,^\circ\mathrm{C}$ using CM-GAI, along with predictions for the training data ($20\,^\circ\mathrm{C}$, $400\,^\circ\mathrm{C}$, $650\,^\circ\mathrm{C}$, $800\,^\circ\mathrm{C}$). The trained neural networks also predict the probability distribution at $950\,^\circ\mathrm{C}$, shown in Figure \ref{fig:chenb}. The stress-strain response exhibits strong nonlinearity across different temperatures. Nevertheless, the proposed CM-GAI method provides good agreement with the test data, demonstrating its capability to handle complex material behavior.

Numerical simulations in comparison with the experimental data in the literature are also carried out, which is shown in Appendix D. Besides, the normalized root mean square error (NRMSE) between the generated stress-strain curves and the target stress-strain curves in all cases are also shown in Table \ref{tab:nrmse_simple} in Appendix D. These numerical simulations prove the capability of the proposed CM-GAI. A comparison with regression method is presented in Appendix E. It is worth mentioning that our method also demonstrates its potential in image generation tasks, as discussed in Appendix F.

\subsection{Generation of temperature-dependent stress fields of a cantilever beam}
 When the research focus shifts from the material itself to the actual engineering components made from it, the complexity of the problem increases significantly. At this point, we are no longer concerned solely with the constitutive relationship at a single temperature, but rather with the overall mechanical response of the structure under varying temperature fields. In this section, the effectiveness of the CM-GAI is tested for generating the temperature-dependent stress fields of a cantilever beam with the temperature difference between the upper and lower surface. The evolution of the internal thermal stress is key to mechanical performance assessment such as fatigue. The model geometry is given in appendix G. The temperature of the lower surface is fixed while the temperature of the upper surface may vary. 

In real engineering applications, the materials for the cantilever beam may exhibit both geometric and material nonlinearity under different temperatures. The geometric and material nonlinearity of beam structures under extreme temperature conditions are difficult to measure, even in offline settings. Even when measurement is possible, the associated experimental costs are prohibitively high, limiting widespread practical application. In contrast, at common temperatures such as room temperature, geometric and material nonlinearity—at least at the material level—can be readily characterized. Under such conditions, conventional finite element analysis is capable of providing high-fidelity solutions for complete stress fields. For extreme temperature environments, however, the scarcity of data on geometric and material nonlinearity prevents reliable numerical simulations from being conducted. This raises a very interesting issue: is it possible to use the data computed by the finite element analysis under several common temperatures to generate the full stress fields of the cantilever beam under an extreme level of the temperature?

Because the exact experimental measurement is difficult, we perform the finite element simulations at the common temperatures, considering the temperature-dependent material nonlinearity and geometric nonlinearity. The details are given in appendix G. Then the full stress fields of the cantilever beam at the common temperatures are assumed to be known. Specifically, the method first treats the obtained stress fields under seven common temperatures as the means of high-dimensional Gaussian distributions, generates synthetic samples by adding Gaussian noise, and then performs PCA dimensionality reduction on this sample set to $d=6$ dimensions \citep{Vol2003Principal} in order to train the transport mapping in CM-GAI. It should be stressed that the temperature is treated as the time variable for the transportation analysis. Based on the probability density distributions of the training set, the probability density distribution at target time is generated using the CM-GAI. Finally, perform a reverse reconstruction of the mean of the generated probability density distribution from the low-dimensional representation space back to the original high-dimensional data space to obtain the stress field.

\begin{figure}[H]
    \centering
        \begin{subfigure}[t]{0.39\textwidth}
        \centering
        \includegraphics[width=\textwidth]{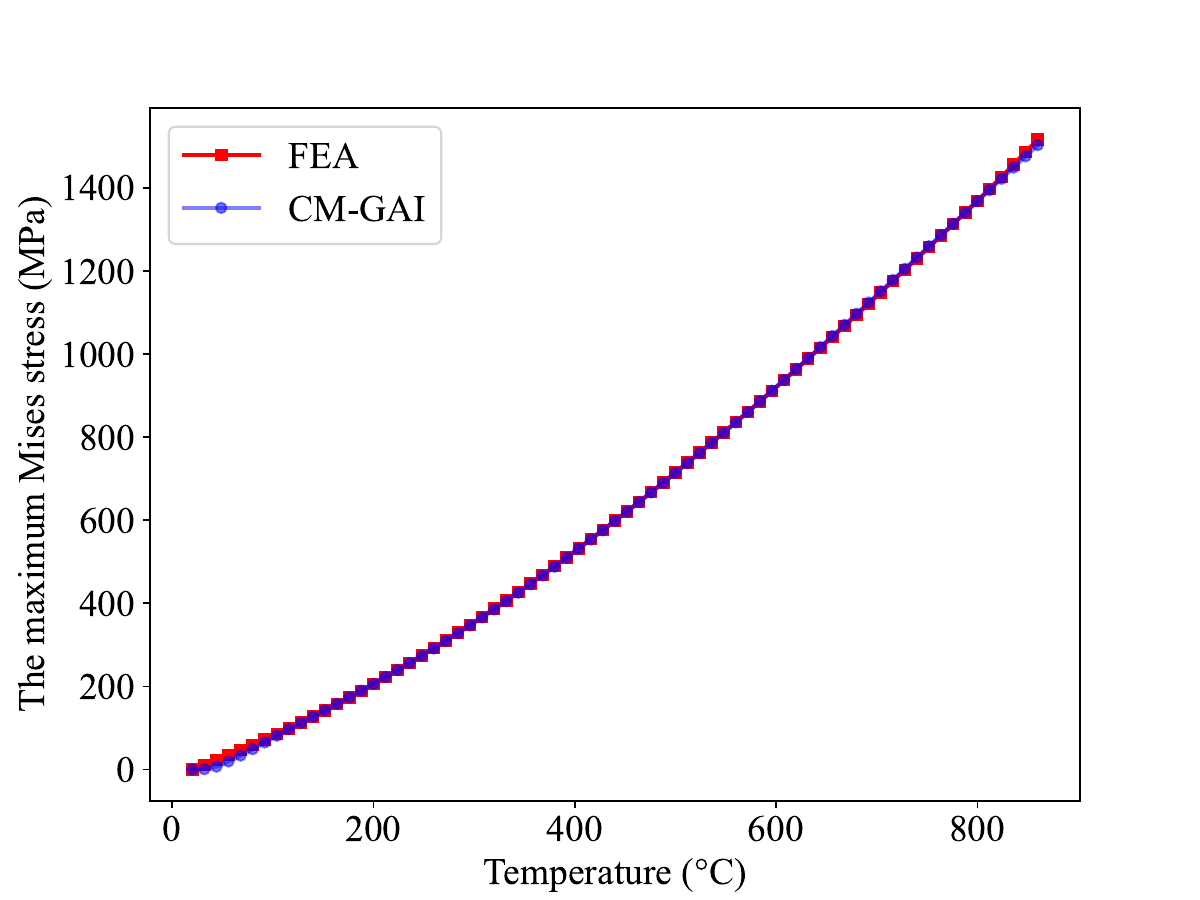}
        \caption{Comparison of the maximum Mises stress varies with temperature calculated by FEA and CM-GAI.}
        \label{scatter}
    \end{subfigure}    \begin{subfigure}[t]{0.60\textwidth}
        \centering
        \includegraphics[width=\textwidth]{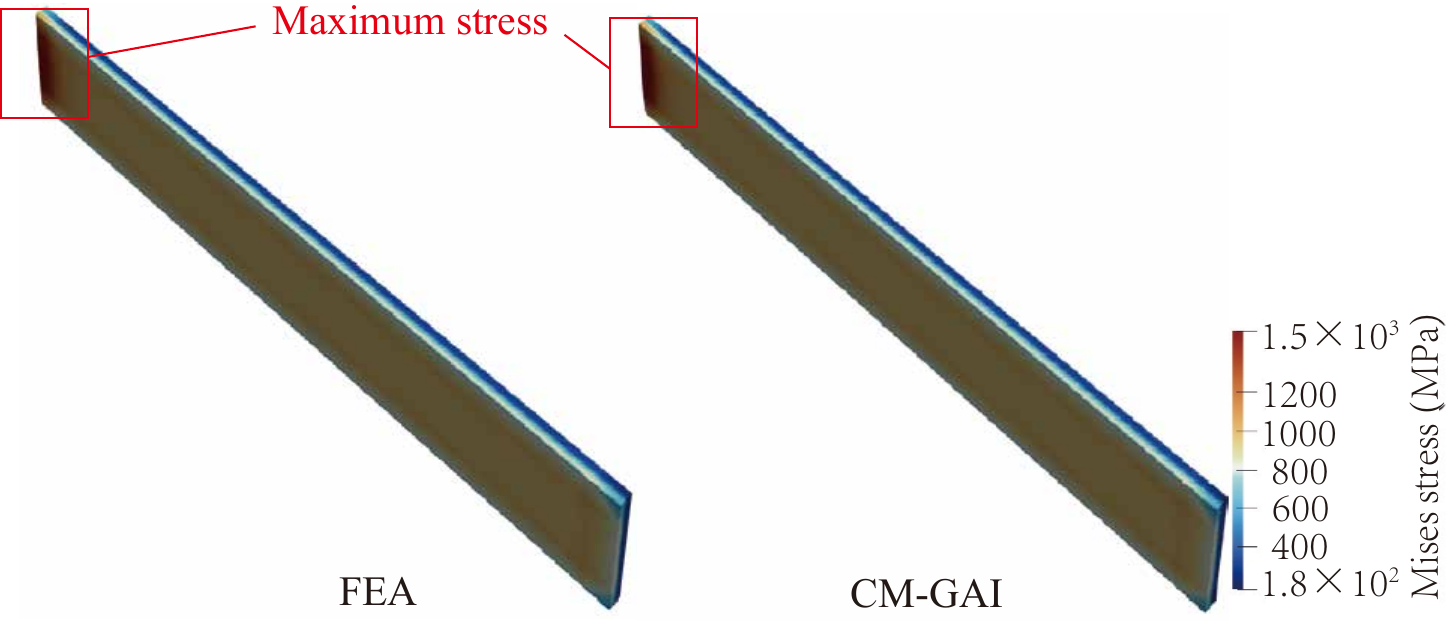}
        \caption{Thermal bending stress cloud diagram of cantilever beam calculated by FEA and CM-GAI.}
        \label{fea}
    \end{subfigure}
    \caption{Comparison of Mises stress calculated by FEA and CM-GAI.}
    \label{beam}
\end{figure}

The generations of the proposed CM-GAI method are compared with the results from finite element simulations at the extreme temperatures, where the finite element results at the extreme temperature serve as the benchmark solution for evaluating the accuracy of the proposed method (Strictly speaking, the experimental measurement at the high temperature during the operation should be taken as the benchmark). Figure \ref{scatter} shows a comparison between the curve of maximum stress versus temperature in the cantilever beam calculated by the finite element method and the corresponding temperature-dependent curve generated by the proposed method. It can be observed that during periods with a relatively smaller temperature difference between the upper and lower surfaces of the cantilever beam (0°C–720°C, corresponding to the range covered by the training dataset), the proposed method closely aligns with the finite element results, demonstrating good agreement in both the rising trend and magnitude of stress. During periods with a larger temperature difference between the upper and lower surfaces (720°C–840°C, corresponding to ranges not covered by the training dataset), the proposed method still accurately captures the variation characteristics of stress and generally aligns with the finite element results. At an upper surface temperature of 860°C (assumed as the extreme temperature in this case) and a lower surface temperature of 20°C, a comparison of the stress field contours between the generated results and the finite element calculations is shown in Figure \ref{fea}, with an NRMSE error of only 0.56\%.

Overall, the proposed method achieves accuracy comparable to that of high-fidelity finite element benchmarks in generating key physical variables of the temperature-dependent stress field of the cantilever beam under the temperature loading. If the stress response of the system under a set of known temperature fields is available, the response under a target temperature field can be generated by CM-GAI. This approach eliminates the need for a full thermomechanical coupling simulation at the extreme level of the temperature, which is hard to be carried out. CM-GAI offers an efficient new strategy for evaluating the thermo-mechanical behavior of engineering structures at the extreme temperature.

\subsection{Generation of plastic strain fields for a copper Taylor rod impacting on a rigid wall}
In this section, the effectiveness of CM-GAI is tested for generating the deformation fields for a copper Taylor rod impacting on a rigid wall with a initial velocity. It is a nonlinear transient dynamic process, involving the geometric nonlinearity due to large plastic deformations, material nonlinearity from the elastic-plastic behavior of copper and contact nonlinearity arising from the sudden imposition of fixed boundary conditions at the impact interface. The plastic strain field should be evaluated because it can assess the capability of energy absorbing during the impacting process. The model geometry is also given in appendix G. 

In real engineering applications, the copper Taylor rod may impact on the rigid wall with different initial velocities. This transient dynamics process of impacting is hard to be experimentally measured for the full fields of stress, strain and plastic strain. Even if it could be measured, it would result in prohibitively high experimental costs. However, when the initial impact velocity is relatively low, it is possible to record the deformation process by high speed camera, although the stress, strain and plastic strain inside the rod is still difficult to measure. At the material level, the mechanical response under the relatively low impacting speed can be measured through Hopkinson rod etc. The measured mechanical response may be embedded into the finite element analysis through user defined material. Then finite element simulations may provide high-fidelity solutions for the full stress, strain, plastic strain fields under the low impacting velocity. However, in high-velocity impact scenarios, finite element analysis may encounter convergence difficulties due to mesh distortion, necessitating specialized solution techniques. Then we are asking if it is possible to use the data computed by the finite element analysis under the relatively low impacting velocities to generate the full plastic strain fields of the rod under the extreme initial velocity.

As discussed in the last paragraph, the exact experimental measurement is difficult. We perform the finite element simulations with the lower initial impacting velocities, considering the geometric, material, and contacting nonlinearity. It should be stressed that this example is taken out from ABAQUS example manual. It is discussed in the manual that even if the impacting velocity is relatively low, the simulations may encounter many numerical issues such as mesh distortion and numerical instability. The interested reader can refer to the ABAQUS Example Manual and Appendix G for more details.

Now, it is assumed that the full equivalent plastic strain fields of the copper rod under the lower impacting velocity is known. The proposed CM-GAI method first treats the obtained equivalent plastic strain (PEEQ) fields under the lower impacting initial velocity as the means of high-dimensional Gaussian distributions, generates synthetic samples by adding Gaussian noise, and then performs PCA dimensionality reduction on this sample set to $d=6$ dimensions \citep{Vol2003Principal} in order to train the transport mapping in CM-GAI. It should be stressed that the initial velocity is now treated as the time variable for the transportation analysis. Based on the probability density distribution of the training set at a time (low initial velocity), the probability density distribution at target time (high initial velocity) is generated using CM-GAI. Finally, perform a reverse reconstruction of the mean of the generated probability density distribution from the low-dimensional (dimension $d=6$) representation space back to the original high-dimensional data space to obtain the plastic strain field.

\begin{figure}[H]
    \centering
        \begin{subfigure}[t]{0.35\textwidth}
        \centering
        \includegraphics[width=\textwidth]{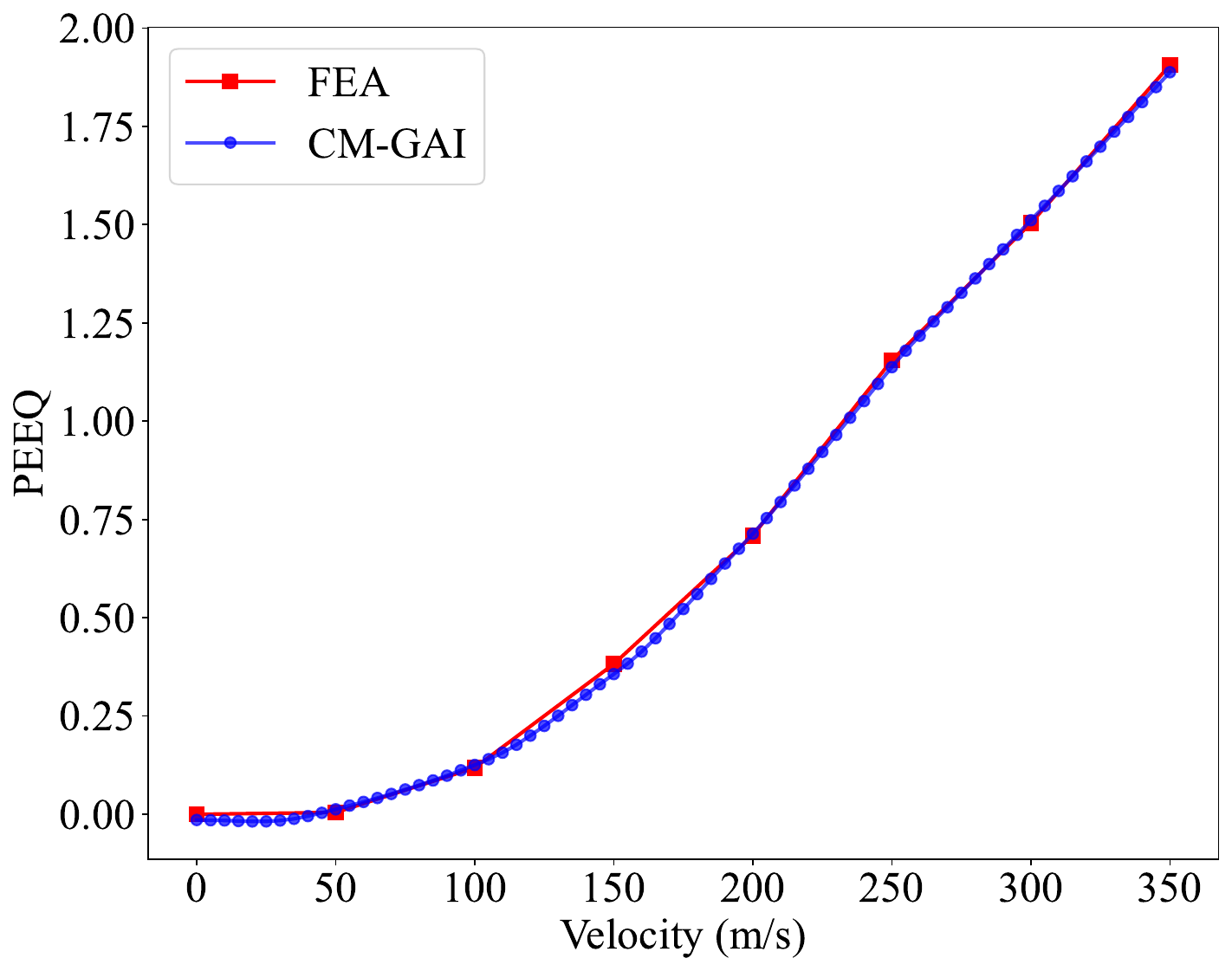}
        \caption{Comparison of the PEEQ of the selected point varies with impact velocity calculated by FEA and CM-GAI.}
        \label{peeqcurve}
    \end{subfigure}    \begin{subfigure}[t]{0.64\textwidth}
        \centering
        \includegraphics[width=\textwidth]{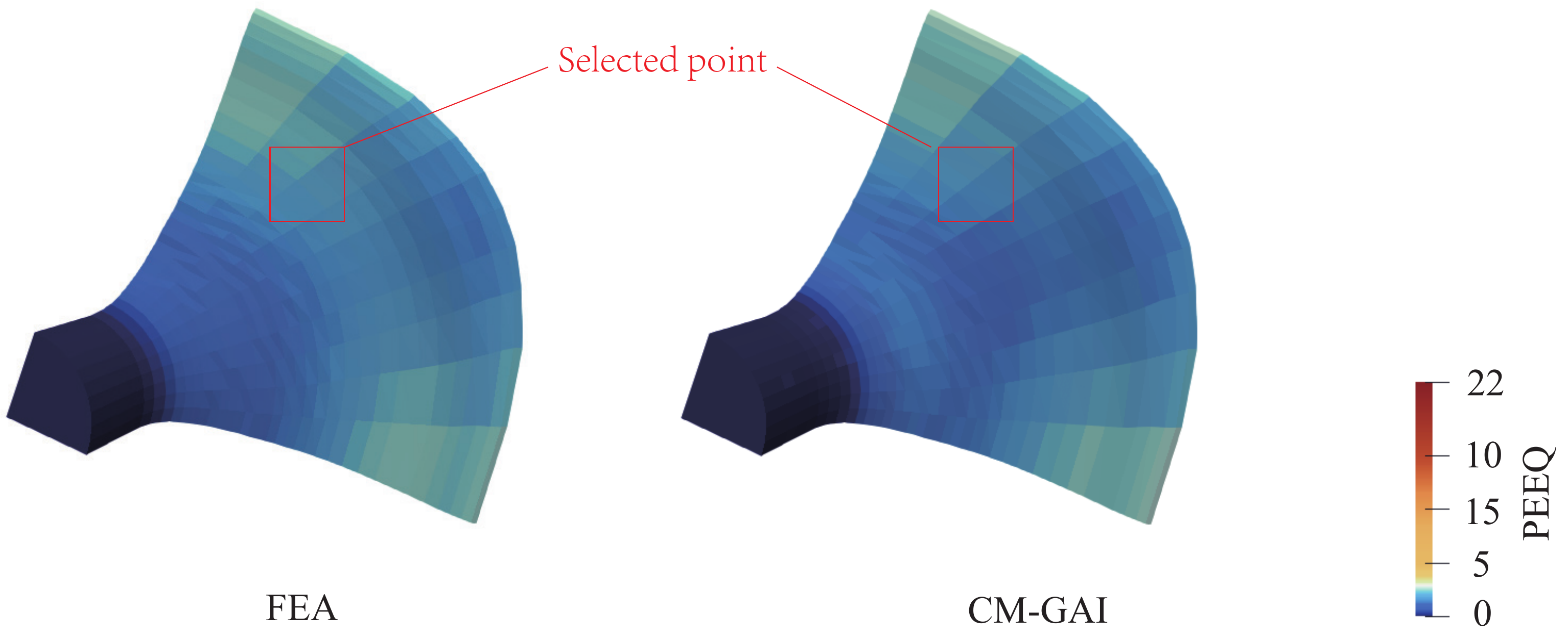}
        \caption{PEEQ cloud diagram of rod calculated by FEA and CM-GAI.}
        \label{peeqcontour}
    \end{subfigure}
    \caption{Comparison of PEEQ calculated by FEA and CM-GAI.}
    \label{peeq}
\end{figure}

The generation results of the proposed CM-GAI method are compared with those obtained from finite element simulations, where the finite element results serve as the benchmark solution for evaluating the accuracy of the proposed method. Figure \ref{peeqcurve} presents a comparison between the variation of PEEQ at a certain point at the final stage with respect to the impact velocity via finite element analysis and the corresponding curve generated by the proposed method. It can be observed that during the relative small impact velocity stage (0–300 m/s, corresponding to the time period covered by the training dataset), the results of the proposed method are in excellent agreement with the finite element results, showing good consistency in both the rising trend and amplitude of PEEQ. In the larger impact velocity stage (300–350 m/s, corresponding to the time period not covered by the training dataset), the proposed method still accurately captures the variation characteristics of PEEQ, remaining largely consistent with the finite element results. At the 350 m/s impact velocity stage, contour plots comparing the generated PEEQ fields with those from the finite element simulation are shown in Figure \ref{peeqcontour}, respectively, with NRMSE errors of only 1.10\%.

Overall, the proposed method achieves accuracy comparable to high-fidelity finite element benchmarks in generating key physical variables of the nonlinear transient dynamic response of the rod impacting on the rigid wall. Once the dynamic characteristics of the system are captured through some efforts under the low impact velocity, the more complicated behavior under some extreme high impacting velocity may be reliably extrapolated. This approach partially circumvents the need for high experimental cost and complicated explicit dynamic finite element simulations, thereby avoiding the convergence issues inherent in explicit dynamics simulations. It may offer a new computational way for evaluating the dynamic response of mechanical systems.

\section{Conclusion}
In this paper, we leverage the well-established continuum mechanics theory to apply optimal transport in engineering applications. The continuum mechanics theory for optimal transport is proposed and generalized based on the continuum mechanics concepts under the finite deformation framework, which can perform the generative tasks for engineering applications where training data can be very limited. Specifically, we take the concepts related to probability space from optimal transport and correspond them one to one with the concepts in continuum mechanics. The continuum mechanics formulation is developed in the reference configuration, where mass conservation becomes simpler than conventional mathematical treatments, while allowing natural application of Hamilton's principle.

The external body forces are introduced, which can be related to the intrinsic geometry of the probability space where the training data are available. The probability space can be non-Euclidean. As discussed in general relativity \citep{einstein2019relativity} and Amari \citep{amari2016}, the connections on the manifold can induce such external body forces. Constitutive laws governing the deformation of feature space can also be incorporated, with different laws leading to distinct deformation modes. While simple hyperelastic material laws are introduced here to demonstrate the method's capability, the framework can be extended via model-based or data-driven approaches. Importantly, this work presents the first theoretical consideration of the deformation of the feature space.

The dynamics of data transport and mass conservation are solved using Physics-Informed Neural Networks (PINNs). Multilayer perceptrons can parameterize the displacement field as a function of reference coordinates and time, enabling efficient solution of the governing equations for three benchmark problems. The resulting probability distribution facilitates effective generative tasks. PINNs offer a powerful numerical tool for high-dimensional mechanics problems. While traditional methods like Finite Element Method remain viable for 2D/3D transport problems, data transport applications provide new motivation for developing PINNs in high-dimensional settings—an area rarely explored in literature.

In the present work, CM-GAI is applied to 2D and 6D data transport, though extension to higher dimensions is necessary to fully validate the proposed theory and numerical methods. Dimension reduction techniques, crucial for mitigating the curse of dimensionality in machine learning, may benefit from mechanical insights \citep{huang2022}. However, formulating constitutive laws for feature space lacks the physical intuition available for 3D continua. Developing effective constitutive models for generative tasks remains challenging. Additionally, manifold learning should be pursued to identify the intrinsic geometry of data manifolds. These aspects present rich opportunities for future research.

\bigskip
\noindent \textbf{Acknowledgment} S.T. appreciates the financial support from the Liaoning Key Science and Innovation Program (2024JH1/11700046), the National Key Research and Development Program of China (2024YFB3310402, 2024YFB3-310401) and the National Natural Science Foundation of China (12372194).
\medskip
\newpage
\section*{Appendix A: Variations by the Hamilton principle}
In this section, the variational principle of the Lagrangian in the high-dimensional space is reproduced.

The variation of the total kinetic energy is:
\begin{eqnarray}
&&\int_{0}^{1}\delta Kdt\nonumber\\
	&=&\int_{0}^{1}\int_{\Omega _{0}}\frac{\partial \mathbf{u}\left( \mathbf{X}%
		,t\right) }{\partial t}\frac{\partial \delta \mathbf{u}\left( \mathbf{X}%
		,t\right) }{\partial t}\rho _{0}\left( \mathbf{X}\right) dVdt\nonumber\\
	&=&\int_{\Omega _{0}}\rho _{0}\left( \mathbf{X}\right) \int_{0}^{1}\frac{%
		\partial }{\partial t}\left[ \frac{\partial \mathbf{u}\left( \mathbf{X}%
		,t\right) }{\partial t}\delta \mathbf{u}\left( \mathbf{X},t\right) \right] -%
	\frac{\partial ^{2}\mathbf{u}\left( \mathbf{X},t\right) }{\partial t^{2}}%
	\delta \mathbf{u}\left( \mathbf{X},t\right) dtdV\nonumber\\
	&=&\int_{\Omega _{0}}\rho _{0}\left( \mathbf{X}\right) \left[ \left. \frac{%
		\partial \mathbf{u}\left( \mathbf{X},t\right) }{\partial t}\right\vert
	_{t=1}\delta \mathbf{u}\left( \mathbf{X},t=1\right) -\left. \frac{\partial 
		\mathbf{u}\left( \mathbf{X},t\right) }{\partial t}\right\vert _{t=0}\delta 
	\mathbf{u}\left( \mathbf{X},t=0\right) \right] dV\nonumber\\
	&-&\int_{\Omega _{0}}\int_{0}^{1}\rho _{0}\left( \mathbf{X}\right) \frac{%
		\partial ^{2}\mathbf{u}\left( \mathbf{X},t\right) }{\partial t^{2}}\delta 
	\mathbf{u}\left( \mathbf{X},t\right) dtdV\nonumber\\
	&=&\int_{\Omega _{0}}\rho _{0}\left( \mathbf{X}\right) \left[ \left. \frac{%
		\partial \mathbf{u}\left( \mathbf{X},t\right) }{\partial t}\right\vert
	_{t=1}\delta \mathbf{u}\left( \mathbf{X},t=1\right) -\left. \frac{\partial 
		\mathbf{u}\left( \mathbf{X},t\right) }{\partial t}\right\vert _{t=0}\delta 
	\mathbf{u}\left( \mathbf{X},t=0\right) \right] dV\nonumber\\
	&-&\int_{\Omega _{0}}\int_{0}^{1}\rho _{0}\left( \mathbf{X}\right) \frac{%
		\partial ^{2}\mathbf{u}\left( \mathbf{X},t\right) }{\partial t^{2}}\delta 
	\mathbf{u}\left( \mathbf{X},t\right) dtdV
\end{eqnarray}%

To derive the variation of the strain energy, the first Piola-Kirchoff (PK) stress can be defined:
\begin{equation}
P_{ki} = \frac{\partial W}{\partial F_{ki}}
\label{PKstress}
\end{equation}

Then the variation of the strain energy density is given:
\begin{equation}
\delta W = P_{ki}\delta F_{ki}
\end{equation}
where the variation of the deformation gradient is:
\begin{equation}
\delta F_{ki} = \frac{\partial \delta u_{k}}{\partial X_{i}}
\end{equation}

Then the variation of the total strain energy is:
\begin{eqnarray}
	&&\int_{0}^{1}\delta Udt\nonumber\\
	&=&\int_{0}^{1}\int_{\Omega _{0}}\delta W\rho _{0}\left( \mathbf{X}\right)
	dVdt\nonumber\\
	&=&\int_{0}^{1}\int_{\Omega _{0}}P_{ki}\frac{\partial \delta
		u_{k}}{\partial X_{i}}\rho _{0}\left( \mathbf{X}\right) dVdt\nonumber\\
	&=&\int_{0}^{1}\int_{\Omega _{0}}-\frac{\partial P_{ki}}{%
		\partial X_{i}}\delta u_{k}\rho _{0}\left( \mathbf{X}\right)
	dVdt+\int_{0}^{1}\int_{\partial \Omega _{0}}P_{ki}\delta
	u_{k}\rho _{0}\left( \mathbf{X}\right) dSdt
\end{eqnarray}%

Substituting $\delta K$ and $\delta U$ into Eq. \ref{Hamilton}, we obtain:
\begin{equation}
\frac{\partial ^{2}u_{k}\left( \mathbf{X},t\right) }{\partial t^{2}}-\frac{\partial P_{ki}}{\partial X_{i}}-F_{b}^{k}=0
\end{equation}
where $\delta u(X,t)$ is arbitrary inside the body. It is assumed that $\delta u(\mathbf{X},t=0)=0$, $\delta u(\mathbf{X},t=1)=0$ and $\delta u(\mathbf{X},t)=0$ on the boundary $\partial \Omega_{0}$. When the strain energy density $W$ is given by Eq. \ref{W-neoHooke}, the first PK stress is:
\begin{equation}
P_{ki} = G F_{ki}
\label{PKstress-neohooke}
\end{equation}

Substituting it into the above equation, we obtain Eq. \ref{equilirium}.

\section*{Appendix B: Dynamics on the Lobachevsky plane}
In this appendix, we analyze the dynamics of data on the Lobachevsky plane, which serves as a simple example of a non-Euclidean space. This section aims to provide insight into how to achieve generative tasks through manifold learning. On the Lobachevsky plane with coordinate $(x^{1},x^{2})$, the measure of distance ($dl$, the differential length) is defined by:
\begin{equation}
dl^{2}=\frac{\left( dx^{1}\right) ^{2}+\left( dx^{2}\right) ^{2}}{\left(
	x^{2}\right) ^{2}}
\end{equation}
The components of the metric tensor are:
\begin{eqnarray}
	g_{11} &=&\frac{1}{\left( x^{2}\right) ^{2}}\qquad g_{22}=\frac{1}{\left(
		x^{2}\right) ^{2}}\qquad g_{12}=g_{21}=0 \\
	g^{11} &=&\left( x^{2}\right) ^{2}\qquad g^{22}=\left( x^{2}\right)
	^{2}\qquad g^{12}=g^{21}=0
\end{eqnarray}%
According to Eq. (\ref{riemannn connection equation}),
\begin{eqnarray}
	\Gamma _{11}^{1} &=&0\qquad \Gamma _{12}^{1}=\Gamma _{21}^{1}=-\frac{1}{x^{2}%
	} \\
	\Gamma _{22}^{1} &=&0\qquad \Gamma _{12}^{2}=\Gamma _{21}^{2}=0\qquad \Gamma
	_{22}^{2}=-\frac{1}{x^{2}}
\end{eqnarray}%
Substituting into Eq. (\ref{motion-non-Elucidean}), the equations of the dynamics of transportation can be simplified without considering the strain energy:
\begin{equation}
\frac{d^{2}x^{1}}{dt^{2}}+\left( -\frac{2}{x^{2}}\right) \frac{dx^{1}}{dt}%
\frac{dx^{2}}{dt}=0
\end{equation}
\begin{equation}
\frac{d^{2}x^{2}}{dt^{2}}+\frac{1}{x^{2}}\frac{dx^{1}}{dt}\frac{dx^{1}}{dt}-%
\frac{1}{x^{2}}\frac{dx^{2}}{dt}\frac{dx^{2}}{dt}=0
\end{equation}
To solved the above differential equations, the energy conservation on the geodesic trajectory should be satisfied:
\begin{equation}
g_{11}\left( \frac{dx^{1}}{dt}\right) ^{2}+g_{22}\left( \frac{dx^{2}}{dt}%
\right) ^{2}=C
\end{equation}
Substituting $g_{11}$ and $g_{22}$ and letting $C=1$ without losing generality, it obtains:
\begin{equation}
\left(\frac{dx^{1}}{dt}\right) ^{2}+\left( \frac{dx^{2}}{dt}\right)
^{2}=\left( x^{2}\right) ^{2}
\end{equation}
\begin{figure}[htbp]
	\centering
	\includegraphics[width=0.45\textwidth]{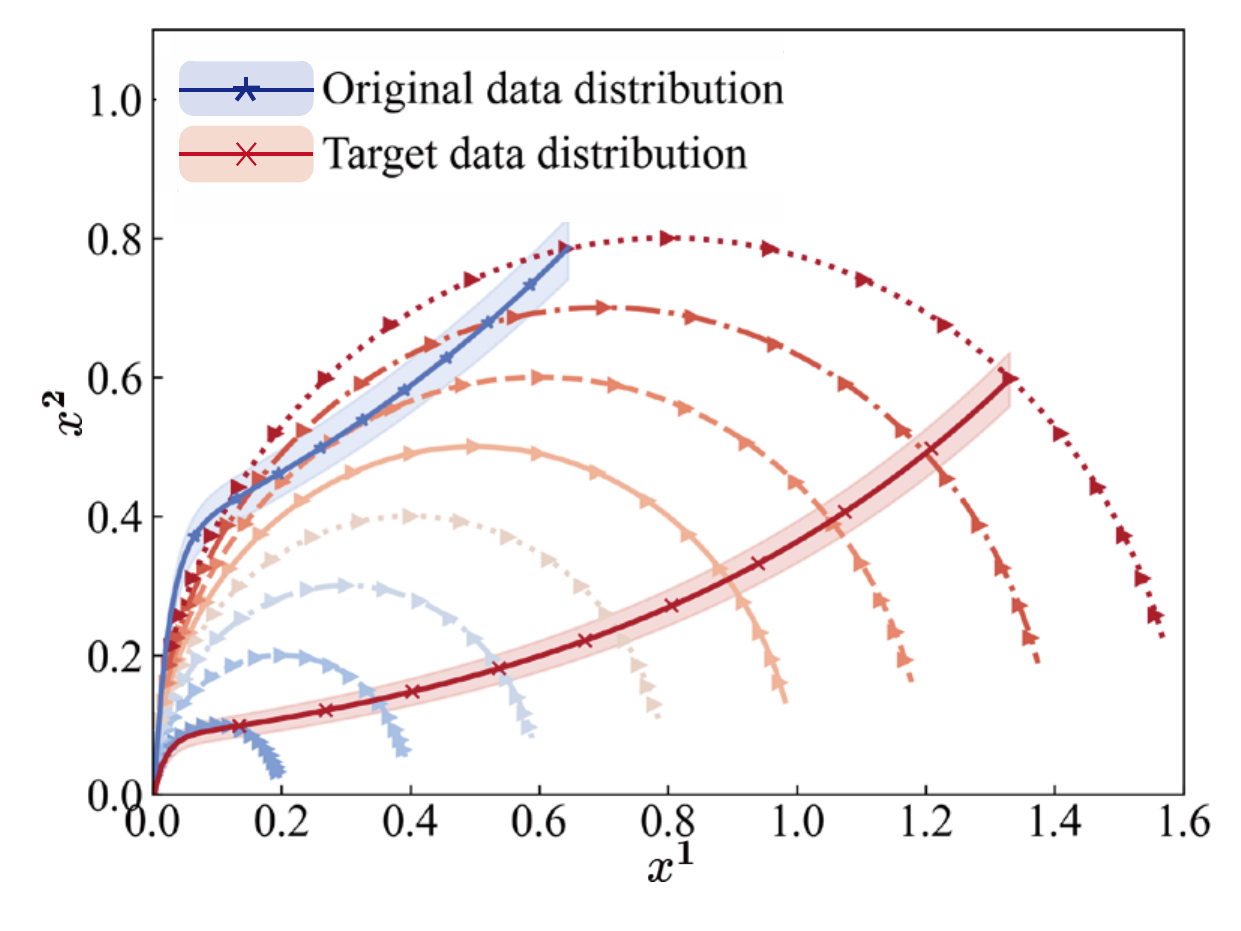}
	\setlength{\abovecaptionskip}{1mm}
	\setlength{\belowcaptionskip}{-2mm}
	\caption{The solutions of the transportation dynamics on the Lobachevsky plane.}
	\label{fig:Fig 17}
\end{figure}

By employing symbolic computation software, we obtain the following solutions: 
\begin{eqnarray}
	x^{1} &=&X_0-2Y_{0}\left( \exp (-2t)+1\right) \frac{\exp \left( 2t\right) }{%
		\left( 1+\exp (2t)\right) ^{2}}+Y_{0} \\
	x^{2} &=&\frac{2\exp (t)Y_{0}}{1+\exp (2t)}
\end{eqnarray}%
where $X_{0}$ and $Y_{0}>0$ are arbitrary constants. When $t=0$, $x^{1}=X_{0}$ and $x^{2}=Y_{0}$. If the transportation originates from a position other than ($X_{0},Y_{0}$), the time in the solution should be shifted. The semi-circular arc in Figure \ref{fig:Fig 17} represents the solutions, illustrating the optimal transport trajectory from the original data distribution to the target data distribution. Another solution is: 
\begin{eqnarray}
	x^{1} &=&X_{0} \\
	x^{2} &=&\sqrt{C_{1}t+C_{2}}
\end{eqnarray}%
where $x_0$, $C_1$ and $C_2$ are constant. In fact, it is a line perpendicular to the axis of $x^1$.

However, given the probability distribution, it can be seen that the data may not be on the Lobachevsky plane because the related metric tensor cannot satisfy the mass conservation (Eq. \ref{nonE-mass conservation}). To align the data with the manifold, the metric tensor must be learned adaptively. This opens significant avenues for future research.

\section*{Appendix C: Neural network architecture and training hyperparameters}{\label{appendix c}}

This section presents the neural network architectures and the hyperparameters utilized in the proposed CM-GAI framework. In the first case in results and discussion, being relatively straightforward, employs a standard four-layer fully connected neural network with an input dimension of 3, an output dimension of 2, and three hidden layers of dimensions 80, 120, and 160, each utilizing the SELU activation function. In the remaining two more complex cases, to simultaneously achieve lower training loss and enhanced generalization capability, D-NN adopts a Fourier feature-enhanced spectral neural network architecture. It first receives the input vector $\mathbf{X}$ and the pseudo time $t$, then performs a linear transformation and scaling of the input using a learnable spectral weight matrix and a Fourier scaling parameter. Subsequently, it computes the sine and cosine values respectively, and concatenates these sinusoidal features, cosine features, and the original vector along the feature dimension to form a high-dimensional feature representation. This feature is fed into a four-layer fully connected network: the first layer maps the feature dimension to 256 dimensions; the second and third layers maintain a 256-dimensional hidden layer, each followed by a Softplus activation function with $\beta=10$; the fourth layer maps the 256 dimensions back to the feature dimension of the input vector $\mathbf{X}$. The weights of this layer are initialized with a normal distribution of standard deviation 0.001, and the biases are initialized to zero, ensuring that the initial output is close to zero. The network output is then element-wise multiplied by a learnable per-dimension scaling parameter. This design ensures that the network approximates a zero mapping at the beginning of training and gradually learns a smooth displacement field as training progresses. F-NN consists of an input layer, 7 hidden layers, and an output layer stacked sequentially. The network employs linear layers for dimensional transformation and utilizes the SELU activation function to enhance nonlinear representation capabilities, while incorporating Dropout (with a rate of 0.1) after each hidden layer to mitigate overfitting. The overall structure is connected sequentially, achieving a mapping from the inputs $\mathbf{x}$ and the pseudo time $t$ to the output body force. Our machine learning model was implemented using the PyTorch framework \citep{pytorch}, with the architectural parameters outlined in Table \ref{model parameter}. The model was trained utilizing the Adam optimizer \citep{adam} with a learning rate of $1\times10^{-3}$. The specific data and implementation code can be provided upon request.

\begin{table}[H]
\centering
\caption{Neural network architectures and training hyperparameters. }
\label{tab:model-parameter}
\begin{tabular}{c|ccccc}
\hline
Model  & Number of Neurons in Hidden Layer & Activation Function& Learning Rate\\
\hline
D-NN  & 256   & Softplus  &$1\times10^{-3}$\\
F-NN  & 300   & SELU  &$1\times10^{-3}$\\
\hline
\end{tabular}
\label{model parameter}
\end{table}

\section*{Appendix D: Further numerical examples}{\label{appendix d}}
\begin{figure}[H]
    \centering
    \begin{subfigure}[t]{0.45\textwidth}
        \centering
        \includegraphics[width=\textwidth]{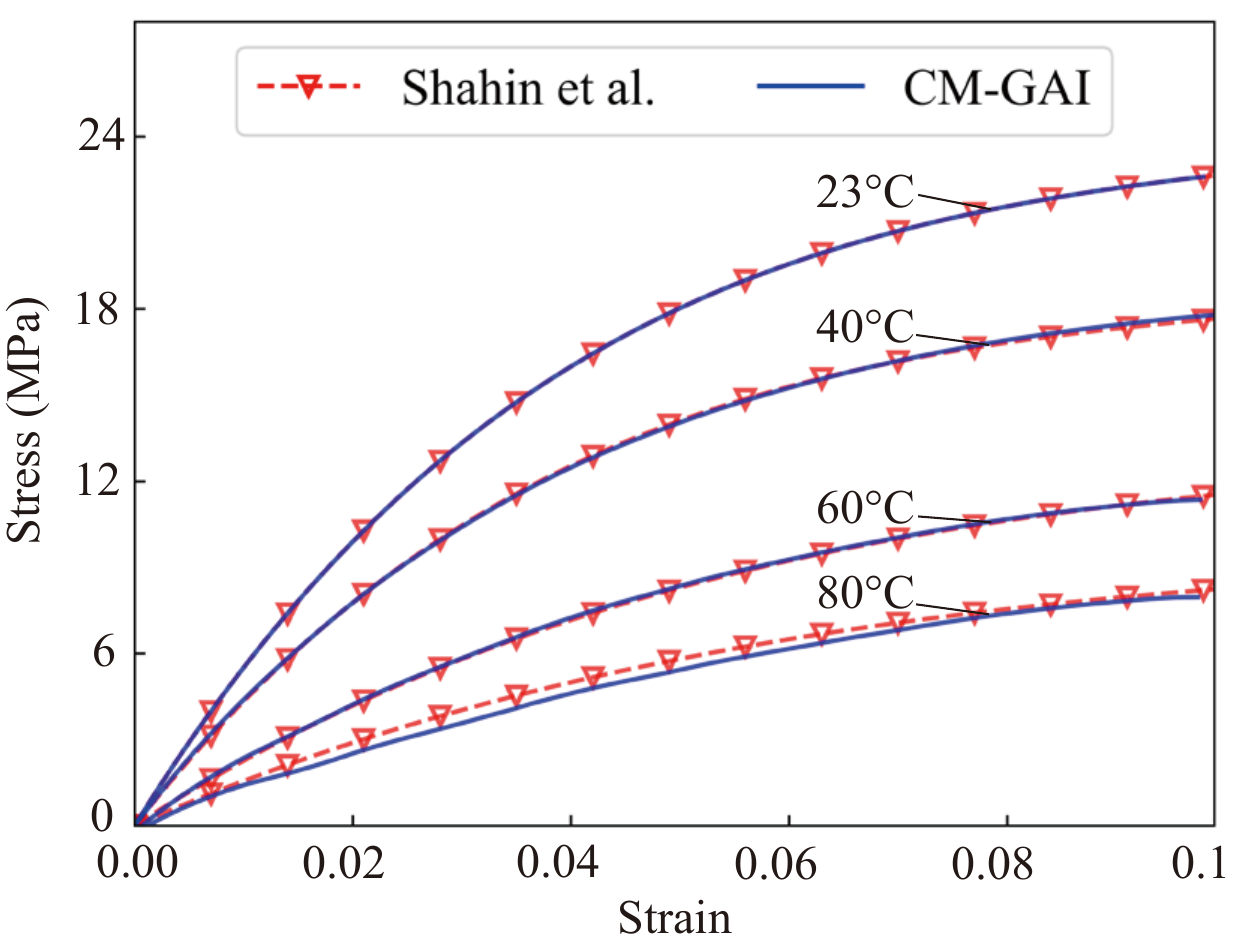}
        \caption{The stress-strain curves at different temperatures.}
        \label{fig:shahina}
    \end{subfigure}
    \begin{subfigure}[t]{0.45\textwidth}
        \centering
        \includegraphics[width=\textwidth]{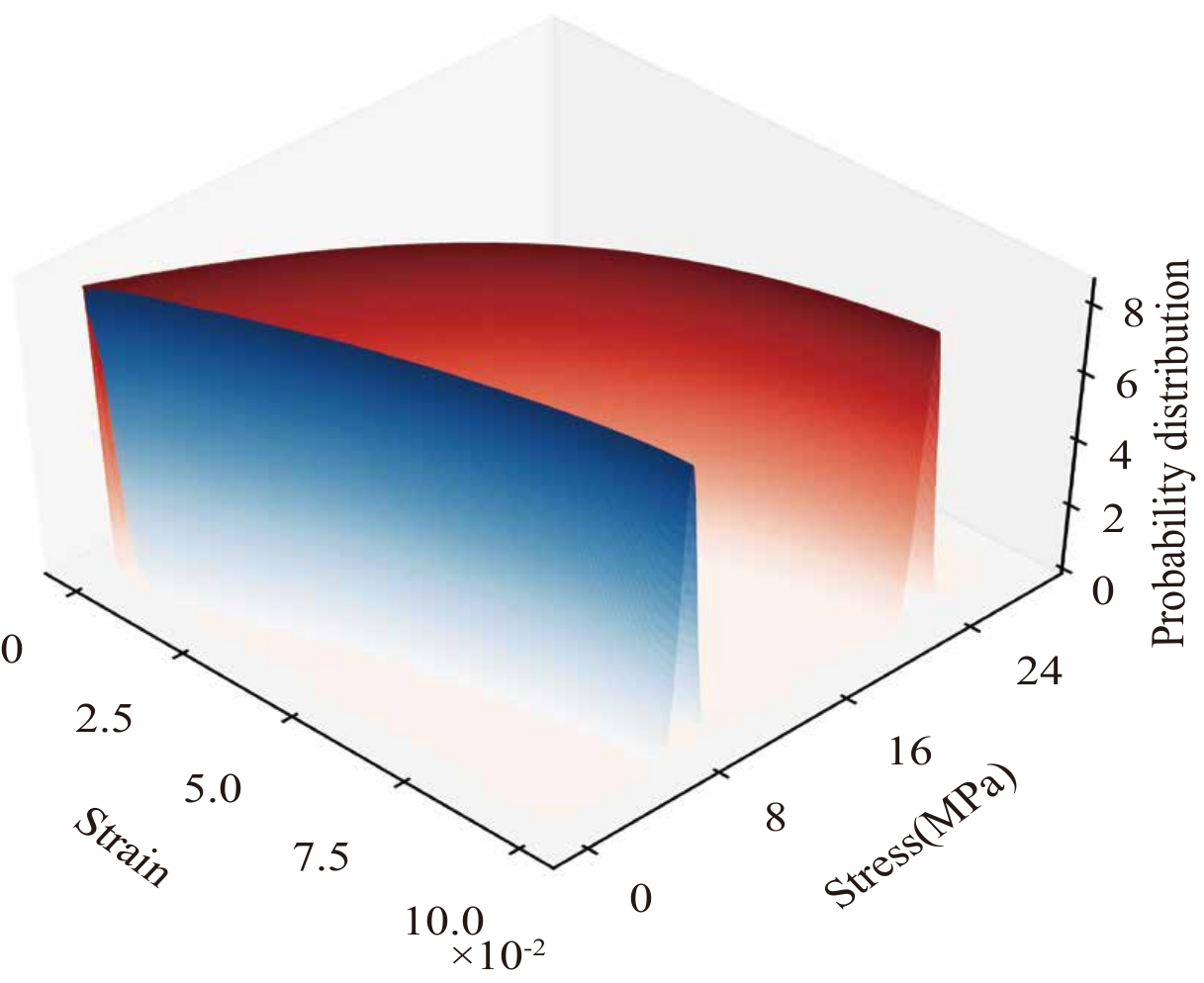}
        \caption{Original probability distribution and the final probability distribution after transportation calculated by CM-GAI.}
        \label{fig:shahinb}
    \end{subfigure}
    \caption{The comparison between Shahin et al.'s experiment \citep{shahin2020} and CM-GAI predictions.}
    \label{fig:shahin}
\end{figure}

High-density polyethylene (HDPE) is used in aerospace applications where materials are subjected to thermal cyclic loadings. Its mechanical behavior is highly dependent on temperature. The temperature-dependent stress-strain response of HDPE has been investigated \citep{shahin2020}. Based on their work, CM-GAI is employed to predict the stress-strain behavior of HDPE at $80\,^\circ\mathrm{C}$ using data from other temperatures ($23\,^\circ\mathrm{C}, 40\,^\circ\mathrm{C}, 60\,^\circ\mathrm{C}$). The results are shown in Figure \ref{fig:shahin}.

\begin{figure}[H]
    \centering
    \begin{subfigure}[t]{0.45\textwidth}
        \centering
        \includegraphics[width=\textwidth]{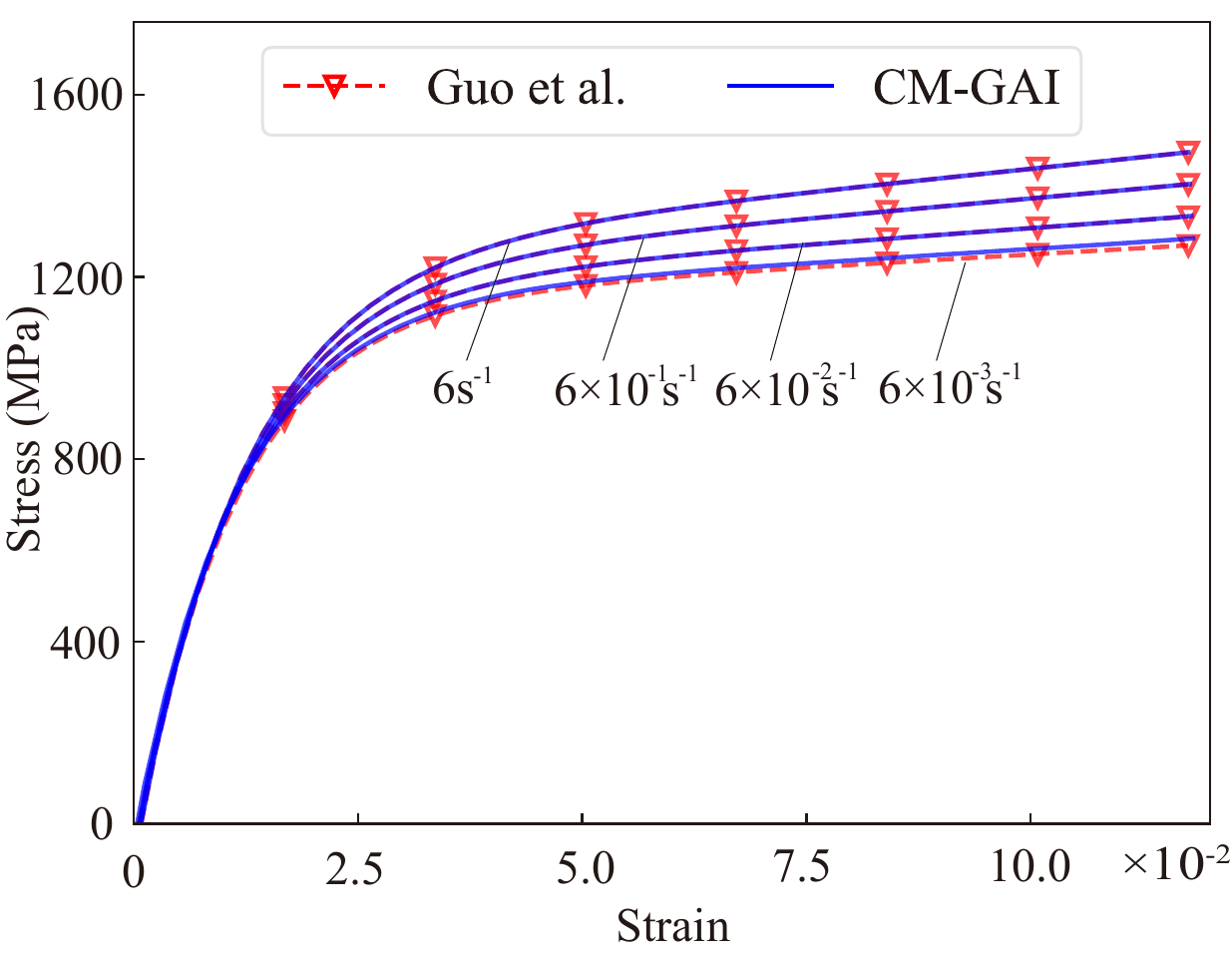}
        \caption{The stress-strain curves at the different strain rates.}
        \label{fig:guo1a}
    \end{subfigure}
    \begin{subfigure}[t]{0.45\textwidth}
        \centering
        \includegraphics[width=\textwidth]{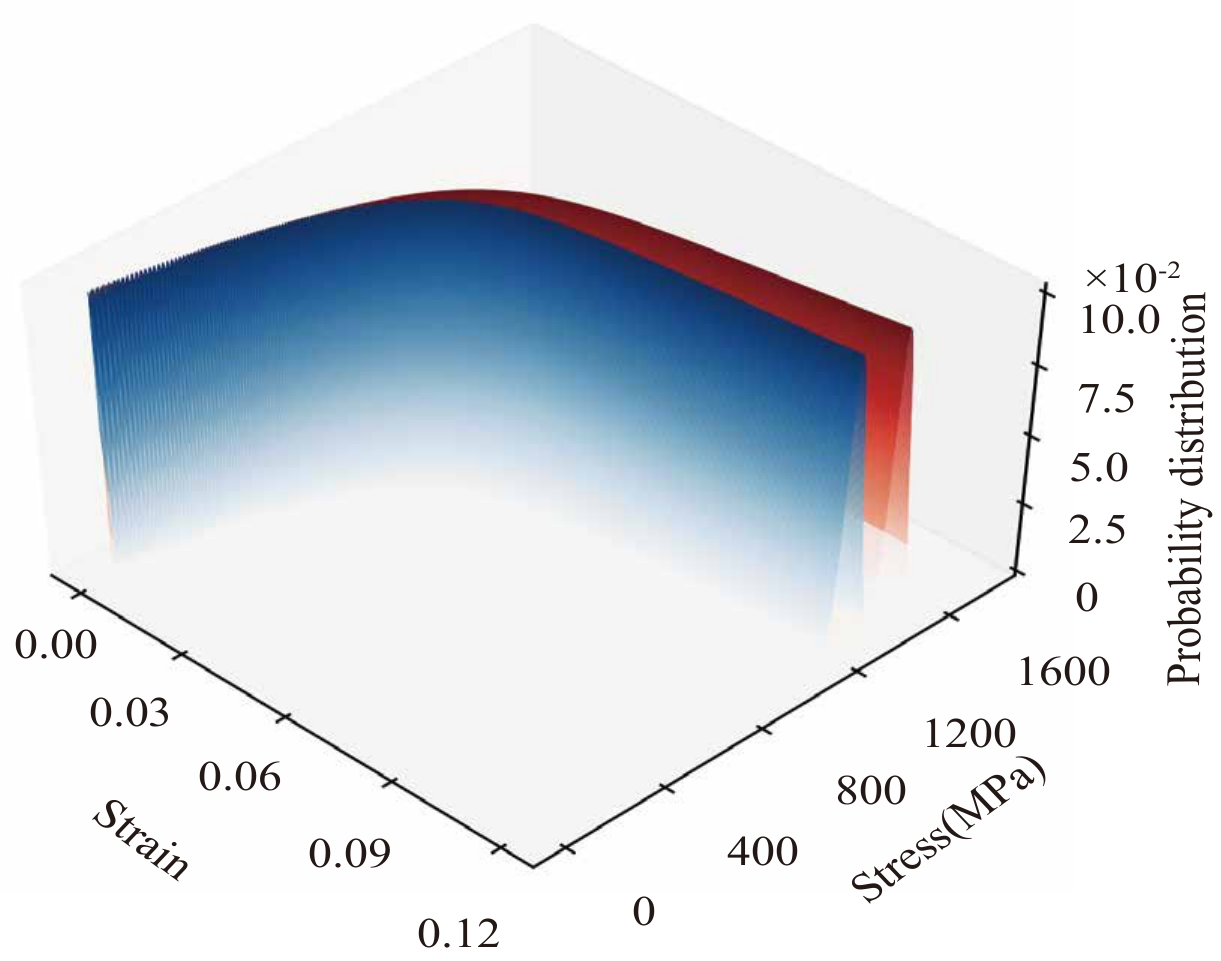}
        \caption{Original probability distribution and the final probability distribution after transportation calculated by CM-GAI.}
        \label{fig:guo1b}
    \end{subfigure}
    \caption{The comparison between Guo et al.'s model \citep{guo2020} and CM-GAI predictions.}
    \label{fig:guo1}
\end{figure}

\begin{figure}[H]
    \centering
    \begin{subfigure}[t]{0.45\textwidth}
        \centering
        \includegraphics[width=\textwidth]{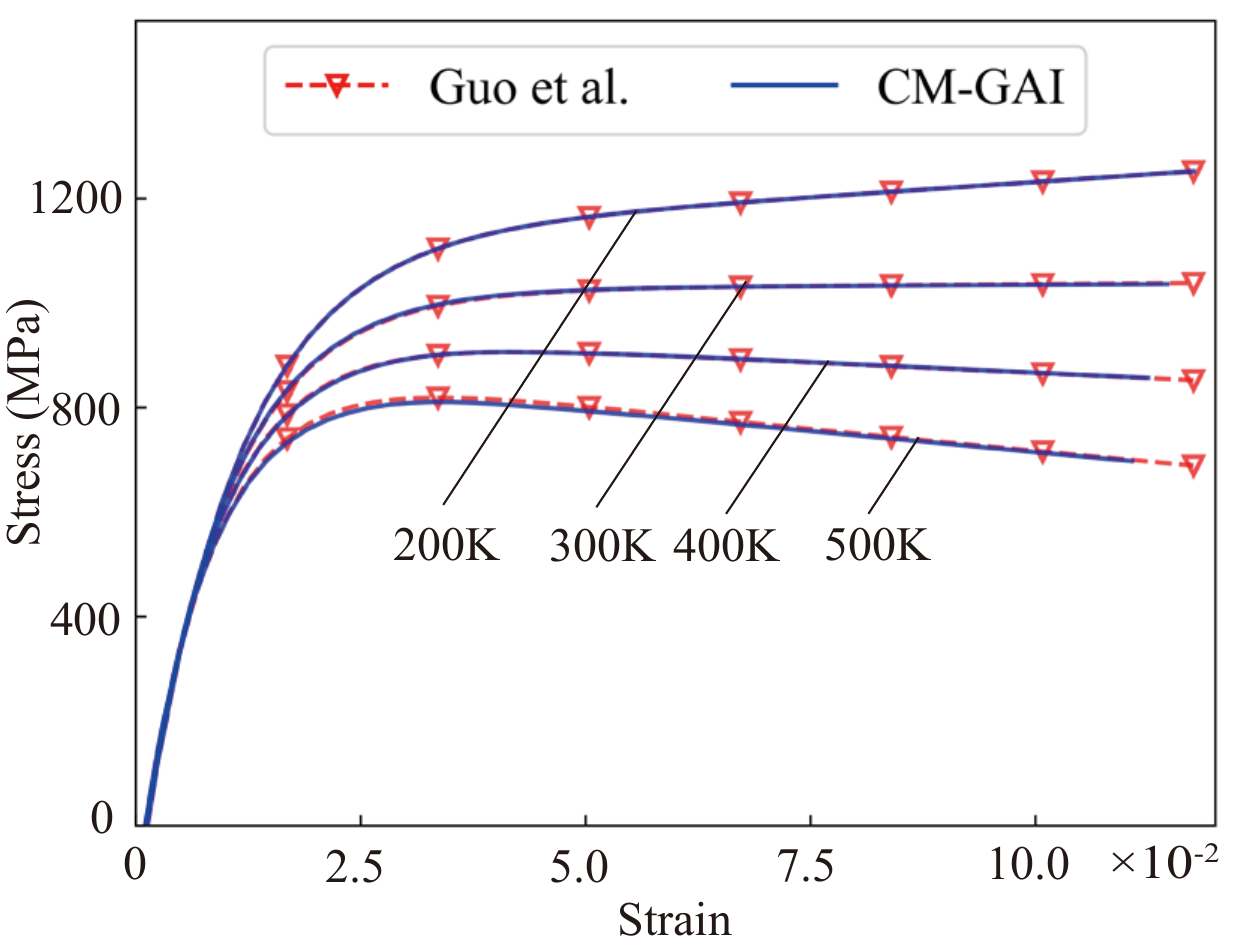}
        \caption{The stress-strain curves at the different temperatures.}
        \label{fig:guo2a}
    \end{subfigure}
    \begin{subfigure}[t]{0.45\textwidth}
        \centering
        \includegraphics[width=\textwidth]{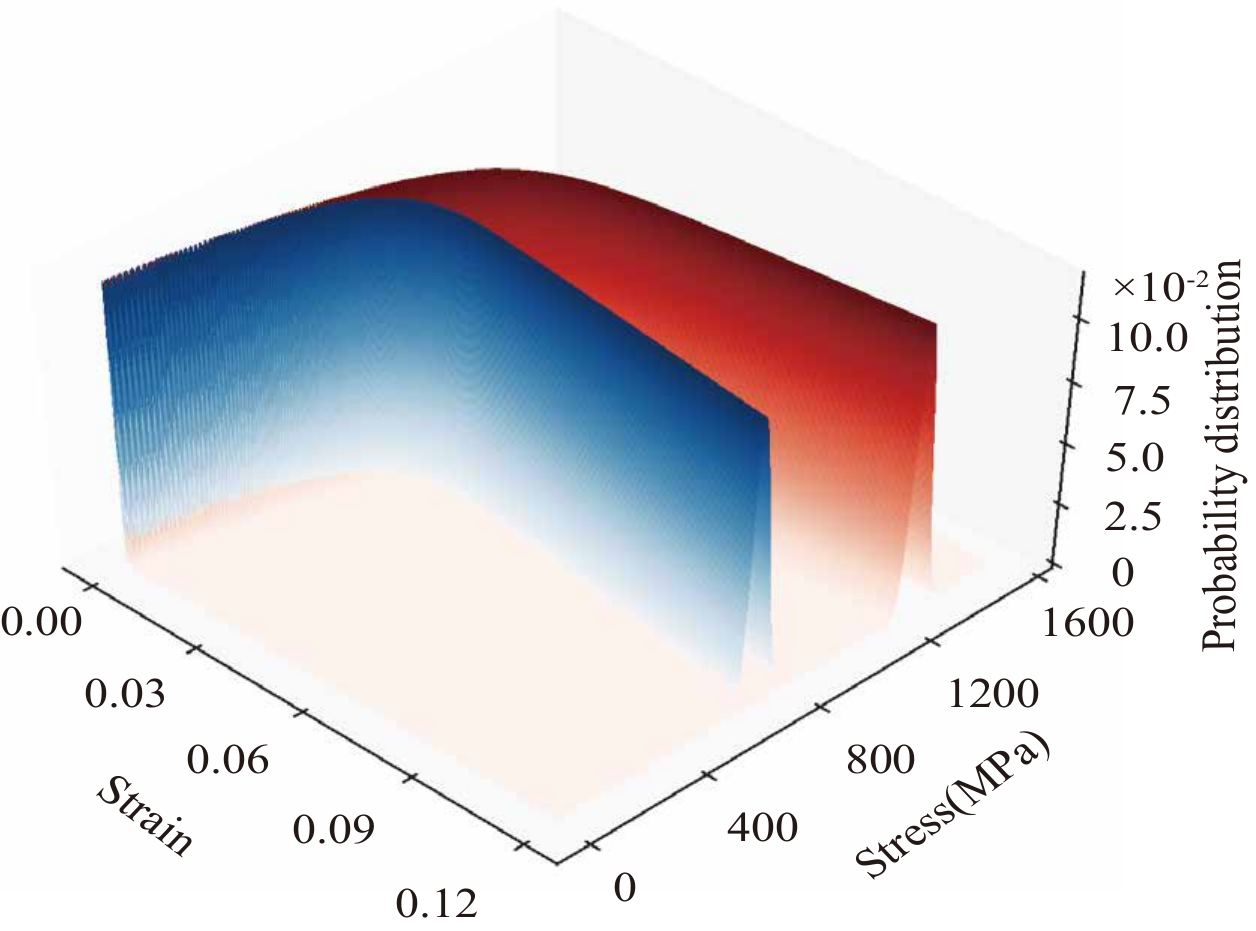}
        \caption{Original probability distribution and the final probability distribution after transportation calculated by CM-GAI.}
        \label{fig:guo2b}
    \end{subfigure}
    \caption{The comparison between Guo et al.'s model \citep{guo2020} and CM-GAI predictions.}
    \label{fig:guo2}
\end{figure}

Nanotwinned copper (ntCu) is widely used in microelectronic and micro-electromechanical systems devices, where both temperature fluctuations and high strain rate loading can significantly impact mechanical performance. The stress - strain behavior of ntCu, which depends on temperature and strain rate, has been studied \citep{guo2020}. Based on their work, CM-GAI is used to predict the stress - strain response of ntCu under target conditions of a strain rate of $6\text{s}^{-1}$ and a temperature of $500$K, using data obtained from other measurements. This is shown in Figure \ref{fig:guo1} and Figure \ref{fig:guo2}, respectively. (In the investigation of temperature-dependent mechanical properties of ntCu, it was observed that the use of SELU activation function led to a sharp increase in loss values during the initial training phase, which rendered the training process entirely ineffective. Consequently, in this case study, we switched to the LeakyReLU activation function as an alternative solution.)

\begin{figure}[H]
    \centering
    \begin{subfigure}[t]{0.45\textwidth}
        \centering
        \includegraphics[width=\textwidth]{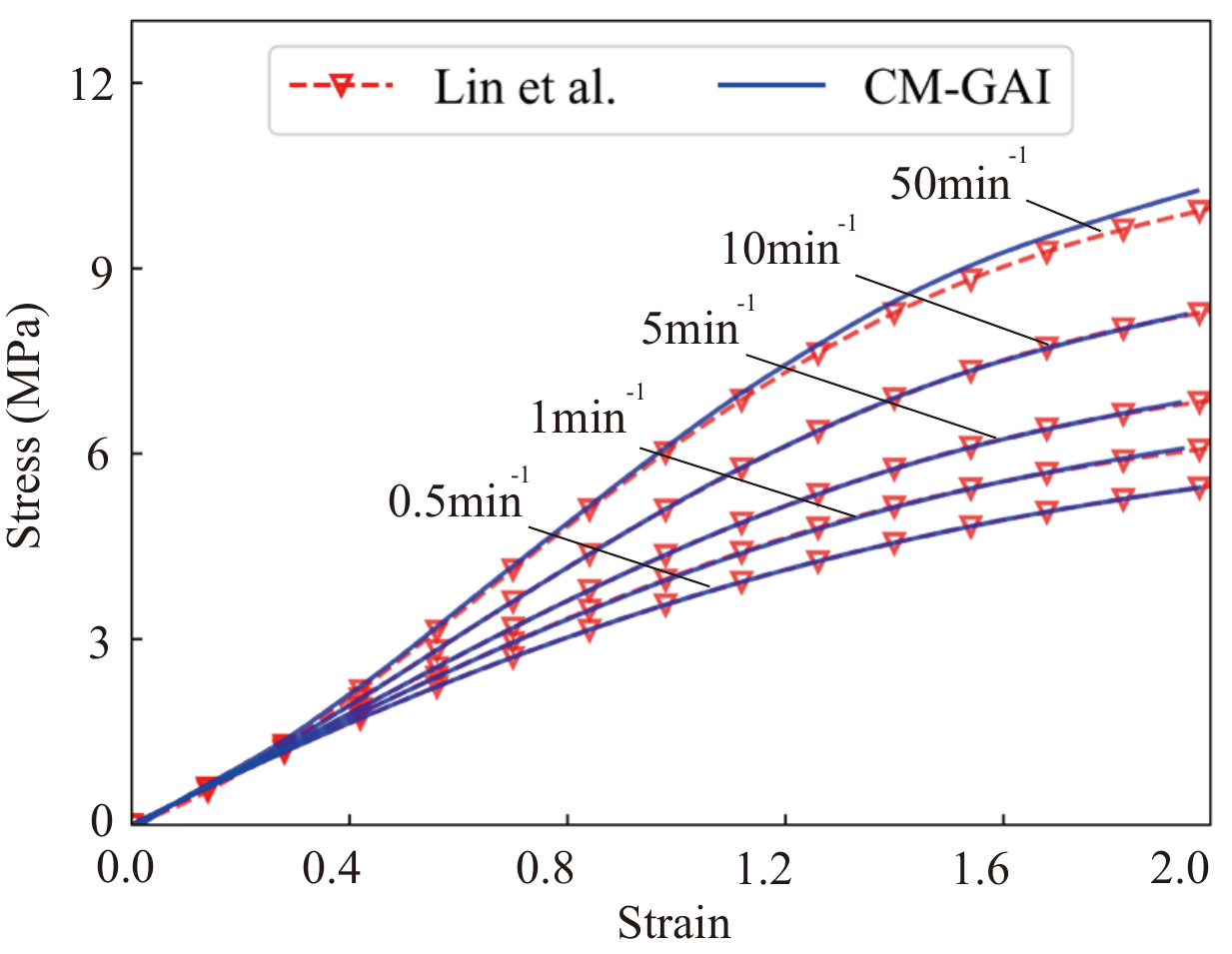}
        \caption{The stress-strain curves at the different strain rates.}
        \label{fig:lina}
    \end{subfigure}
    \begin{subfigure}[t]{0.45\textwidth}
        \centering
        \includegraphics[width=\textwidth]{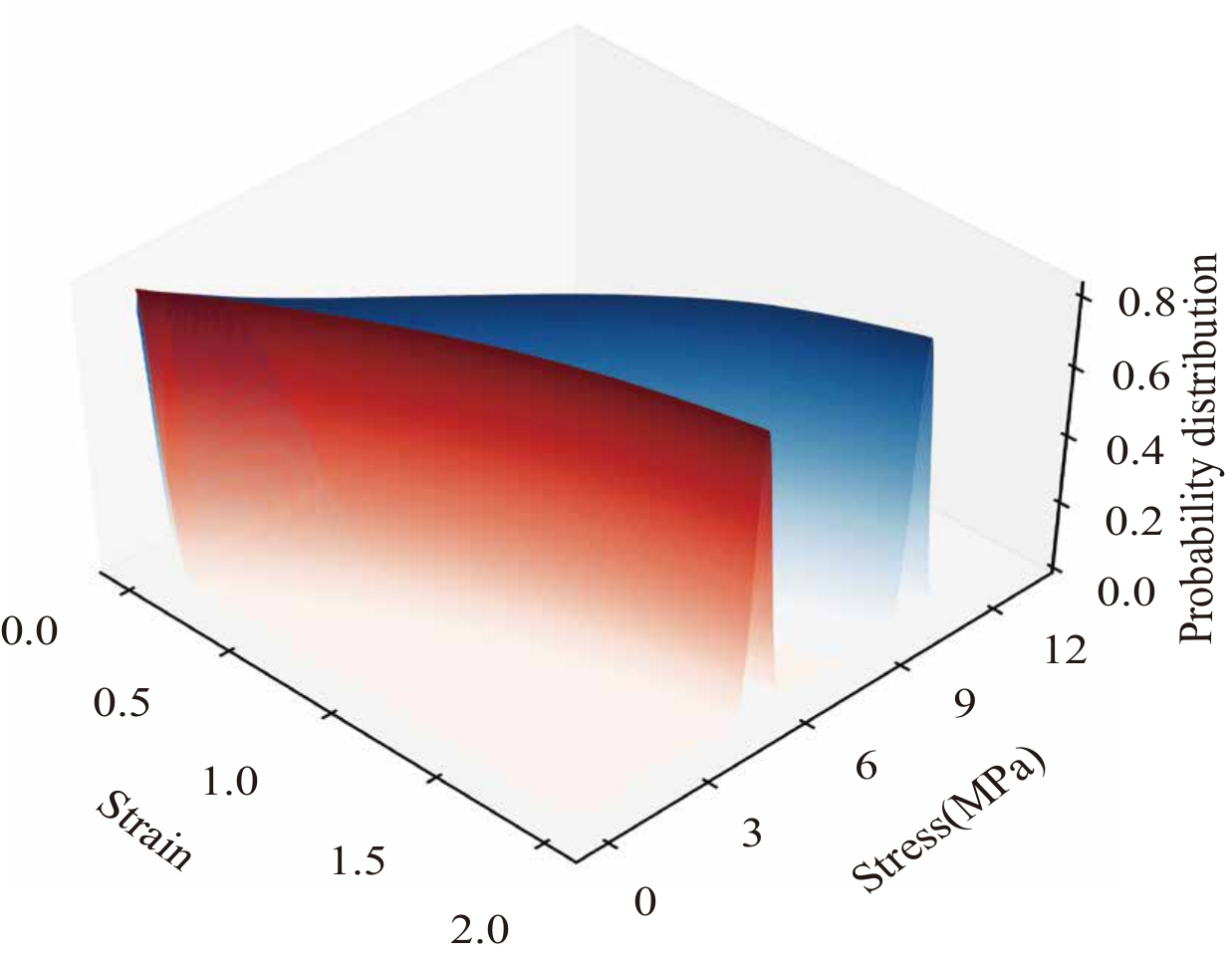}
        \caption{Original probability distribution and the final probability distribution after transportation calculated by CM-GAI.}
        \label{fig:linb}
    \end{subfigure}
    \caption{The comparison between Lin et al.'s experiment \citep{lin2025} and CM-GAI predictions.}
    \label{fig:lin}
\end{figure}

Glycerol gel is commonly used in biomedical applications where deformation can be induced by varying strain rates, such as impact or dynamic loading. The strain rate-dependent stress - strain behavior of glycerol gel has been studied \citep{lin2025}. Based on their work, CM-GAI used to predict the stress - strain response of glycerol gel at a target strain rate of $50 \text{min}^{-1}$, using data from lower strain rate conditions ($0.5 \text{min}^{-1}, 1 \text{min}^{-1}, 5 \text{min}^{-1}, 10 \text{min}^{-1}$), as shown in Figure \ref{fig:lin}.

\begin{figure}[H]
    \centering
    \begin{subfigure}[t]{0.45\textwidth}
        \centering
        \includegraphics[width=\textwidth]{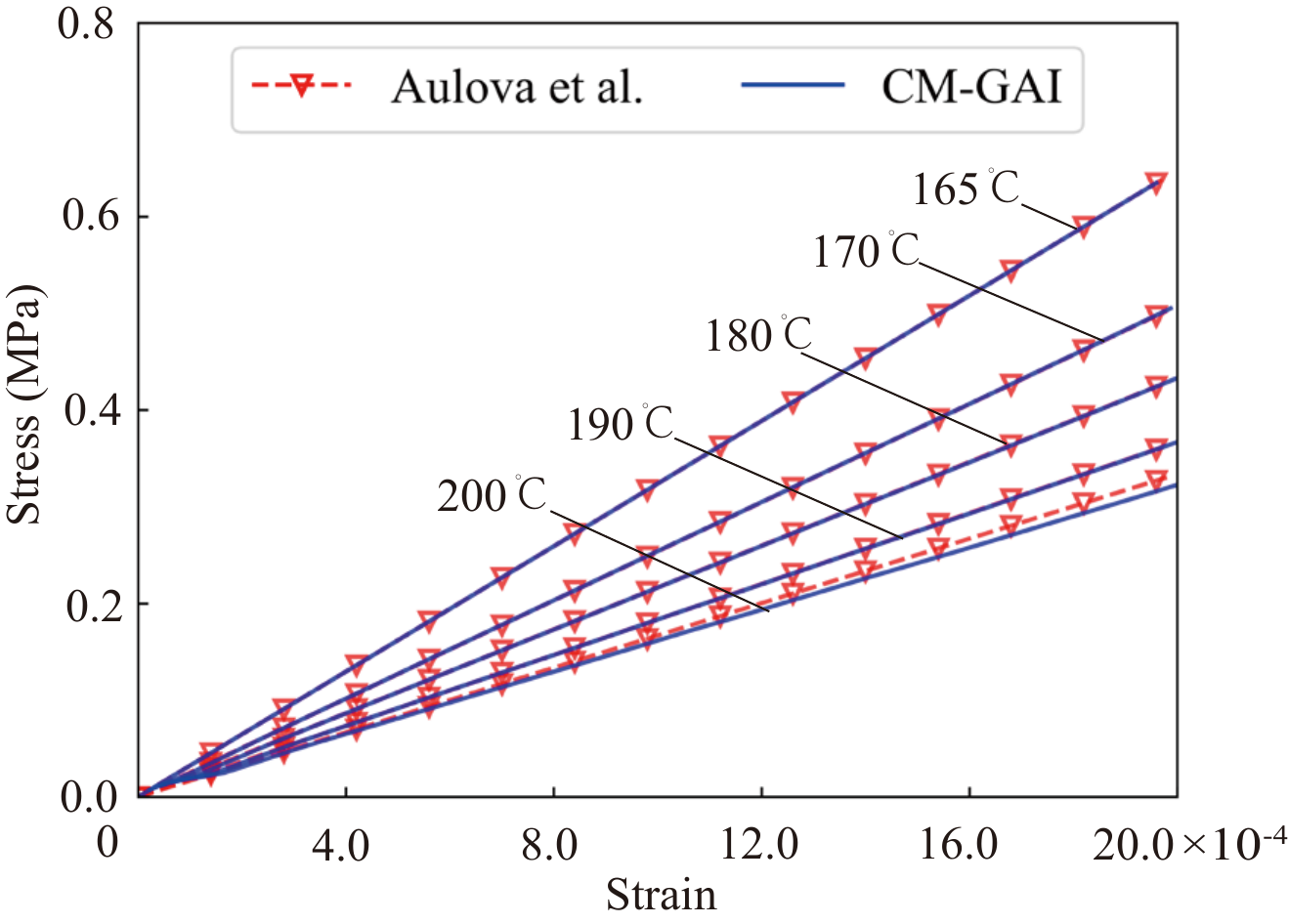}
        \caption{The stress-strain curves at the different temperatures.}
        \label{fig:aulovaa}
    \end{subfigure}
    \begin{subfigure}[t]{0.45\textwidth}
        \centering
        \includegraphics[width=\textwidth]{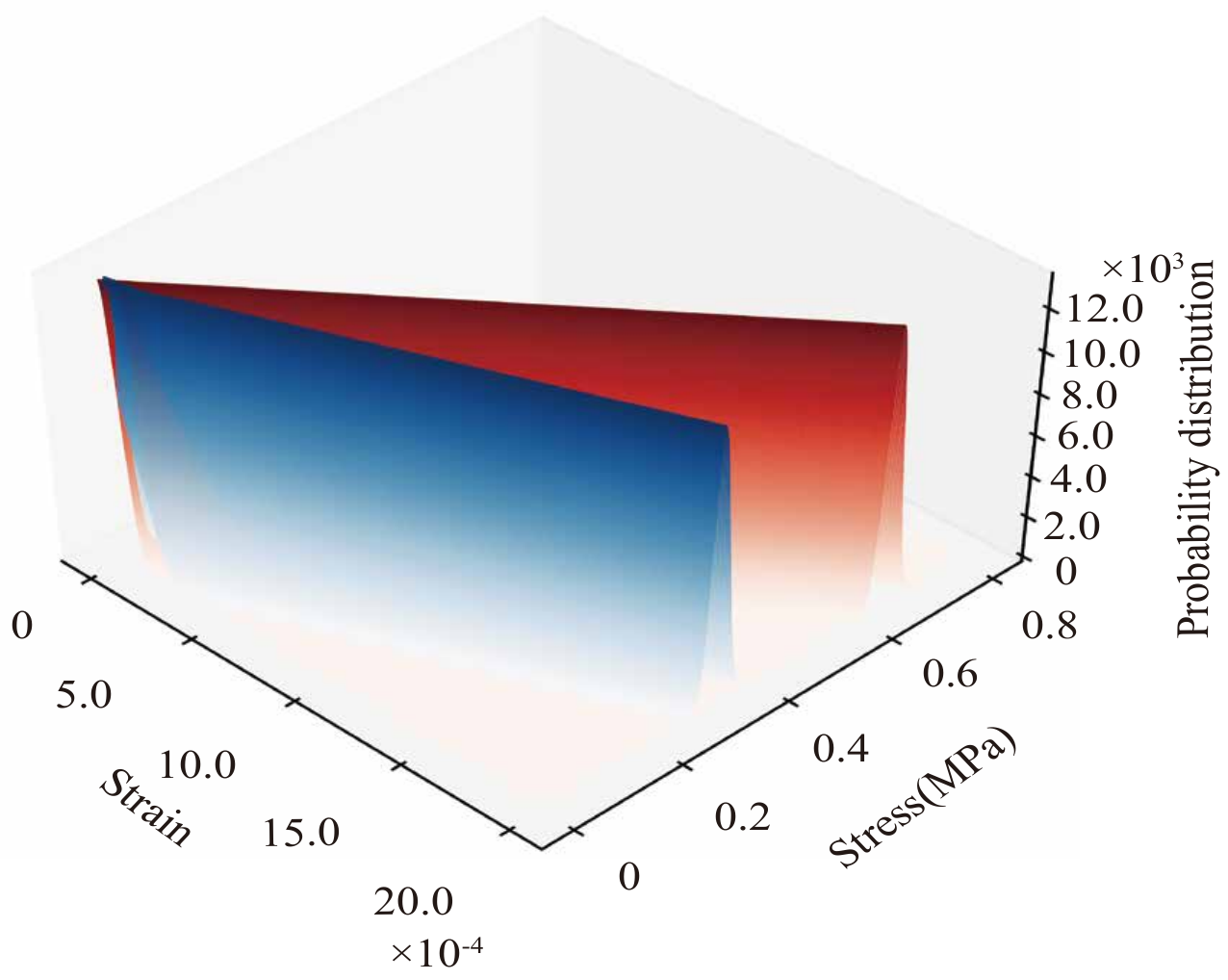}
        \caption{Original probability distribution and the final probability distribution after transportation calculated by CM-GAI.}
        \label{fig:aulovab}
    \end{subfigure}
    \caption{The comparison between Aulova et al.'s experiment \citep{aulova2021} and CM-GAI predictions.}
    \label{fig:aulova}
\end{figure}

Polyether Ether Ketone (PEEK) is widely used in aerospace and biomedical applications due to its excellent thermal and mechanical stability. In these environments, PEEK components often operate across a broad temperature range, which significantly affects their mechanical behavior. The temperature-dependent stress-strain response of PEEK has been studied \citep{aulova2021}. Therefore, in this work, CM-GAI is used to predict the stress - strain response of PEEK at the target temperature of $200~^\circ\mathrm{C}$, using data from other thermal conditions ($165~^\circ\mathrm{C}, 170~^\circ\mathrm{C},180~^\circ\mathrm{C},190~^\circ\mathrm{C}$). The results obtained are shown in Figure \ref{fig:aulova}.

The normalized root mean square error of stress-strain generation for all the cases are summarized in Table 3.
\begin{table}[H]
\centering
\caption{The NRMSE values between the generated stress-strain curves and the target stress-strain curves}
\label{tab:nrmse_simple}
\begin{tabular}{lccc}
\toprule
\textbf{Case} & \textbf{NRMSE}\\
\midrule
Adhesive & 7.99\%\\
Glassy polymers & 1.92\%\\
Polymer foams & 3.21\%\\
Concrete material & 3.06\%\\
High-density polyethylene & 3.02\%\\
Nanotwinned copper (strain-rate dependent) & 1.16\%\\
Nanotwinned copper (temperature dependent) & 0.84\%\\
Glycerol gel & 2.46\%\\
Polyether ether ketone & 1.23\%\\\bottomrule
\end{tabular}
\end{table}

\section*{Appendix E: Comparison of stress-strain relationship prediction based on Functional Principal Component Analysis and Gaussian Process Regression (FPCA-GPR) with CM-GAI}

For the prediction of function curves under specific working conditions, functional principal component analysis (FPCA) \citep{rao1958some} achieves dimensionality reduction and reconstruction by extracting the common modes of variation from a set of known curves. In contrast, Gaussian process regression (GPR) \citep{williams1995gaussian} directly establishes a nonlinear mapping from working condition parameters to the entire function curve by defining kernel functions and priors in the function space. Combining these two regression methods can effectively accomplish this task. Therefore, we attempt to employ FPCA and GPR to predict stress-strain relationships under different temperatures or strain rates.

This appendix compares the performance of a stress-strain relationship prediction method based on FPCA and GPR with CM-GAI, using the two aforementioned case studies. The core of the FPCA-GPR method lies in mapping experimental data onto a common strain grid, extracting the principal feature patterns of the curves through FPCA, and then establishing a mapping relationship between the principal component coefficients and temperature using GPR. This method not only enables the prediction of stress-strain responses at specific temperatures based on sparse temperature data but also provides uncertainty quantification for the prediction results.  

The method primarily consists of four steps.  

First, data preprocessing is conducted. A common strain grid \(\{\varepsilon_j\}_{j=1}^m\) is defined, covering the range from zero to the maximum strain. The original stress-strain curves at each temperature point \(T_i\ (i=1,2,\dots,n_T)\) are interpolated onto this grid to obtain the corresponding stress values.  

Second, functional basis decomposition is performed. The interpolated curves are constructed into an observation matrix \(\mathbf{Y} \in \mathbb{R}^{m \times n_T}\), where each column corresponds to a discrete curve at a specific temperature. FPCA is applied to the strain dimension of \(\mathbf{Y}\), and dimensionality reduction is achieved through Singular Value Decomposition (SVD) \citep{van1996matrix}:  
\begin{equation}
\mathbf{Y} \approx \sum_{k=1}^r a_k \Phi_k
\end{equation} 
Here, \(\Phi_k(\varepsilon)\) represents the \(k\)-th principal component mode, describing the variation pattern of the curves. \(a_k\) is the corresponding coefficient, reflecting the influence of temperature on this mode. The number of modes that capture over 99\% of the variance is selected to achieve effective dimensionality reduction of the data.  

Next, coefficient-temperature mapping is established. GPR is employed to build a mapping relationship between each principal component coefficient \(a_k\) and temperature \(T\). The statistical model is as follows:  
\begin{equation}
a_k = f_k(T) + \epsilon, \quad \text{where} \quad \epsilon \sim \mathcal{N}(0, \sigma_{\text{noise}}^2)
\end{equation}
The function \(f_k(T)\) is modeled as a Gaussian process:  
\begin{equation}
f_k(T) \sim \mathcal{GP}\big(0, k(T, T')\big)
\end{equation}  
Here,$\sim$ means follows the distribution. \(T\) and \(T'\) are any two temperature input points, and \(k(T, T')\) is the covariance function defining the correlation between function values at different temperature points. \(\epsilon\) is an observation noise term independent of \(f_k(T)\), and its variance \(\sigma_{\text{noise}}^2\) is estimated during the training process. This process learns a smooth mapping between coefficients and temperature within a Bayesian framework.  

Finally, stress-strain curve prediction at specific temperatures is performed. For a new temperature \(\hat{T}\), GPR provides the posterior distribution of the coefficients \(a_k(\hat{T})\), which is a Gaussian distribution. The mean \(\hat{a}_k(\hat{T})\) serves as the predicted value, and the variance \(\hat{\sigma}^2_k(\hat{T})\) quantifies the prediction uncertainty. The stress-strain curve is then reconstructed as follows:  
\begin{equation}
\hat{\sigma}(\varepsilon, \hat{T}) = \sum_{k=1}^r \hat{a}_k(\hat{T}) \Phi_k(\varepsilon)
\end{equation}  

To systematically compare the performance of CM-GAI with FPCA-GPR, tests were conducted on the two aforementioned case studies.  

The first case study utilized data from the temperature-dependent deformation behavior of glassy polymers measured by Clarijs and Govaert \citep{clarijs2019}. The training set included stress-strain curves at 5 temperature points ($T= -25^{\circ} $C, $0^{\circ}$C, $22^{\circ}$C, $50^{\circ}$C, $100^{\circ}$C), and the test set included one temperature point ($T = 150^{\circ}$C). The second case study was based on data from the temperature-dependent stress-strain behavior of ntCu studied by Guo et al \citep{guo2020}. The training set included stress-strain curves at 4 temperature points ($T = 200$K, $300$K, $400$K), and the test set included one temperature point ($T = 500$K).

\begin{figure}[htbp]
	\centering
	\begin{minipage}[t]{0.45\textwidth}
		\centering
		\includegraphics[width=\textwidth]{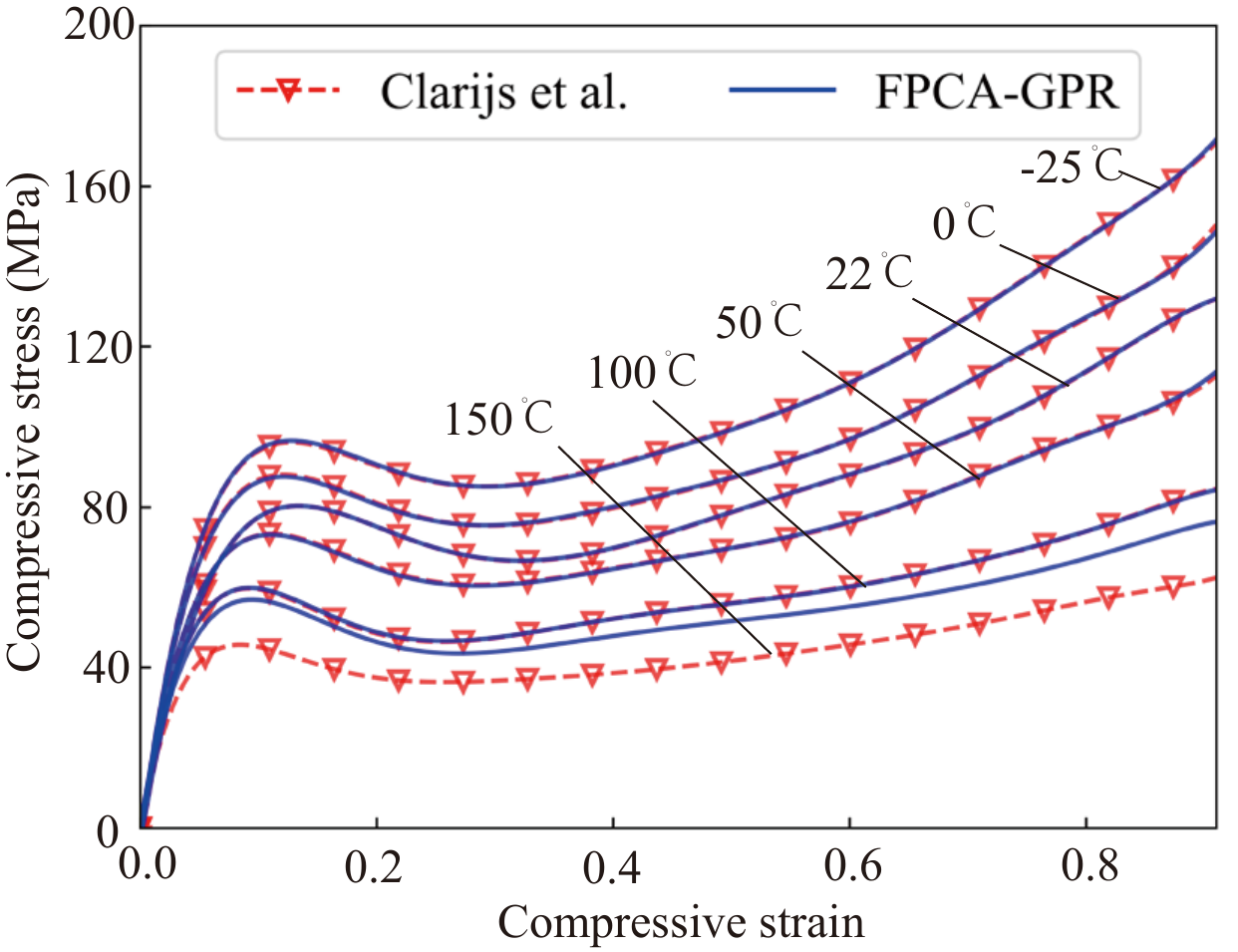}
		\setlength{\abovecaptionskip}{1mm}
		\setlength{\belowcaptionskip}{-2mm}
		\caption{The comparison between Clarijs et al.'s experiment \citep{clarijs2019} and FPCA-GPR predictions}
		\label{fig:Fig 20}
	\end{minipage}
	\hfill 
	\begin{minipage}[t]{0.45\textwidth}
		\centering
		\includegraphics[width=\textwidth]{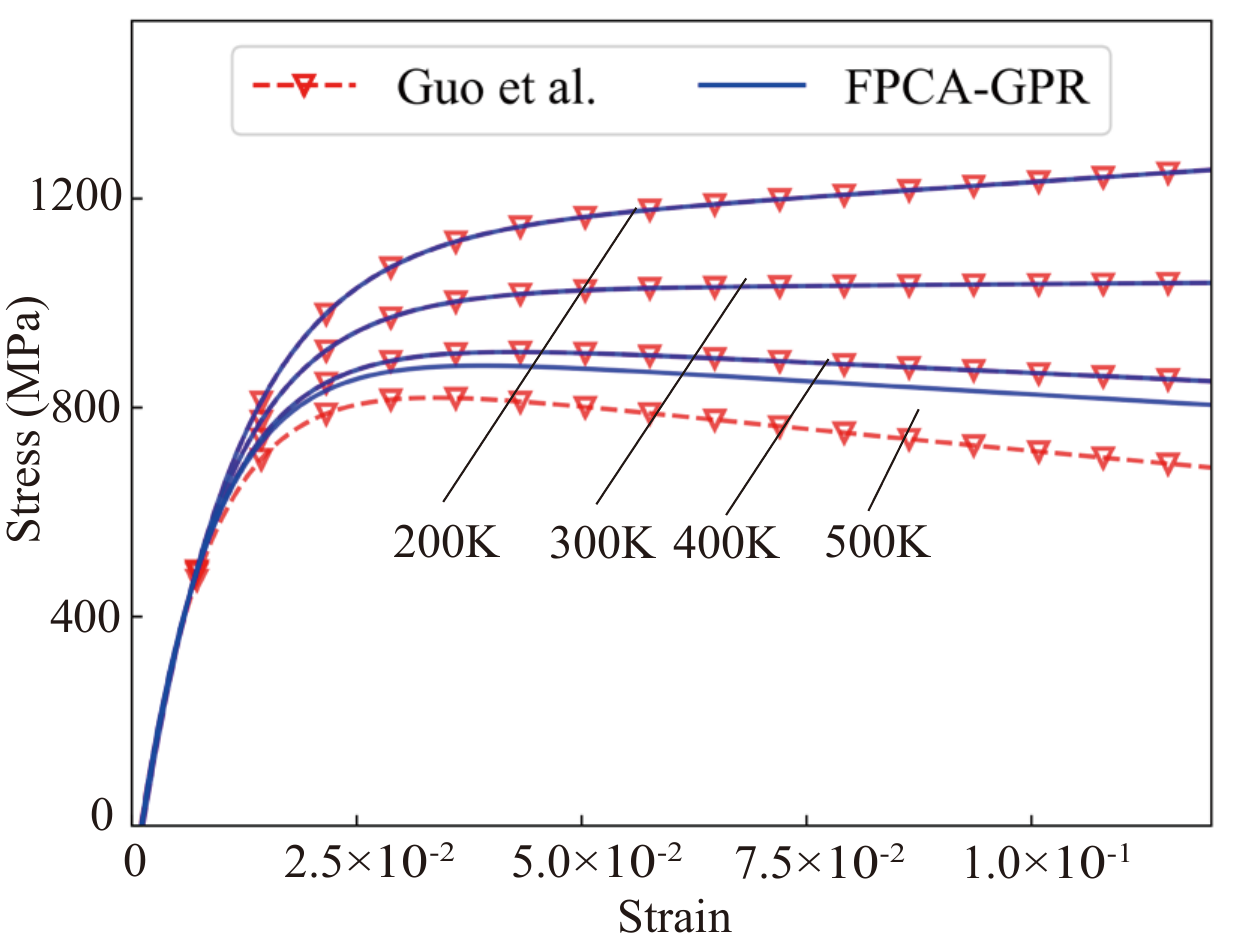}
		\setlength{\abovecaptionskip}{1mm}
		\setlength{\belowcaptionskip}{-2mm}
		\caption{The comparison between Guo et al.'s model \citep{guo2020} and FPCA-GPR predictions}
		\label{fig:Fig 21}
	\end{minipage}
\end{figure}

Figure \ref{fig:Fig 20} and Figure \ref{fig:Fig 21} present the prediction results of the FPCA-GPR method at representative test temperatures ($T = 150^\circ$C and $T = 500$K) for the two case studies, respectively.

It can be observed that, in both cases, the FPCA-GPR prediction curves exhibit noticeable deviations from the true curves. In contrast, CM-GAI (as shown in Figures \ref{fig:clarijs} and \ref{fig:guo2}) demonstrates excellent agreement with the true curves in both case studies.

We used the full-field NRMSE as evaluation metrics. In the two case studies, the NRMSE values of FPCA-GPR are 15.67\% and 9.98\%, respectively, compared to the NRMSE values of CM-GAI - 1.92\% and 0.84\%, respectively. From these data, we can see that CM-GAI outperforms FPCA-GPR in prediction accuracy in these two cases.

The primary reason for this performance disparity lies in the difficulty for GPR to accurately learn the complex mapping relationship between the principal component coefficients and temperature when training data are scarce. Although GPR can maintain a certain level of accuracy for training data, its reliability in predicting coefficients deteriorates significantly during extrapolation. The predicted means tend to revert towards the prior mean, and the associated uncertainty increases drastically. This systematic bias in coefficient prediction leads to reconstructed stress-strain curves that, while capturing the general shape patterns, exhibit a clear "drift" phenomenon along the stress axis. More importantly, this method suffers from poor interpret-ability. Although it can provide prediction results and uncertainty estimates, it fails to clearly reveal the physical mechanisms through which temperature changes specifically affect each principal component coefficient. The complex transformation process of the kernel function and the abstract mathematical meaning of the principal component coefficients make the model's decision-making process lack transparency, making it difficult to explain systematically why predictions exhibit systematic deviations in certain temperature ranges from the perspective of physical or mathematical essence. In comparison, CM-GAI does not rely on explicit basis function decomposition and coefficient regression. Instead, it achieves a more precise characterization of the complex temperature-strain-stress relationship through studying the transportation of probability distribution, maintaining superior generalization capability even under data-scarce conditions.

The experimental results indicate that the predictive accuracy of CM-GAI is superior to that of the FPCA-GPR method in the two typical case studies, further validating its advantages in different material. Although the FPCA-GPR framework offers a probabilistic uncertainty quantification capability and can roughly reconstruct the shape characteristics of the curves, its core GPR regression step suffers from significant coefficient prediction bias under data scarcity, which limits its overall performance. CM-GAI provides a more reliable and accurate solution for modeling material behavior in data-scarce scenarios.

\section*{Appendix F: Handwritten digit generation via Probability Flow Ordinary Differential Equations (Probability Flow ODE) and CM-GAI}

This appendix briefly discusses the connection between the score matching and probability flow ODE-based generative modeling method proposed by \citep{song2020score} and the proposed CM-GAI. The probability flow ODE is used to derive the body force term in CM-GAI in the image generation problem (see \citep{song2020score} for details on score matching and probability flow ODE). Furthermore, the finite difference method was employed to solve the kinematic equation (Eq. \ref{equilirium}) derived in this paper. Thereby, it enables efficient sampling using CM-GAI to generate handwritten digit images. Correspondingly, the approximate probability distribution model for such images can also theoretically be obtained through mass conservation equation Eq. \ref{Monge-Ampere Equation-Equi}. This appendix presents the relevant theoretical derivations and showcases the training and generation results on the MNIST dataset, ultimately demonstrating the effectiveness of the proposed method.

Score-based generative modeling is built upon the framework of continuous-time stochastic differential equations (SDEs). Specifically, the forward diffusion process of this method can be described by the following SDE: 
\begin{equation}
dx = f(x,t)dt + g(t)dw
\end{equation}
where \(x\) represents the image data, \(t \in [0,t_f]\) is the time variable, \(f(x,t)\) is the drift coefficient, \(g(t)\) is the diffusion coefficient, and \(dw\) is the increment of standard Brownian motion. This equation describes a process that starts from the original data distribution \(p_0(x)\) and gradually transforms it into a Gaussian distribution by continuously adding noise.

\citep{song2020score} proved that the reverse process of any diffusion process can be represented by a probability flow ODE:  
\begin{equation}
dx = \left[ f(x,t) - \frac{1}{2}g(t)^2 \nabla_x \log p_t(x) \right] dt
\end{equation}

This deterministic ODE reverse process describes a continuous deterministic path from the noise distribution to the data distribution, enabling the use of numerical ODE solvers for sampling, which greatly improves generation efficiency. In the method of this paper, we adopt the variance exploding (VE) SDE configuration, where the drift coefficient \(f(x,t)\equiv 0\) and the diffusion coefficient \(g(t)\) is a hyper-parameter. 

To obtain the complete form of the probability flow ODE, we need the score function, i.e., the gradient of the log probability of the data distribution $\nabla_x \log p_t(x)$. The core idea of score-based generative models is to train a parameterized model $s_{\bm{\theta}}(x,t)$ to estimate the true score function. This model is trained using a denoising score matching objective, with the loss function designed as:  
\begin{equation}
\mathcal{L}(\bm{\theta}) = \mathbb{E}_{t \sim \mathcal{U}(0,t_f]} \mathbb{E}_{x(0) \sim p_0} \mathbb{E}_{x(t) \sim p_{0t}(x(t)|x(0))} \left[ \| s_{\bm{\theta}}(x(t),t) - \nabla_{x(t)} \log p_{0t}(x(t)|x(0)) \|_2^2 \right]
\end{equation}

$\mathcal{U}(0,t_f]$ represents sampling in the uniform distribution over the interval $(0,t_f]$. Under the VE configuration, the forward SDE admits an analytic solution, and the conditional transition probability is given by:
\begin{equation}
p_{0t}(x(t)|x(0)) = \mathcal{N}\left(x(t); x(0), \sigma(t)^2 I\right)
\end{equation}
where $\sigma(t) = \sqrt{\int_0^t g(s)^2 ds}$ represents the cumulative noise intensity, and $I$ is the identity matrix with the same dimensionality as the data.

In practical implementation, considering that \(\nabla_{x(t)} \log p_{0t}(x(t)|x(0)) = -z/\sigma(t)\), where \(z \sim \mathcal{N}(0,I)\), this loss function simplifies to: 
\begin{equation}
\mathcal{L}(\bm{\theta}) = \mathbb{E}_{t,x(0),z} \left[ \| \sigma(t)s_{\bm{\theta}}(x(0) + \sigma(t)z, t) + z \|_2^2 \right]
\end{equation}

The theory related to score matching and probability flow ODE has been clarified. Next, we will elucidate the connection between the probability flow ODE and CM-GAI. The probability flow ODE defines the velocity field governing the transition from noise to meaningful image data, while in CM-GAI, the equation of motion Eq. \ref{equilirium} defines the body force field in the transportation of data probability, which is the acceleration field per unit mass. To obtain the body force field in the picture generation problem using the present method, we take the total derivative with respect to time of probability flow ODE, thereby deriving the body force field in the equation of motion. The specific formulation is as follows:
\begin{equation}
\frac{\partial u(x(0),t)}{\partial t}=\frac{\partial (x(0)+u(x(0),t))}{\partial t}=\frac{\partial x(x(0),t)}{\partial t}=\frac{dx(t)}{dt}=v(x,t)
={f(x, t)}-\frac{1}{2} {g(t)^2}\cdot{\nabla_x \log p_t(x)}
\end{equation}
\begin{eqnarray}
      &&\frac{\partial ^2u(x(0),t)}{dt^2}
      =\frac{dv(x,t)}{dt}
      = \frac{\partial}{\partial t}\left[ f(x, t) - \frac{1}{2} g(t)^2 \nabla_x \log p_t(x) \right]\nonumber\\
      &+&\left( \frac{\partial}{\partial x}\left[ f(x, t) - \frac{1}{2} g(t)^2 \nabla_x \log p_t(x) \right] \right) \cdot \left( f(x, t) - \frac{1}{2} g(t)^2 \nabla_x \log p_t(x) \right)
\end{eqnarray}

Consequently, the body force field guiding the transportation of data distribution in image generation is given by the right-hand side of the above equation, highlighting the connection between the proposed method and probability flow ODE.

Based on the theoretical framework presented in this paper, the finite difference method needs to be employed to inversely solve the kinetic equation (Eq. \ref{equilirium}) for image generation. First, initial noise is sampled from a standard Gaussian distribution (since the method proposed in this paper defines a deterministic path between noise and meaningful clear images, each noisy image $x(t_f)$ corresponds to a clear image $x(0)$). Then, the time interval $[t_f, \epsilon]$ (where $\epsilon$ is a small positive quantity) is discretized into $N=100$ steps, with $\Delta t = (t_f - \epsilon)/N$. For each time step $t_i$, differentiate the probability flow ODE in order to calculate the current value of body force field (Since calculating derivatives in high-dimensional spaces is computationally complex, this derivative is also approximated using the finite difference method.). Then, the state at the previous time step is calculated by reverse iteration based on the iterative formula of the discretized dynamic equations Eq.18. The specific iterative formula is as follows:

\begin{equation}
u(x(t_f),t-\Delta t) = 2u(x(t_f),t) - u(x(t_f),t+\Delta t) + \frac{\Delta t}{2} (v(x(t+\Delta t),t+\Delta t) - v(x(t-\Delta t),t-\Delta t))
\end{equation}

After obtaining the displacement field at each time step, the image can be calculated using the following formula:

\begin{equation}
x(\epsilon)=x(t_f)-u(t_f)
\end{equation}

\begin{figure}[H]
	\centering
	\includegraphics[width=0.45\textwidth]{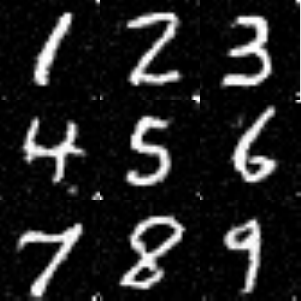}
	\setlength{\abovecaptionskip}{1mm}
	\setlength{\belowcaptionskip}{-2mm}
	\caption{The handwritten digit generation via Probability Flow ODE and CM-GAI.}
	\label{fig:digits}
\end{figure}


Numerical experiments were conducted on the MNIST handwritten digit dataset, which contains 60,000 $28 \times 28$ grayscale training images. In the training process, the loss function decreases steadily with the number of training epochs, without the appearance of mode collapse or training divergence. The generated digit images shown in Figure \ref{fig:digits} exhibit sharp edges and structurally coherent shapes, demonstrating the effectiveness of the proposed method.

This appendix derives the body force term using probability flow ODE and provides an image generation framework based on second-order partial differential equations by solving the kinetic equations formulated in this paper. Experimental analysis demonstrates that the proposed second-order kinetic equation method, similar to probability flow ODE, can effectively generate image, offering an alternative form of sampling that strikes a good balance between generation quality and efficiency.

\section*{Appendix G: Finite element parameter settings of the examples in result and discussion}
\noindent\textit{G.1 Temperature-dependent response simulation of a cantilever beam}

\begin{figure}[H]
	\centering
	\includegraphics[width=1\textwidth]{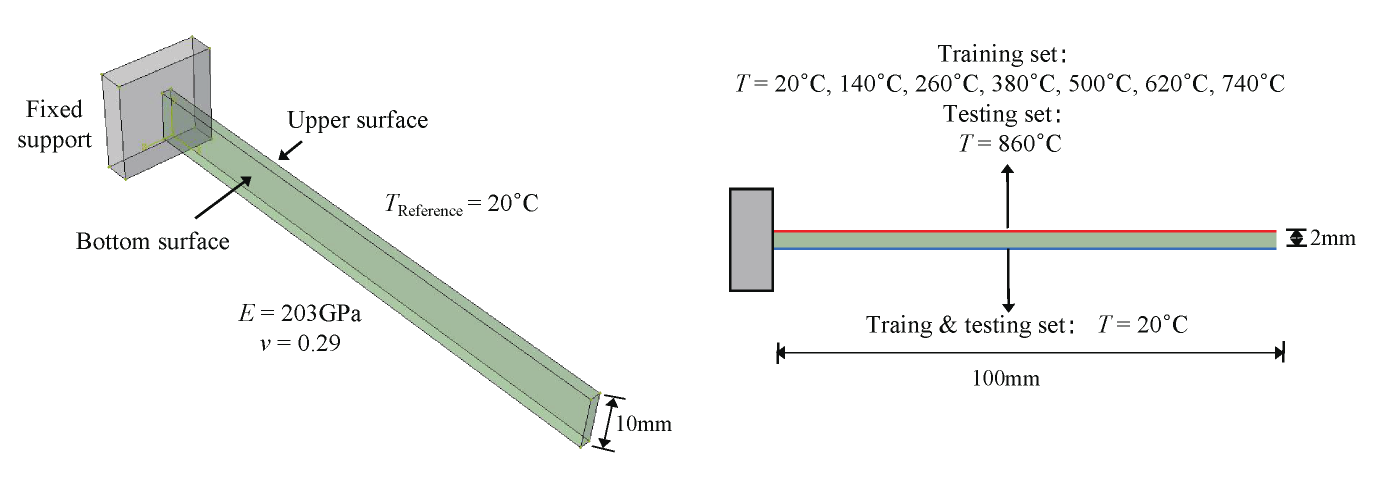}
	\setlength{\abovecaptionskip}{1mm}
	\setlength{\belowcaptionskip}{-2mm}
	\caption{Schematic diagram of the model geometry, materials, and boundary conditions.}
	\label{fig:geomodel}
\end{figure}

The specific finite element settings for this example are as follows. First, the geometric model of the problem needs to be defined. Specifically, the cantilever beam has geometric dimensions of 100 mm × 10 mm × 2 mm, with one end fixed, as shown in Figure \ref{fig:geomodel}. The key material parameters are set as follows: elastic modulus $E$=203GPa, Poisson’s ratio $v$=0.29, thermal expansion coefficient $\alpha$ shown in Table \ref{tab:cte_data}, and reference temperature 20°C. Temperature boundary conditions of 860°C and 20°C are applied on the upper and lower surfaces, respectively. 

\begin{table}[H]
\centering
\caption{Thermal expansion coefficient varies with temperature}
\label{tab:cte_data}
\begin{tabular}{cc}
\toprule
$T$ ($^\circ\mathrm{C}$) & $\alpha$ ($\times 10^{-6}/^\circ\mathrm{C}$) \\
\midrule
20.5 & 10.3 \\
70.5  & 11.8 \\
120.5  & 13.3 \\
170.5  & 14.7 \\
220.5  & 16.0 \\
270.5    & 17.2 \\
320.5   & 18.3 \\
370.5   & 19.3 \\
420.5   & 20.2 \\
470.5   & 21.2 \\
520.5  & 22.3 \\
570.5  & 23.6 \\
620.5  & 25.1 \\
670.5  & 27.2 \\
720.5  & 30.2 \\
770.5  & 34.6 \\
820.5  & 40.9 \\
\bottomrule
\end{tabular}
\end{table}

Regarding the mesh generation, the cantilever beam was discretized using 8-node reduced integration elements (C3D8R), resulting in a model with 2500 elements. 

The analysis was performed using the ABAQUS/Standard solver. The analysis type was a static, general analysis considering geometric nonlinearity to accurately simulate the potential large-displacement response of the structure. A fixed time increment solution control strategy was employed. The total analysis time was set to 1, and the full load was applied progressively through 100 equal increments to complete the simulation of the entire loading process.

Considering geometric nonlinearity due to large deformation, a finite element solution is performed, and the full-field displacement responses in three directions at all nodes were extracted as the mean of the probability distribution of the training set, with a series of temperatures of 20°C, 140°C, 260°C, 380°C, 500°C, 620°C, and 740°C applied on the upper surface respectively. The full-field displacement responses in three directions at all nodes were extracted as the mean of the probability distribution of the testing set, with temperature of 860°C applied on the upper surface.

\vspace{\baselineskip} 
\noindent\textit{G.2 Nonlinear transient dynamic simulation of Taylor rod's high-velocity impact on a rigid wall.}

First, the problem is defined as the simulation of a cylindrical copper rod impacting a rigid wall axially at various initial velocities \citep{abaqus2016}. This problem involves high strain rates and large plastic deformation, serving as a typical benchmark for validating material constitutive models and element formulations.

\begin{figure}[H]
	\centering
	\includegraphics[width=0.7\textwidth]{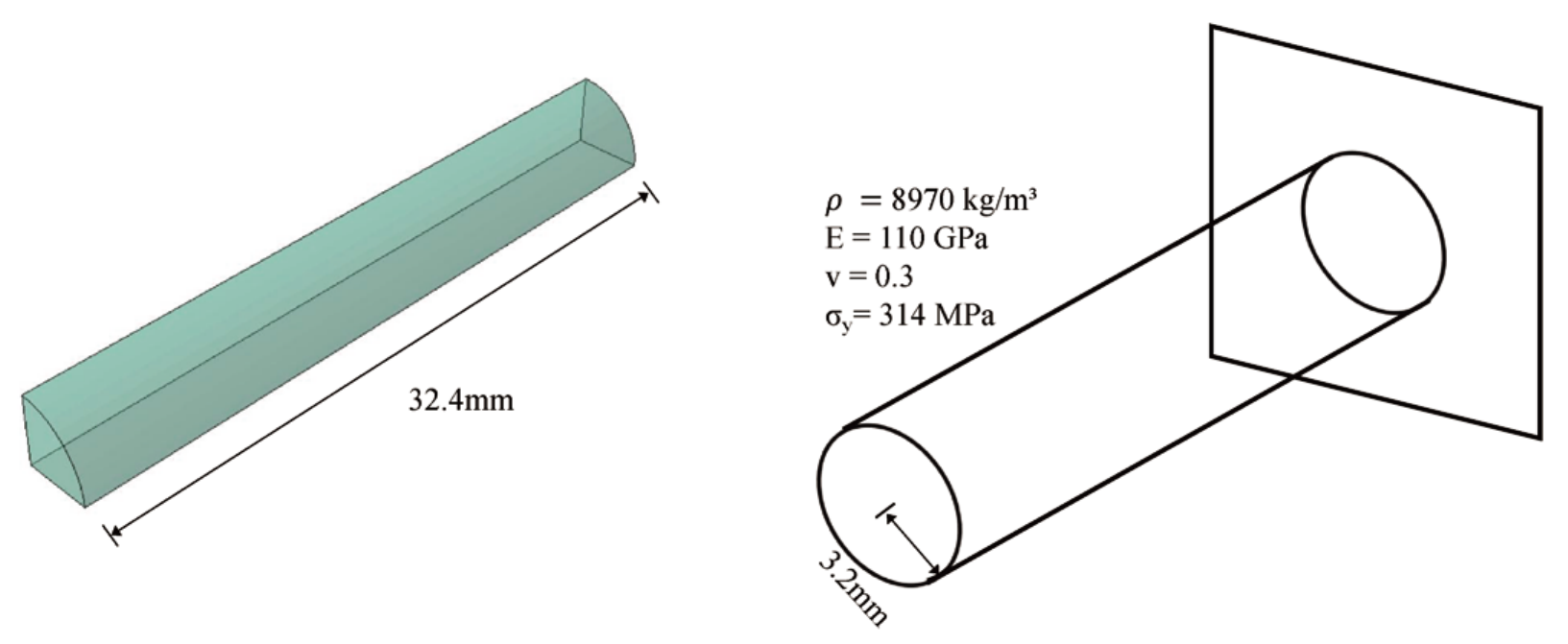}
	\setlength{\abovecaptionskip}{1mm}
	\setlength{\belowcaptionskip}{-2mm}
	\caption{Schematic diagram of the model geometry, materials, and boundary conditions.}
	\label{fig:rodgeomodel}
\end{figure}

The specific finite element settings for this example are as follows. The geometry is a standard cylindrical rod with a length of 32.4 mm and a radius of 3.2 mm (Figure \ref{fig:rodgeomodel}). To improve computational efficiency, a quarter-symmetry model is established by exploiting its two symmetry planes. This quadrant is spatially discretized using C3D8R (8-node linear reduced-integration hexahedron) elements, with a total of 2700 elements in a structured mesh to ensure good initial quality.

The material model selected is the classical elastic-perfectly plastic von Mises model to simulate the behavior of copper. The specific parameters are: density 8970 kg/m³, Young's modulus 110 GPa, Poisson's ratio 0.3, and a constant yield stress of 314 MPa.

The analysis type is an explicit dynamic analysis, with a total duration of 80 microseconds to fully capture the post-impact dynamics. A core aspect of the element configuration in this example is the modification of the section controls: the orthogonal kinematic formulation combined with the combined hourglass control is adopted, aiming to effectively suppress hourglass modes while maintaining computational efficiency.

The application of boundary conditions and loads must accurately simulate the actual physical process. Normal displacement constraints are applied on the two symmetry planes to enforce symmetry. On all nodes at the rod end representing the impact with the rigid wall, the axial displacement is fully constrained (U1=0). The most critical load is the initial velocity field applied as an initial condition: all model nodes, except those at the impact end, are assigned an axial initial velocity of a specified magnitude directed towards the wall.

Upon completion of the finite element computation, the PEEQ at all nodes are extracted under the impact velocity of 0m/s, 50m/s, 100m/s, 150m/s, 200m/s, 250m/s and 300m/s as the mean of the training set data probability distribution, and the impact velocity of 350m/s as the mean of the testing set data probability distribution, respectively.

\bibliographystyle{elsarticle-harv}
\biboptions{round,sort&compress}
\addcontentsline{toc}{section}{References}
\bibliography{reference}

\end{document}